\documentclass{aa}

\usepackage[normalem]{ulem}
\usepackage{graphicx}
\begin{document}

%
\def\cm{\,{\rm cm}}
\def\cmcube{\,{\rm cm^{-3}}}
\def\kms{\,{\rm km\,s^{-1}}}
\def\Jy{\,{\rm Jy}}
\def\Jyb{\,{\rm Jy/beam}}
\def\mJy{\,{\rm mJy}}
\def\muJy{\,\mu{\rm mJy}}
\def\mJyb{\,{\rm mJy/beam}}
\def\muJyb{\,\mu{\rm mJy/beam}}
\def\K{\,{\rm K}}
\def\kpc{\,{\rm kpc}}
\def\Mpc{\,{\rm Mpc}}
\def\mkG{\,\mu{\rm G}}
\def\MHz{\, {\rm MHz}}
\def\GHz{\, {\rm GHz}}
\def\pc{\,{\rm pc}}
\def\kpc{\,{\rm kpc}}
\def\Mpc{\,{\rm Mpc}}
\def\HI{\rm H\,{\scriptstyle I}}
\def\HII{\rm H\,{\scriptstyle II}}
\def\nelectron{n_{\rm e}}
\def\RM{{\rm RM}}
\def\radm{\,{\rm rad\,m^{-2}}}
\def\s{\,{\rm s}}
\def\yr{\,{\rm yr}}
\def\pheins{\phantom{1}}

\title{Magnetic fields in barred galaxies}
       \subtitle{I. The atlas}

\author{ R. Beck\inst{1}, V. Shoutenkov\inst{2}, M. Ehle\inst{3,4},
         J.~I. Harnett\inst{5}, R.~F. Haynes\inst{6},
         A. Shukurov\inst{7}, D.~D. Sokoloff\,\inst{8},\\
         M. Thierbach\inst{1} }

\offprints{R. Beck}

\institute{
         Max-Planck-Institut f\"ur Radioastronomie,
             Auf dem H\"ugel 69,
             D-53121 Bonn, Germany
\and     Pushchino Radioastronomy Observatory, Astro Space Center,
             142292 Pushchino, Russia, \& Isaac Newton Institute of Chile,
             Pushchino Branch
\and     XMM-Newton Science Operations Centre, Apdo. 50727,
             28080 Madrid, Spain
\and     Research and Scientific Support Department of ESA, ESTEC, Postbus 299,
             2200 AG Noordwijk, The Netherlands
\and     University of Technology Sydney, P.O. Box 123, Broadway 2007,
             NSW, Australia
\and     School of Mathematics and Physics, University of Tasmania,
         GPO Box 252-21, Hobart, Tas 7001, Australia
\and     Department of Mathematics, University of Newcastle,
             Newcastle upon Tyne NE1 7RU, UK
\and     Department of Physics, Moscow University,
             119899 Moscow, Russia
}

%

\date{Received 13 Feb 2002 / Accepted 29 April 2002}

\abstract{
The total and polarized radio continuum emission of 20 barred galaxies was
observed with the Very Large Array (VLA) at $\lambda$3, 6, 18 and 22~cm and
with the Australia Telescope Compact Array (ATCA) at $\lambda$6~cm and
13~cm. Maps at 30\arcsec\ angular resolution are presented here.
Polarized emission (and therefore a large-scale regular magnetic field)
was detected in 17 galaxies. Most galaxies of our sample are similar to
non-barred galaxies with respect to the radio/far-infrared flux correlation
and equipartition strength of the total magnetic field.
Galaxies with highly elongated bars are not always radio-bright.
We discuss the correlation of radio properties with the aspect ratio
of the bar and other measures of the bar strength. We introduce a new measure
of the bar strength, $\Lambda$, related to the quadrupole moment of the bar's
gravitational potential. The radio surface brightness {\rm I} of the
barred galaxies in our sample is correlated with $\Lambda$,
$I\propto\Lambda^{0.4\pm0.1}$, and thus is highest in galaxies with a long bar
where the velocity field is distorted by the bar over a large fraction of the disc.
In these galaxies, the pattern of the regular field is significantly
different from that in non-barred galaxies. In particular,
field enhancements occur upstream of the dust lanes
where the field lines are oriented at large angles to the bar's major axis.
Polarized radio emission seems to be a good indicator of large-scale
non-axisymmetric motions.
 }

\titlerunning{Magnetic fields in barred galaxies. I. The atlas}
\authorrunning{R. Beck et al.}

\maketitle
%

\section{Introduction}

Polarization observations in radio continuum have revealed basic
properties of interstellar magnetic fields
in a few dozen spiral galaxies (Beck et al.\ \cite{beck+96},
Beck\ \cite{beck00}). Large-scale regular fields form spiral patterns
with pitch angles similar to those of the optical spiral arms.
The strongest {\it regular\/} fields usually occur between the optical arms,
sometimes concentrated in `magnetic arms' (Beck \& Hoernes \cite{beck+hoernes96}).
The {\it total} (= polarized + unpolarized) nonthermal (synchrotron)
radio emission is
a tracer of the {\it total\/} field which comprises both regular and random
field components. It generally
peaks on the optical arms because the random field is strongest there.
This distinction implies that the regular and random magnetic
fields are maintained and affected by different physical processes.

Spiral patterns of the regular magnetic field are believed to be
generated by dynamo action in a differentially rotating disc
(Beck et al.\ \cite{beck+96}). The dynamo reacts or interacts with
non-axisymmetric disturbances like density waves (Mestel \& Subramanian\
\cite{mestel+91}, Rohde et al.\ \cite{rohde+99}), but little is known about the
effects of bar-like distortions.
Chiba \& Lesch (\cite{chiba+lesch94})
suggested that a bar may excite higher dynamo modes, while Moss et al.
(\cite{moss+98}) found from their models a mixture of modes
with rapidly changing appearance.

Radio observations of barred galaxies are rare. The angular resolution
of the maps in Condon's (\cite{condon87}) atlas was
insufficient to distinguish emission from the bar, the spiral arms and
the halo. Another survey of barred galaxies in radio continuum by
Garc\'{\i}a-Barreto et al.\ (\cite{garcia+93}) had even lower angular
resolution; neither survey included polarization. The first
high-resolution radio map of a barred galaxy, NGC~1097 (Ondrechen \&
van der Hulst\ \cite{ondrechen+83}), showed narrow ridges
in total intensity coinciding
with the dust lanes, which are tracers of compression regions along the leading
(with respect to the sense of rotation) edge of the bar. A similar
result was obtained for M83 (Ondrechen\ \cite{ondrechen85})
which hosts a bar of smaller size than NGC~1097. No polarization
could be detected in NGC~1097 by Ondrechen \& van der Hulst
(\cite{ondrechen+83}).
Radio observations of NGC~1365 at $\lambda\lambda20,\ 6$ and $2\cm$,
restricted to a central region, have revealed similar features
(J\"ors\"ater \& van Moorsel\ \cite{joersaeter+moorsel95}). The first
detection of polarized radio emission from the bar region was
reported by Ondrechen (\cite{ondrechen85}) for M83, with a mean
fractional polarization at $\lambda6$~cm of 25\%.  Neininger et al.\
(\cite{neininger+91}) mapped the polarized emission from M83 at
$\lambda2.8$\,cm. They showed that the regular magnetic field in the
bar region is aligned with the bar's major axis.  Observed with
higher resolution, the regular field is strongest at the leading
edges of the bar of M83 (Beck\ \cite{beck00}).

Another barred galaxy which has been studied in detail in radio polarization
is NGC~3627 (Soida et al.\ \cite{soida+01}). The regular field
in the bar region is again
aligned parallel to the bar's major axis, being strongest at the leading
edges of the bar.
However, east of the bar the field behaves anomalously,
forming a `magnetic arm' crossing the gaseous arm.

The first high-resolution polarization observations of a galaxy with
a massive bar, NGC~1097, were presented by Beck et al.
(\cite{beck+99}). The magnetic field lines in and around the bar
appear to follow the velocity field of the gas
expected from a generic gas dynamic model (Athanassoula \cite{atha92}).
The regular magnetic field outside the bar region has a spiral
pattern similar to that seen optically.
A narrow ridge of greatly reduced polarized intensity
indicates the deflection of the field lines in a shear shock
(the dust lane), but the magnetic field lines turn more smoothly
than the gas streamlines (Moss et al.\ \cite{moss+01}, hereafter Paper II).
Velocity fields are available from HI observations only for the
outer parts of NGC~1097 (Ondrechen et al.\ \cite{ondrechen+89})
and from CO observations only for the circumnuclear ring (Gerin et al.\
\cite{gerin+88}).

NGC~1097 is one of the objects in our sample of
barred galaxies observed with the VLA and the ATCA.
In this paper we present the full set of radio maps of our survey,
smoothed to a common resolution of 30\arcsec, and give an
overview of their salient properties.
Higher-resolution maps of NGC~1097, 1365 and 7479 will be
presented and discussed in subsequent papers.
New dynamo models for barred galaxies are discussed in Paper II.
Further details on the magnetic fields in NGC1672, 2442 and 7552
will be given by Harnett et al.\ (\cite{harnett+02},
hereafter Paper III).

\scriptsize
\begin{table*}[htb]
\caption{The VLA sample of barred galaxies}
\label{tab:vla-sample}
\scriptsize
\begin{tabular}{llllllllllllllll} \hline
\noalign{\smallskip}
NGC &Hubble&Lum.&RC3 &R.A. &Dec. &$d_{25}$ &$q_{25}$
    &$v_{\rm GSR}$ &$D$ &$i$ &PA &$b/a$ &$2a/$ &$S_{60\mu \rm m}$
    &$S^{\rm tot}_{20\rm cm}$ \\
    &type&class&class&(2000)&(2000)&[\arcmin ]
    & &[km/s]&[M&[\degr ] &[\degr ] & &$d_{25}$ &[Jy] &[mJy]\\
    &(1)&(1)&(2) &[h m s]&[\degr\ \arcmin\ \arcsec]
    &(2) &(2) &(2) &pc]& & & & &(3) &(4)\\
\noalign{\smallskip}
\hline\noalign{\smallskip}
1097  &SBbc(rs) &I-II &SBS3 &02 46 19.0 &$-$30 16 21 &\pheins 9.3
      &1.48 &        1193 & 16 & 45 & 135  & [0.4] & 0.37 & 45.9 & 415\\
1300  &SBb(s) &I.2     &SBT4 &03 19 40.9 &$-$19 24 41 &\pheins 6.2
      &1.51 &        1496 & 20 & 35 &\pheins  86 & [0.3] & 0.41 &\pheins 2.4 &\pheins  35\\
1365  &SBb(s) &I      &SBS3 &03 33 36.7 &$-$36 08 17 &    11.2
      &1.82 &        1541 & 19 & 40 &\pheins  40  & 0.51 & 0.47 & 78.2 & 530\\
2336  &SBbc(r) &I      &SXR4 &07 27 04.4 &$+$80 10 41 &\pheins 7.1
      &1.82 &        2345 & 31 & 59 & 178  & 0.41 & 0.17 &\pheins 1.0 &\pheins  18\\
3359  &SBc(s) &I.8     &SBT5 &10 46 37.8 &$+$63 13 22 &\pheins 7.2
      &1.66 &        1104 & 15 & 55 & 170  & 0.32 & 0.25 &\pheins 4.1 &\pheins  50\\
3953  &SBbc(r) &I-II  &SBR4 &11 53 49.6 &$+$52 19 39 &\pheins 6.9
      &2.00 &        1122 & 15 & 61 &\pheins  13 & 0.89 & 0.17 &\pheins 2.9 &\pheins 41\\
3992  &SBb(rs) &I      &SBT4 &11 57 36.3 &$+$53 22 31 &\pheins 7.6
      &1.62 &        1121 & 15 & 59 &\pheins  67 & 0.58 & 0.27 &\pheins $\simeq$3 &\pheins 21\\
4535  &SBc(s) &I.3     &SXS5 &12 34 20.4 &$+$08 11 53 &\pheins 7.1
      &1.41 &        1892 & 16 & 26 &\pheins  28  &[0.6] &[0.1] &\pheins 6.5 &\pheins 65\\
5068  &SBc(s) &II-III &SXT6 &13 18 55.4 &$-$21 02 21 &\pheins 7.2
      &1.15 &\pheins  550 &\pheins 7 & 29 & 110  & 0.44 & 0.16 &\pheins 2.3 &\pheins 39\\
7479  &SBbc(s) &I-II  &SBS5 &23 04 57.2 &$+$12 19 18 &\pheins 4.1
      &1.32 &        2544 & 34 & 45 &\pheins  25  & 0.41 & 0.46 & 12.1 & 109 \\
\noalign{\smallskip}
\hline\noalign{\smallskip}
\end{tabular}
\vbox{
\noindent References: (1) Sandage \& Tammann (\cite{sandage+tammann81});
(2) de Vaucouleurs et al.\ (\cite{vaucouleurs+91});
(3) Fullmer \& Lonsdale (\cite{fullmer+lonsdale89});
(4) Condon (\cite{condon87}).
}
\end{table*}

\begin{table*}[htb]
\caption{The ATCA sample of barred galaxies}
\label{tab:atca-sample}
\scriptsize
\begin{tabular}{llllllllllllllll} \hline
\noalign{\smallskip}
NGC &Hubble&Lum.&RC3 &R.A. &Dec. &$d_{25}$ &$q_{25}$
    &$v_{\rm GSR}$ &$D$ &$i$ &PA &$b/a$ &$2a/$ &$S_{60\mu \rm m}$
    &$S^{\rm tot}_{6\rm cm}$ \\
    &type&class&class&(2000) &(2000) &[\arcmin ]
    & &[km/s]&[M&[\degr ]&[\degr ]& &$d_{25}$ &[Jy] &[mJy] \\
    &(1)&(1)&(2) &[h m s]&[\degr\ \arcmin\ \arcsec]
    &(2) &(2) &(2) &pc]& & & & &(3) &(4)\\
\noalign{\smallskip}
\hline\noalign{\smallskip}
\pheins 986   &SBb(rs) &I--II  &SBT2  &02 33 34.3 &$-$39 02 43 &\pheins 3.9
      &1.32 &       1907 & 25 &?&\pheins?& [0.5]& 0.46 & 23.1 &\pheins 40\\
1313  &SBc(s) &III-IV &SBS7  &03 18 15.5 &$-$66 29 51 &\pheins 9.1
      &1.32 &\pheins 292 &\pheins4 & 38 & 170 & 0.63 & 0.31 & 10.4 &\pheins 59\\
1433  &SBb(s) &I-II   &PSBR2 &03 42 01.4 &$-$47 13 17 &\pheins 6.5
      &1.10 &\pheins 920 & 12 & 27 &\pheins  17 & 0.33 & 0.36 &\pheins 3.3 &\pheins --\\
1493  &SBc(rs) &III    &SBR6  &03 57 27.9 &$-$46 12 38 &\pheins 3.5
      &1.07 &\pheins 900 & 12 & 30 &\pheins ?  & 0.32 & 0.18 &\pheins 2.2 &\pheins -- \\
1559  &SBc(s) &II.8    &SBS6  &04 17 37.4 &$-$62 47 04 &\pheins 3.5
      &1.74 &       1115 & 15 & 55 &\pheins  65 &[0.3] &[0.2] &23.8 & 120\\
1672  &SBb(rs) &II     &SBS3  &04 45 42.2 &$-$59 14 57 &\pheins 6.6
      &1.20 &       1155 & 15 & 39 & 170  & 0.41 & 0.68 & 34.8 & 100\\
2442  &SBbc(rs) &II    &SXS4P &07 36 23.9 &$-$69 31 50 &\pheins 5.5
      &1.12 &       1236 & 16 & 24 &\pheins  40 & [0.5]& 0.42 &$\simeq$22&\pheins 70\\
3059  &SBc(s) &III     &SBT4  &09 50 08.1 &$-$73 55 17 &\pheins 3.6
      &1.12 &       1056 & 14 &?&\pheins?&[0.3] &[0.2] &\pheins 9.6 &\pheins --\\
5643  &SBc(s) &II-III &SXT5  &14 32 41.5 &$-$44 10 24 &\pheins 4.6
      &1.15 &       1066 & 14 &?&\pheins?&[0.4] &[0.35] & 18.7 &\pheins 64\\
7552  &SBbc(s) &I-II  &PSBS2 &23 16 11.0 &$-$42 35 01 &\pheins 3.4
      &1.26 &       1568 & 21 & 31 &\pheins\pheins 1  & 0.29 & 0.59 & 72.9 & 140\\
\noalign{\smallskip}
\hline\noalign{\smallskip}
\end{tabular}
\vbox{
\noindent References:
(1)--(3) see Table 1; (4) Whiteoak (\cite{whiteoak70}).
}
\end{table*}
\normalsize

\section{The sample}

Table~\ref{tab:vla-sample} lists the barred galaxies that have been
observed with the VLA
(declinations north of $-39\degr$). These have an optical
extent $\ge 4\arcmin$, to obtain reasonable spatial resolution,
a mean total intensity in radio continuum
of $\ge 3\mJy/\mbox{arcmin}^2$, and were
selected from Condon's (\cite{condon87}) survey.
Our sample of galaxies observed with the ATCA
(declinations south of $-39\degr$)
is shown in Table~\ref{tab:atca-sample}.
The criteria are the same, except for the size limit of 3\arcmin.
The isophotal major-axis diameter in arcmin is denoted by $d_{25}$, and
$q_{25}$ is the ratio of the major-to-minor isophotal diameters;
the subscript 25 refers to the isophotal level of 25~mag/$\mbox{arcsec}^2$.
The systemic velocity $v_{\rm GSR}$ has been transformed into
distance $D$ using $H_0$=75~km s$^{-1}$ Mpc$^{-1}$, except for
NGC~1365 and NGC~4535, for which
Cepheid distances are available (Madore et al.\ \cite{madore+98},
Macri et al.\ \cite{macri+99}). The inclination and position angle of
each projected galaxy disc are denoted $i$ and PA
(Martin \cite{martin95}; Martin \& Friedli \cite{martin+friedli97};
Ma et al.\ \cite{ma+97},
\cite{ma+98}; Wilke et al.\ \cite{wilke+00}; M\"ollenhoff \& Heidt\
\cite{moellenhoff+01});
$i=0$ means face-on and $\mbox{PA}=0$ is the north-south direction.
For several southern galaxies, no values of $i$ and PA are available,
as indicated by `?'.
The deprojected ratio of the bar's minor and major axes is denoted
$b/a$, and $2a/d_{25}$ is its relative length
according to Elmegreen \& Elmegreen (\cite{elmegreen+elmegreen85}) or
Martin (\cite{martin95}), derived from
deprojected images according to the galaxy's inclination and position angle,
assuming that the bars are flat.
Values in brackets are our estimates from optical images.

The far-infrared flux densities at $\lambda60\, \mu$m, denoted
$S_{60\mu \rm m}$, are from Fullmer \& Lonsdale (\cite{fullmer+lonsdale89}).
Their values for NGC~2442 and NGC~3992 are
obviously underestimated and were
recalculated from the flux density at 100\,$\mu$m using
$S_{60\mu \rm m} = fS_{100\mu \rm m}$ with
$f=0.5$ for the luminous galaxy NGC~2442 and
$f=0.3$ for the faint galaxy NGC~3992
(see Fig.~5b in Young et al.\ \cite{young+89}).
The integrated radio
continuum flux densities at $\lambda20\cm$ and $\lambda6\cm$,
$S^{\rm tot}_{20\rm cm}$ in Table~\ref{tab:vla-sample} and
$S^{\rm tot}_{6\rm cm}$ in Table~\ref{tab:atca-sample},
were taken from Condon\ (\cite{condon87}) and
Whiteoak\ (\cite{whiteoak70}), respectively.

\scriptsize
\begin{table*}[htb]
\caption{Summary of the VLA observations. $t$
is the on-source observation time (in hours).
$\sigma_I$ is the rms noise in the final map of total intensity and
$\sigma_{PI}$ the rms noise in the final map of polarized intensity
(both in $\mu$Jy/beam). Column 5 gives only noise levels at $\lambda22$\,cm,
those at $\lambda18$\,cm are about 30\% higher.}
\label{tab:vla-obs}
\begin{tabular}{lllllllllllll} \hline
\noalign{\smallskip}
&\multicolumn{4}{l}{$\lambda$22 cm and $\lambda$18~cm}
&\multicolumn{4}{l}{$\lambda$6.2 cm}
&\multicolumn{4}{l}{$\lambda$3.5 cm}\\
\cline{2-5} \cline{6-9} \cline{10-13}
\noalign{\smallskip}
NGC  &Date  &Conf.  & $t$ &$\sigma_I/\sigma_{PI}$
     &Date  &Conf.  & $t$ &$\sigma_I/\sigma_{PI}$
     &Date  &Conf.  & $t$ &$\sigma_I/\sigma_{PI}$\\
\noalign{\smallskip}
\hline\noalign{\smallskip}
1097  &96 Feb3  &CnB &0.7 & 60/40
      &96 May20  &DnC &0.5 &
      &96 May20  \\
     &&&&&&&& &+97 Oct7+9 &DnC &5.6 &\\
      &&&& &+98Nov13+15 &CnB &9.3 & 40/20
           &+98Nov11+14 &CnB &9.3 & 30/15\\
\noalign{\smallskip}
1300  &--"--  &CnB  &1.0 & 50/40
      &96 May20  &DnC &1.8 & 25/25
      &96 May20  &DnC &2.0 & 25/20\\
\noalign{\smallskip}
1365  &--"--  &CnB  &1.0 & 80/60
      &96 May20  &&&
      &96 May20  \\
     &&&& &+99 Feb19 &DnC  &6.0& 15/15 &+97 Oct7+9 &DnC  &5.6& 20/15\\
\noalign{\smallskip}
2336  &96Mar10+22  &C  &1.1 & 25/25
      &96 Jul22  &D &2.0 & 15/20
      &96 Jul22  &D &2.2 & 20/--\\
\noalign{\smallskip}
3359  &96 Mar10  &C  &0.4 & 30/25
      &--"--       &D  &0.7 & 30/25
      &--"--       &D  &0.9 & 30/35\\
\noalign{\smallskip}
3953  &--"--  &C  &0.4 & 30/20
      &--"--  &D  &0.7 & 25/30
      &--"--  &D  &0.9 & 30/--\\
\noalign{\smallskip}
3992 &--"-- &C  &0.7 & 30/25
      &96 Jul22  &&&
      &96 Jul22  \\
     &&&& &+96 Aug17 &D  &0.7& 20/20 &+96 Aug17 & D  &0.9& 40/25\\
\noalign{\smallskip}
4535  &--"--  &C  &0.4 & 20/20
      &96 Aug17  &D  &0.5 & 20/20
      &96 Aug17  \\
     &&&&&&&& &+99 Apr2 &D  &6.6& 25/15\\
\noalign{\smallskip}
5068  &97 Jun12  &CnB  &2.0 & 30/35
      &96 Jun1   &DnC  &0.8 &
      &96 Aug17  &D    &1.2 & 20/25\\
      &&&& &+96 Aug17   &D  &1.1 & 20/15\\
\noalign{\smallskip}
7479  &96 Mar22  &&&
      &96 Jul22  &D  &0.5 & 25/30
      &96 Jul22  \\
      &+97 Jul2 &C  &2.1& 40/25 &&&& &+98 Jan9+10 &D  &5.0& 20/15\\
\noalign{\smallskip}
\hline \noalign{\medskip}
\end{tabular}
\end{table*}
\normalsize

\bigskip
\begin{table*}[htb]
\caption{Summary of the ATCA observations.
$t$ is the on-source observation time (in hours).
$\sigma_I$ is the rms noise in the final map of total intensity and
$\sigma_{PI}$ the rms noise in the final map of polarized intensity
(both in $\mu$Jy/beam).}
\label{tab:atca-obs}
\begin{center}
\begin{tabular}{lllllllll} \hline
\noalign{\smallskip}
&\multicolumn{4}{l}{$\lambda$13 cm}
&\multicolumn{4}{l}{$\lambda$5.8 cm} \\
\cline{2-5} \cline{6-9}
\noalign{\smallskip}
NGC  &Date  &Conf.  & $t$ &$\sigma_I/\sigma_{PI}$
     &Date  &Conf.  & $t$ &$\sigma_I/\sigma_{PI}$\\
\noalign{\smallskip}
\hline\noalign{\smallskip}
\pheins 986  & & &  & &96 Jan20     &750C &\pheins2 &\\
             & & &  & &+96 Feb10+11  &750B &\pheins3 & 40/25\\
\noalign{\smallskip}
1313         & & &  & &96 Jan20     &750C &\pheins2 &\\
             & & &  & &+96 Feb10+11  &750B &\pheins2 & 70/25\\
\noalign{\smallskip}
1433         & & &  & &96 Jan20     &750C &\pheins2 &\\
             & & &  & &+96 Feb10+11  &750B &\pheins2 & 40/25\\
\noalign{\smallskip}
1493         & & &  & &96 Jan20     &750C &\pheins2 &\\
             & & &  & &+96 Feb10+11  &750B &\pheins2 & 50/25\\
\noalign{\smallskip}
1559   &96 Oct29 &1.5A &\pheins5  & 100/40 &96 Jan20 &750C &\pheins2 &\\
             & & &  & &+96 Feb10+11  &750B &\pheins3 &\\
             & & &  & &+96 Nov1      &750A & 11  & 30/15\\
\noalign{\smallskip}
1672   &93 May19 &1.5B &10 $*$      & 60/-- & 92 Mar15 &375 & 11 &\\
             & & &  & &+93 Jul26     &750D & 10 $**$&\\
             & & &  & &+93 Sep9      &750C & 12 $**$& 50/25\\
\noalign{\smallskip}
2442   &96 Oct29 &1.5A &\pheins5  & 70/60 & 96 Jan19 &750C &\pheins3 &\\
             & & &  & &+96 Feb9+11   &750B &\pheins4 &\\
             & & &  & &+96 Nov2      &750A & 10 &\\
             & & &  & &+98 Mar25     &375  &\pheins 9  &\\
             & & &  & &+00 Dec31     &750C & 11 & 25/25\\
\noalign{\smallskip}
3059   &98 Jan9 &1.5A &\pheins7 $*$ & 60/60 &96 Jan19 &750C &\pheins3 &\\
             & & &  & &+96 Feb9+11   &750B &\pheins4 &\\
             & & &  & &+97 Jan7      &750D &\pheins6 &\\
             & & &  & &+98 Apr9      &750A &\pheins10 & 25/25\\
\noalign{\smallskip}
5643         & & &  & &96 Jan19     &750C &\pheins3 &\\
             & & &  & &+96 Feb9+11   &750B &\pheins3 & 40/25\\
\noalign{\smallskip}
7552         & & &  & &96 Jan19     &750C &\pheins3 &\\
             & & &  & &+96 Feb9+11   &750B &\pheins3 &\\
             & & &  & &+99 Jan3      &375  &\pheins8 &\\
             & & &  & &+00 Dec30     &750C &\pheins7 & 100/25\\
\noalign{\smallskip}
\hline\noalign{\medskip}
\multicolumn{9}{l}{$*$ Simultaneous observations at $\lambda$22~cm (1380~MHz) }\\
\multicolumn{9}{l}{$**$ Simultaneous observations at $\lambda$3.5~cm (8640~MHz) }\\
\end{tabular}
\end{center}
\end{table*}

\section{Observations and results}

The observations were performed with the VLA operated by the
NRAO\footnote{The National Radio Astronomy Observatory is a facility of the
National Science Foundation operated under cooperative agreement
by Associated Universities, Inc.}
and with the ATCA\footnote{The Australia Telescope
Compact Array is part of the Australia Telescope which is funded by the
Commonwealth of Australia for operation as a National Facility managed by
CSIRO.}
operated by the ATNF. Details of the observations are given
in Tables~\ref{tab:vla-obs} and
\ref{tab:atca-obs}. The antenna configurations were chosen to obtain
half-power widths of the synthesized beams of about
20\arcsec.

In the L band, VLA observations were performed at 1365~MHz ($\lambda$22~cm)
and 1665~MHz ($\lambda$18~cm) simultaneously.
In each of the VLA C and X bands, the data from two channels were combined
(4835~MHz + 4885~MHz and 8435~MHz + 8485~MHz).
The four southern galaxies from the VLA sample were observed
with hybrid configurations (CnB and DnC) which allow to synthesize
a more circular beam.
For NGC~1097, the CnB and DnC $uv$ data at the same wavelength
were combined.

Several ATCA 750~m configurations were combined at $\lambda$5.8~cm
(4800~MHz + 5568~MHz)
to achieve higher sensitivity and better coverage of the $uv$ plane.
For three large galaxies, additional observations with the ATCA 375~m
configuration were added. In two observation sessions (1993 May 19 and 1998 January 9),
data at 1380~MHz ($\lambda$22~cm) and at 2368~MHz
($\lambda$13~cm) were recorded simultaneously. No significant
polarization was detected at $\lambda$22~cm. In the session
of 1996 October 29,
frequencies were set to 2240~MHz ($\lambda$13.4~cm) and 2368~MHz
($\lambda$12.7~cm). In the sessions
of 1993 July 26 and September 9, data at 4800~MHz
($\lambda$6.2~cm) and at 8640~MHz ($\lambda$3.5~cm) were recorded simultaneously.

The largest visible structure for full synthesis observations
(that requires an observing time in excess of
8~h with the VLA or 12~h with the ATCA)
is 15\arcmin\ at $\lambda$18~cm and 22~cm (VLA C or CnB arrays), 5\arcmin\ at
$\lambda$6~cm (VLA D or DnC arrays), 3\arcmin\ at $\lambda$3~cm (VLA D or DnC arrays),
3\arcmin\ at $\lambda$13~cm (ATCA 1.5~km arrays),
4\arcmin\ at $\lambda$6~cm (ATCA 750~m arrays)
and 8\arcmin\ at $\lambda$6~cm (ATCA 375~m array).

The data were reduced with the standard AIPS and MIRIAD software packages.
The maps in Stokes parameters {\rm I}, {\rm Q} and {\rm U} were smoothed to a common
resolution of 30\arcsec\ to achieve a higher signal-to-noise ratio.
These were combined to maps of total and polarized surface
brightness\footnote{In
the following we will also use the usual expression `intensity'.},
measured in Jansky per solid angle of the telescope beam (`beam area'),
and polarization angle. The positive bias in {\rm PI} due to noise
was corrected by subtraction of a constant value, which is equal to
1.0-1.4$\times$(rms noise) in the maps of {\rm Q} and {\rm U}.

The rms noise in the final maps in {\rm I} (total intensity)
and {\rm PI} (polarized intensity) is given in Tables~\ref{tab:vla-obs} and
\ref{tab:atca-obs}. Since the noise in the {\rm PI} maps has a non-Gaussian distribution
(even if {\rm Q} and {\rm U} have Gaussian noise) the standard deviation underestimates
the noise. Therefore we assume the noise in {\rm PI} to be the same as that
in {\rm Q} and {\rm U}. The rms noise in the ATCA maps is typically larger by a factor
two in comparison to the VLA maps.

The final maps are displayed in Figs.~5--24,
overlayed onto images from the Digitized Sky Surveys.
\footnote{The Digitized Sky Surveys were produced at the Space Telescope Science
Institute under U.S. Government grant NAG W-2166. The images of these surveys
are based on photographic data obtained using the Oschin Schmidt Telescope on
Palomar Mountain and the UK Schmidt Telescope. The plates were processed into
the present compressed digital form with the permission of these
institutions.}
Contours show the total intensity
at the wavelength indicated near the upper left
corner of each frame, dashes indicate the orientation of the
observed $E$ vector of the polarized emission turned by 90\degr.
These `$B$ vectors' indicate the orientation of the magnetic field
only in case of small Faraday rotation (see below).
Due to missing spacings, the VLA maps at $\lambda$3~cm and the
ATCA maps at $\lambda$13~cm do not show the extended emission in full.

We did not attempt to separate the thermal from the nonthermal
emission because for most galaxies we have only maps at one or two
wavelengths which show the full extended emission. The average thermal
fraction in spiral galaxies is only $\simeq$10\% at $\lambda$20~cm
(Niklas et al.\ \cite{niklas+97}) which corresponds to $\simeq$20\% at
$\lambda$6~cm and $\simeq$30\% at $\lambda$3~cm (assuming a nonthermal
spectral index of 0.85).

Figures 25 and 26 show the distribution of polarized intensity {\rm PI}
and the observed $B$ vectors for the
galaxies with the strongest polarization. We give the accurate
observational wavelengths in the titles, but for ease of reading we
will summarize all C-band observations as `$\lambda$6~cm' and
all X-band observations as `$\lambda$3~cm'. At the shorter wavelengths,
$\lambda$6~cm and $\lambda$3~cm, the $B$ vectors in the Figures
show the approximate orientation of the magnetic
field averaged over the beam.
A correction for Faraday rotation, significant at $\lambda\ge13$~cm,
was not attempted because of insufficient signal-to-noise ratios
of the polarization data at these wavelengths.

In Section~4 we quote Faraday rotation measures (\RM)
between $\lambda$22~cm and $\lambda$6~cm
which, however, can strongly be affected by Faraday depolarization
(Sokoloff et al.\ \cite{sokoloff+98}).
\RM\ values between $\lambda$6~cm and $\lambda$3~cm were computed only
for NGC~1097 and NGC~1365 for which the signal-to-noise ratio at
$\lambda$3~cm is sufficiently high. Correction for Faraday rotation in the
Galactic foreground was not attempted.

We note that polarized emission can also be produced
by anisotropic turbulent magnetic fields (Laing\ \cite{laing81},
Sokoloff et al.\ \cite{sokoloff+98}, Laing\ \cite{laing02})
which can be a result of compression
and/or shearing by streaming velocities.
These turbulent magnetic fields do not produce any Faraday rotation.
The anisotropy of turbulence could be significant
in bars. Further Faraday rotation measures with high accuracy and
good resolution are required to distinguish between 
anisotropic turbulent and coherent regular fields in our sample galaxies.

The total and polarized intensities {\rm I} and {\rm PI} were integrated in
concentric rings (15\arcsec\ wide) defined in each galaxy's plane,
using the inclination
$i$ and position angle PA given in Tables~\ref{tab:vla-sample} and
\ref{tab:atca-sample}. The maximum radius for the integration is the
outer radius of the ring where $I$ reaches the noise level. The
integrated flux density $S_{\lambda}$ is
given in Tables~\ref{tab:vla-flux} and \ref{tab:atca-flux}.
The average degree of polarization $p_{\lambda}$
was obtained from the integrated values of {\rm I} and {\rm PI}.
The errors in $S_{\lambda}$ and $p_{\lambda}$ include
(as a quadratic sum) a 5\% uncertainty in
the absolute flux calibration and the zero-level uncertainty.
In order to determine the zero level, we calculated
the average surface brightness in several rings located outside the galaxy image.
The rms scatter of these averages was adopted as the zero-level uncertainty.

\begin{figure}
\centering
\includegraphics[bb = 130 117 708 473,width=8.8cm,clip]{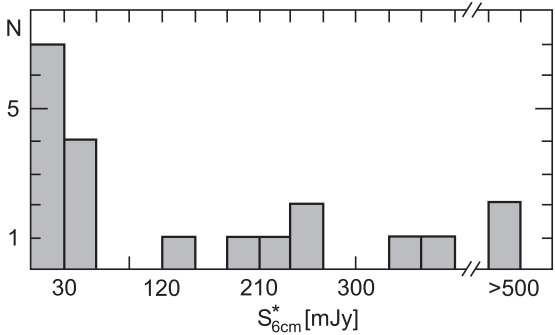}
\caption{Histogram of the total
radio flux densities $S_{6\rm cm}^*$, scaled
to a distance of 10~Mpc, of the sample galaxies at $\lambda$6~cm}
\label{fig:hist}
\end{figure}

The variations of the radio flux density and far-infrared flux density
between the galaxies of our sample cannot be explained by
variations in distance alone.
Having scaled the radio flux density at $\lambda$6~cm to a
common distance of 10~Mpc ($S_{6\rm cm}^*$, a measure of radio luminosity),
we define three groups of galaxies 
(see Fig.~\ref{fig:hist} and column 7 of Tables
\ref{tab:vla-flux} and \ref{tab:atca-flux}):

\begin{itemize}
\item {\bf Radio-weak}: NGC~1300, 1313, 1433, 1493, 2336, 3059,
3359, 3953, 3992, 4535, 5068
($S_{6\rm cm}^* < 60$~mJy).
\item {\bf Moderate}: NGC~986, 1559, 1672, 2442, 5643
(120~mJy $ < S_{6\rm cm}^* \le 270$~mJy).
\item {\bf Radio-bright}: NGC~1097, 1365, 7479, 7552
(${S_{6\rm cm}}^* > 330$~mJy).
\end{itemize}

Our sample is admittedly small and biased towards
radio-bright galaxies. Our preliminary classification 
should be investigated with a
larger sample in radio continuum and/or infrared emission.

The average total surface bightness $I_{6\rm cm}$ was computed by dividing
$S_{6\rm cm}$ by the number of beams in the integration area.
The results are given in column 11 of Tables~\ref{tab:vla-flux} and
\ref{tab:atca-flux}.
The error in $I_{6\rm cm}$ is dominated by the uncertainty in the
integration area which is estimated to be about 25\%.
A classification based on $I_{6\rm cm}$ is similar to that
based on $S_{6\rm cm}^*$.

\section{Individual galaxies}
\subsection{The VLA sample}

{\it NGC~1097\/} (Fig.~5) has one of the intrinsically longest bars
in our sample ($\simeq16$~kpc)
and is the most interesting galaxy concerning radio polarization.
The nucleus (Seyfert~1 type, Storchi-Bergmann et al.\ \cite{storch+97})
and the circumnuclear starburst ring
of 17\arcsec\ ($\simeq1.3$\,kpc) radius (see Hummel et al.\
\cite{hummel+87}, Gerin et al.\ \cite{gerin+88})
are prominent in total radio intensity.
Polarized radio emission in this region reveals a
spiral magnetic field extending from the circumnuclear ring
towards the centre (Beck et al.\ \cite{beck+99}).
NGC~1097 also features strong polarized emission {\it upstream\/}\footnote{As
the gas moves faster than the pattern speed of the bar,
{\it upstream} is defined as the region in front of the shock which
lags behind with respect to the galaxy's rotation.}
of the shock fronts (identified with
the dust lanes offset from the bar major axis),
where the regular magnetic field makes a large angle with the bar's
major axis and then, upstream of the dust lanes, turns to become aligned with
the bar. This strong turning of polarization vectors leads to beam depolarization,
so that the upstream regions appear as two elongated minima of polarized emission
in Fig.~25.
The enhancement of total and polarized emission in the dust lanes is only moderate
and much weaker than the gas density enhancement (Beck et al.\ \cite{beck+99}).
Extended polarized emission with apparent magnetic field orientation
{\it perpendicular\/} to the optical spiral arms (Fig.~25) is visible
in the northeast and southwest.
The spiral arms outside the bar region exhibit only weak radio emission.
Faraday rotation between $\lambda22$~cm and $\lambda6$~cm is generally weak,
with $|\RM|\le 10\radm$, except in the central region
where \RM\ varies between $-35\radm$ and $+35\radm$.
\RM\ between $\lambda6$~cm and $\lambda3$~cm is generally higher, it varies
between $+160\radm$ in the upstream region and $-150\radm$ in the downstream
region of the southern bar, and between $+200\radm$ and $-170\radm$
near the centre. The increase of \RM\ with decreasing wavelength is typical
for spiral galaxies (Sokoloff et al.\ \cite{sokoloff+98}).

{\it NGC~1300\/} (Fig.~6) has one of the most pronounced optical bars in our sample.
Total radio intensity is maximum in the nuclear region and at the ends of the bar
where the spiral arms start (both are sites of strong star formation).
No significant radio emission has been detected in the bar.
The  polarized emission is weak. The northern extensions visible in the maps
at $\lambda$22 and 6~cm are not real and result from poor data coverage in
the $uv$ plane. The $\lambda18$~cm and $\lambda3$~cm maps show less emission
than those at $\lambda22$~cm and $\lambda6$~cm because the signal-to-noise
ratios are worse.

{\it NGC~1365\/} (Fig.~7) is the intrinsically largest and
radio-brightest galaxy in our sample. Its bar length is $\simeq$29~kpc.
It has a Seyfert~1-type nucleus surrounded by a starburst region
(see review by Lindblad\ \cite{lindblad99}).
Similarly to NGC~1097, it has significant (though weaker)
polarized radio emission upstream of the shock fronts and relatively weak
emission enhancements on the dust lanes.
The turn of magnetic field lines towards the
dust lanes near the bar major axis is much smoother than in NGC~1097.
The magnetic field orientations near the centre also
form a spiral pattern.
Sandqvist et al.\ (\cite{sandqvist+95}) describe a circumnuclear elliptical ring
of about 1\,kpc in radius, visible in radio continuum at $\lambda\lambda20$ 
and $6\cm$ at a resolution of $2.3\arcsec\times1.0\arcsec$,
and note its similarity to that in NGC~1097.
Kristen et al.\ (\cite{kristen+97}) have revealed, in the optical range, 
a large number of bright spots arranged along the ring and
suggest that the continuous ring structure
might be obscured by dust absorption.
The ring is not resolved in our observations.
Faraday rotation between $\lambda22$~cm and
$\lambda6$~cm is significant only in the central region ($\RM\simeq
-20\radm$).  \RM\ between $\lambda6$~cm and $\lambda3$~cm jumps
between $\simeq+600\radm$ and $\simeq-600\radm$ near the centre.  The
spiral arms outside the bar region are bright in radio and the
regular magnetic field is well aligned with them (different from
NGC~1097).  A ridge of polarized emission (Fig.~25) is observed on
the {\it inner\/} side of the northwestern spiral arm which may
indicate field compression by a density wave, a `magnetic arm', or
depolarization along the optical spiral arm.  This galaxy has an
extended, almost circular, polarized envelope with the regular field
aligned with the optical spiral arms.

{\it NGC~2336\/} (Fig.~8) has a small optical bar. Two extended regions
of total radio emission are prominent, but they do not coincide well
with star-forming regions. No emission has been detected from
the bar and the nucleus. No polarization has been detected.
The northern extensions at $\lambda$18~cm and 22~cm are probably not real
but result from poor data coverage in the $uv$ plane.

{\it NGC~3359\/} (Fig.~9) has enhanced total radio emission from
the bar where star formation is strong as well.
The total emission from the southern spiral arm peaks
in star-forming regions, which are best visible at $\lambda6$~cm. Weak
polarized emission has been detected in the {\it interarm\/}
regions with field lines parallel to the adjacent optical arm.
No polarized emission has been detected from the bar, possibly due
to insufficient angular resolution and depolarization.

{\it NGC~3953\/} (Fig.~10) is notable for its extended, diffuse,
polarized radio emission from the outer disc.
No significant radio emission from the bar and the nucleus has been detected.
Significant Faraday rotation between $\lambda$22~cm and $\lambda$6~cm
occurs in the southern part of the galaxy ($\RM\simeq +15\radm$).

{\it NGC~3992\/} (Fig.~11) features strong
polarized radio emission from spiral arms with aligned
magnetic fields. Weak emission from the nucleus has only
been detected at $\lambda$6~cm and 3~cm
(the sensitivity was too low at $\lambda$22~cm and 18~cm).
No emission from the bar has been detected.
Faraday rotation is significant ($\RM\simeq +20\radm$) between
$\lambda$22~cm and $\lambda$6~cm.

{\it NGC~4535\/} (Fig.~12) exhibits apparently unpolarized radio
emission from the central region and the bar, which is small and just
resolved in our images.
The polarized emission mainly comes from spiral arms with
aligned magnetic fields and from the northern
{\it interarm} region (Fig.~25). Faraday rotation is weak between
$\lambda$22~cm and $\lambda$6~cm ($|\RM|\le 10\radm$),
but larger between $\lambda$6~cm and $\lambda$3~cm
($\RM\simeq +80\radm$).

{\it NGC~5068\/} (Fig.~13) has strong radio emission from the
small bar and star-forming regions. Weak polarization has been
detected at $\lambda$6~cm in the southern part.

{\it NGC~7479\/} (Fig.~14) has strong polarized radio emission,
mainly due to the nuclear `jet' which was discovered by Laine \& Gottesman
(\cite{laine+gottesman98}). The jet is
not resolved in the maps presented here (Figs.~14 and 25).
Total emission is enhanced in the western spiral arm.
The polarized emission and Faraday rotation in this galaxy
will be discussed in detail elsewhere.

\subsection{The ATCA sample}

{\it NGC~986\/} (Fig.~15) has strong radio emission from the central
star-forming region and the bar. Polarized emission is observed
in the inner bar.

{\it NGC~1313\/} (Fig.~16) exhibits strong radio emission from the bar
and star-forming regions. No polarization has been detected.

{\it NGC~1433\/} (Fig.~17) has weak radio emission from the
central region which hosts an irregular star-forming ring of
5\arcsec\ or $\simeq0.3$\,kpc radius (Maoz et al.\ \cite{maoz+96}).
Star formation has not been detected in the bar but is
noticeable in the ring-like spiral arms, in particular at the ends
of the bar (see H$\alpha$ image by Crocker et al.\ \cite{crocker+96})
where weak radio emission is seen. No polarization has been detected.

{\it NGC~1493\/} (Fig.~18) shows weak radio emission from the outer spiral arms.
Polarization has only been detected in a small region in the southeast.

{\it NGC~1559\/} (Fig.~19) possesses massive spiral arms with strong
star formation. The small bar of about 40\arcsec\ length is oriented
almost east-west. Very strong radio emission originates in the bar
and the disc. Polarized emission is strongly asymmetric with
peaks near the ends of the bar
and magnetic field lines at large angles to the bar (Fig.~26).
The region where the magnetic field is strongly aligned and almost perpendicular
to the bar's major axis is larger than in NGC~1097.

{\it NGC~1672\/} (Fig.~20) has the second largest bar in our sample
($\simeq$20~kpc). Its radio emission is very strong in the nucleus, the bar
and the inner part of the spiral arm region. The nucleus is known to be of Seyfert~2
type and is surrounded by a starburst region (Evans et al.\ \cite{evans+96}).
Polarized emission is strongest in the northeastern region {\it upstream\/} of
the dust lanes, with magnetic field lines
at large angles to the bar, smoothly turning towards the centre
as in NGC~1097 and 1365 (Fig.~26).

{\it NGC~2442\/} (Fig.~21), a member of the Volans Group,
has an asymmetric appearance which may be
a result of tidal interaction (Mihos \& Bothun \cite{mihos+bothun97}) or
ram pressure stripping (Ryder et al.\ \cite{ryder+01}). The H$\alpha$ image by
Mihos \& Bothun (\cite{mihos+bothun97}) shows an unresolved central source
and a circumnuclear star-forming ring of 8\arcsec\ ($\simeq0.6$\,kpc) radius.
NGC~2442 exhibits strong radio emission from the nucleus and the ends
of the bar. Very strong and polarized emission has been detected
in the northern arm (hosting a massive dust lane) with aligned field lines,
possibly a signature of field compression and/or shearing.
Diffuse radio emission is visible in the eastern part (Fig.~21),
with a blob of highly polarized emission (Fig.~26).
This galaxy will be discussed in Paper III (Harnett et al.\ \cite{harnett+02}).

{\it NGC~3059\/} (Fig.~22) has a small optical bar.
Diffuse, polarized radio emission has been detected in
the whole disc, indicating a widespread, spiral, regular magnetic field
similar to that of non-barred spiral galaxies.
This is not surprising as the bar is small and under-resolved.
We cannot exclude that this galaxy has a magnetic field component
aligned with the bar.

{\it NGC~5643\/} (Fig.~23) has strong, diffuse radio emission from the nucleus,
the bar and the disc. The central region and outer disc are
weakly polarized with some indication of a spiral pattern.

{\it NGC~7552\/} (Fig.~24), a member of the Grus Quartet,
has a starburst circumnuclear ring of 8\arcsec\ ($\simeq0.8$\,kpc) radius,
observed in radio continuum and near-infrared,
a nuclear bar (observed in radio continuum and near-infrared)
lying perpendicular to the primary bar,
but no active nucleus (Forbes et al.\ \cite{forbes+94a},b).
It is notable for strong, highly polarized radio emission from the centre,
the bar and the inner parts of the spiral arms.
Our resolution is insufficient to resolve the radio ring.
Polarized emission is strong
upstream of the dust lanes, with magnetic field lines oriented at large
angles to the bar major axis (Fig.~26); these features make this galaxy
similar to NGC~1097 and NGC~1672.
The outer extensions in Fig.~24 are artifacts due to unsufficient
$uv$ coverage. A detailed discussion of this galaxy will be given in Paper III.

\begin{table*}[htb]
\caption{The flux density $S_{\lambda}$, the flux density $S_{6\rm cm}^*$ scaled
to a distance of 10~Mpc, the average degree of polarization $p_{\lambda}$
(based on the maps at 30\arcsec\ resolution),
the average total surface brightness $I_{6\rm cm}$ (in mJy/beam area),
the total equipartition field strength $B_{\rm tot}$, and its resolved
regular component $B_{\rm reg}$ for the VLA sample. Values in brackets
are lower limits where low signal-to-noise ratios or insufficient $uv$ coverage
preclude better estimates.}
\label{tab:vla-flux}
\begin{tabular}{lr@{$\pm$}lr@{$\pm$}lr@{$\pm$}lr@{$\pm$}lr@{$\pm$}lrr@{$\pm$}
lr@{$\pm$}lr@{$\pm$}lrr@{$\pm$}lr@{$\pm$}l}
\hline
\noalign{\smallskip}
NGC  &\multicolumn{2}{c}{$S_{22\rm cm}$} &\multicolumn{2}{c}{$p_{22\rm cm}$}
     &\multicolumn{2}{c}{$S_{18\rm cm}$} &\multicolumn{2}{c}{$p_{18\rm cm}$}
     &\multicolumn{2}{c}{$S_{6\rm cm}$}  &$S_{6\rm cm}^*$
     &\multicolumn{2}{c}{$p_{6\rm cm}$}
     &\multicolumn{2}{c}{$S_{3\rm cm}$}    &\multicolumn{2}{c}{$p_{3\rm cm}$}
     &$I_{6\rm cm}$
     &\multicolumn{2}{c}{$B_{\rm tot}$} &\multicolumn{2}{c}{$B_{\rm reg}$}\\
     &\multicolumn{2}{c}{[mJy]} &\multicolumn{2}{c}{[\%]}
     &\multicolumn{2}{c}{[mJy]} &\multicolumn{2}{c}{[\%]}
     &\multicolumn{2}{c}{[mJy]}   & [mJy] &\multicolumn{2}{c}{[\%]}
     &\multicolumn{2}{c}{[mJy]}    &\multicolumn{2}{c}{[\%]} & [mJy/
     &\multicolumn{2}{c}{[$\mu$G]} &\multicolumn{2}{c}{[$\mu$G]}\\
     &\multicolumn{2}{c}{} &\multicolumn{2}{c}{}
     &\multicolumn{2}{c}{} &\multicolumn{2}{c}{}
     &\multicolumn{2}{c}{}   & &\multicolumn{2}{c}{}
     &\multicolumn{2}{c}{}    &\multicolumn{2}{c}{} & b.a.]
     &\multicolumn{2}{c}{} &\multicolumn{2}{c}{}\\
\noalign{\smallskip}
\hline\noalign{\smallskip}
1097  &350&25  &1.5&0.9  &262&30  &1.3&1 &142&12 &359 &8.5&1.0
      &\pheins94&18  &5.7&1.7
      &1.67 &13&4 &4&1\\
1300  &29&\pheins4    &10&\pheins3  &\multicolumn{2}{c}{[14]} &\multicolumn{2}{c}{---}
      &11&\pheins2 & 44 &\pheins9&\pheins4  &\pheins5&\pheins1  &\pheins10&\pheins5
      &0.19 &8&2 &3&1\\
1365  &540&31  &1.6&0.4  &423&30  &1.7&1 &206&13 &744 &2.4&0.5
      &116&10  &2.2&0.6
      &2.69 &15&5 &3&1\\
2336  &13&\pheins2  &\multicolumn{2}{c}{$<3$} &\multicolumn{2}{c}{[5]}
      &\multicolumn{2}{c}{---} &4.2&0.6 & 41 &\multicolumn{2}{c}{$<4$}
      &1.4&0.3 &\multicolumn{2}{c}{---}
      &0.10 &6&2 &\multicolumn{2}{c}{$<2$}\\
3359  &41&\pheins3  &6&\pheins3  &\multicolumn{2}{c}{[24]} &\multicolumn{2}{c}{---}
      &13&\pheins1  & 28 &\pheins5&\pheins3  &\multicolumn{2}{c}{[5]}
      &\multicolumn{2}{c}{---}
      &0.33 &8&2 &2&1\\
3953  &24&\pheins2  &2&\pheins2  &\multicolumn{2}{c}{[15]} &\multicolumn{2}{c}{---}
      &7&\pheins1 & 16 &\pheins7&\pheins3  &\multicolumn{2}{c}{[2]}  &\multicolumn{2}{c}{---}
      &0.21 &7&2 &2&1\\
3992  &12&\pheins2  &7&\pheins6 &\multicolumn{2}{c}{---} &\multicolumn{2}{c}{---}
      &4.5&0.8 & 10 &11&\pheins4  &\pheins2&\pheins1  &\multicolumn{2}{c}{---}
      &0.09 &6&2 &2&1\\
4535  &43&\pheins4  &9&\pheins5  &\multicolumn{2}{c}{[19]} &\multicolumn{2}{c}{---}
      &13&\pheins2 &33 &14&\pheins4  &\pheins8&\pheins1  &\pheins8&\pheins4
      &0.21 &8&2 &3&1\\
5068  &43&\pheins5  &\multicolumn{2}{c}{---} &36&\pheins5 &\multicolumn{2}{c}{---}
      &14&\pheins1 &8 &\pheins4&\pheins2  &\multicolumn{2}{c}{[4]} &\multicolumn{2}{c}{---}
      &0.23 &8&2 &2&1\\
7479  &110&\pheins6  &4&\pheins2  &\multicolumn{2}{c}{[55]} &\multicolumn{2}{c}{---}
      &33&\pheins4  & 379 &6.5&1.5  &22&\pheins4  &\pheins9&\pheins3
      &0.83 &11&3 &3&1\\
\noalign{\smallskip}
\hline
\end{tabular}
\end{table*}

\begin{table*}[htb]
\caption{As in Table~\ref{tab:vla-flux}, but for the ATCA sample.}
\label{tab:atca-flux}
\begin{tabular}{lr@{$\pm$}lr@{$\pm$}lr@{$\pm$}lr@{$\pm$}lr@{$\pm$}lrr@{$\pm$}
lr@{$\pm$}lr@{$\pm$}lrr@{$\pm$}lr@{$\pm$}l}\hline
\noalign{\smallskip}
NGC  &\multicolumn{2}{c}{$S_{22\rm cm}$} &\multicolumn{2}{c}{$p_{22\rm cm}$}
     &\multicolumn{2}{c}{$S_{13\rm cm}$} &\multicolumn{2}{c}{$p_{13\rm cm}$}
     &\multicolumn{2}{c}{$S_{6\rm cm}$}   & $S_{6\rm cm}^*$ &\multicolumn{2}{c}{$p_{6\rm cm}$}
     &\multicolumn{2}{c}{$S_{3\rm cm}$}    &\multicolumn{2}{c}{$p_{3\rm cm}$}
     &$I_{6\rm cm}$
     &\multicolumn{2}{c}{$B_{\rm tot}$} &\multicolumn{2}{c}{$B_{\rm reg}$}\\
     &\multicolumn{2}{c}{[mJy]} &\multicolumn{2}{c}{[\%]}
     &\multicolumn{2}{c}{[mJy]} &\multicolumn{2}{c}{[\%]}
     &\multicolumn{2}{c}{[mJy]}  & [mJy]  &\multicolumn{2}{c}{[\%]}
     &\multicolumn{2}{c}{[mJy]}    &\multicolumn{2}{c}{[\%]} & [mJy/b.a.]
     &\multicolumn{2}{c}{[$\mu$G]} &\multicolumn{2}{c}{[$\mu$G]}\\
\noalign{\smallskip}
\hline\noalign{\smallskip}
\pheins 986  &\multicolumn{2}{c}{---} &\multicolumn{2}{c}{---}
             &\multicolumn{2}{c}{---} &\multicolumn{2}{c}{---}
             &38&\pheins 3 &246 &\pheins2&\pheins1
             &\multicolumn{2}{c}{---} &\multicolumn{2}{c}{---}
             &0.86 &12&4 &2&1\\
1313         &\multicolumn{2}{c}{---} &\multicolumn{2}{c}{---}
             &\multicolumn{2}{c}{---} &\multicolumn{2}{c}{---}
             &20&\pheins 3 & 3&\multicolumn{2}{c}{$< 3$}
             &\multicolumn{2}{c}{---} &\multicolumn{2}{c}{---}
             &0.37 &9&3 &\multicolumn{2}{c}{$<2$}\\
1433         &\multicolumn{2}{c}{---} &\multicolumn{2}{c}{---}
             &\multicolumn{2}{c}{---} &\multicolumn{2}{c}{---}
             &3&\pheins 1 & 5&\multicolumn{2}{c}{$< 3$}
             &\multicolumn{2}{c}{---} &\multicolumn{2}{c}{---}
             &0.06 &6&2 &\multicolumn{2}{c}{$<1$}\\
1493         &\multicolumn{2}{c}{---} &\multicolumn{2}{c}{---}
             &\multicolumn{2}{c}{---} &\multicolumn{2}{c}{---}
             &3&\pheins 1 & 4&\pheins6&\pheins4
             &\multicolumn{2}{c}{---} &\multicolumn{2}{c}{---}
             &0.09 &7&2  &2&1\\
1559         &\multicolumn{2}{c}{---} &\multicolumn{2}{c}{---}
             &\multicolumn{2}{c}{[150]} &0.3&0.1 &105&\pheins6 & 232 &3.9&0.4
             &\multicolumn{2}{c}{---} &\multicolumn{2}{c}{---}
             &4.13 &16&5 &4&1\\
1672         &\multicolumn{2}{c}{[190]} &\multicolumn{2}{c}{---}
             &\multicolumn{2}{c}{[170]} &\multicolumn{2}{c}{---}
             &106&\pheins 6 & 250 &3.8&0.9
             &47&\pheins 3  &5.7&1.3
             &1.95 &14&4 &3&1\\
2442         &\multicolumn{2}{c}{---} &\multicolumn{2}{c}{---}
             &\multicolumn{2}{c}{[65]} &\multicolumn{2}{c}{---}
             &74&\pheins 4 & 201 &2.6&0.5
             &\multicolumn{2}{c}{---} &\multicolumn{2}{c}{---}
             &1.17 &13&4 &2&1\\
3059         &\multicolumn{2}{c}{[45]} &11&8
             &\multicolumn{2}{c}{[16]} &\multicolumn{2}{c}{---}
             &29&\pheins 2 & 57 &\pheins7&\pheins1
             &\multicolumn{2}{c}{---} &\multicolumn{2}{c}{---}
             &0.65 &11&3 &3&1\\
5643         &\multicolumn{2}{c}{---} &\multicolumn{2}{c}{---}
             &\multicolumn{2}{c}{---} &\multicolumn{2}{c}{---}
             &64&\pheins 4 & 129 &\pheins2&\pheins1
             &\multicolumn{2}{c}{---} &\multicolumn{2}{c}{---}
             &0.92 &13&4 &2&1\\
7552         &\multicolumn{2}{c}{---} &\multicolumn{2}{c}{---}
             &\multicolumn{2}{c}{---} &\multicolumn{2}{c}{---}
             &\multicolumn{2}{c}{125$\pm$20} & 546 &2.0&0.5
             &\multicolumn{2}{c}{---} &\multicolumn{2}{c}{---}
             &2.16 &15&5 &2&1\\
\noalign{\smallskip}
\hline
\end{tabular}
\end{table*}

\section{Discussion}

\subsection{The radio--infrared correlation}\label{RFIR}

\begin{figure}
\centering
\includegraphics[bb = 126 126 699 477,width=8.8cm,clip]{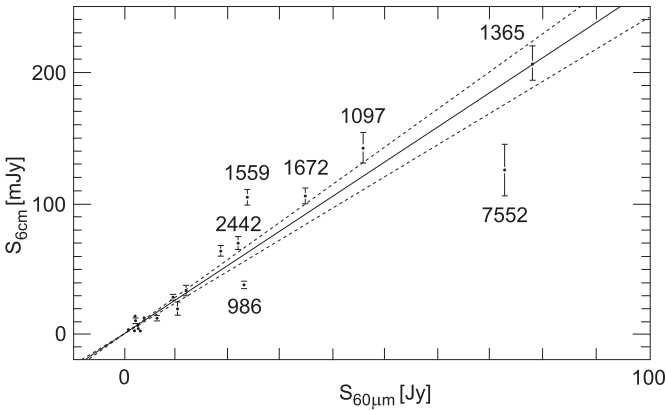}
\caption{The radio--far-infrared correlation for the sample of barred
galaxies: integrated radio continuum flux density at $\lambda6\cm$ versus
the integrated far-infrared flux density at $\lambda60\,\mu$m.
NGC names are indicated for bright galaxies}
\label{fig:fir}
\end{figure}

The integrated $\lambda$6~cm radio flux density $S_{6\rm cm}$
(Tables~\ref{tab:vla-flux} and \ref{tab:atca-flux})
is correlated with the integrated $\lambda60\,\mu$m far-infrared flux
density $S_{60\mu \rm m}$ (Tables~\ref{tab:vla-sample} and
\ref{tab:atca-sample}) as shown in Fig.~\ref{fig:fir}.
The correlation coefficient is $0.97\pm0.02$.
NGC~1559 lies well above the fitted line,
i.e., its radio emission is `too high' compared with its far-infrared
emission (cf.\ Section~5.2). This is possibly also true for NGC~1097.
NGC~986 and NGC~7552 are `too radio-faint', possibly due to
the incomplete $uv$ coverage of our observations.

The radio--far-infrared correlation has been studied in detail for large
samples of barred and non-barred galaxies. The average {\it flux density} ratio
$S_{6\rm cm}/S_{60\mu \rm m}$ (where $S_{6\rm cm}$ is measured in mJy and
$S_{60\mu \rm m}$ in Jy) is $3.0\pm0.3$ for the RC2 galaxies in the sample of
de Jong et al.\  (\cite{jong+85}) and $2.3\pm0.1$ for the 134 galaxies observed by
Unger et al.\ (\cite{unger+89}, scaled to $\lambda$6~cm).  The slope of the fitted
line in our Fig.~\ref{fig:fir} is $2.6\pm0.3$, in agreement with these results.
The radio continuum {\it luminosity} of spiral galaxies is also closely related
to the far-infrared luminosity (Condon\ \cite{condon92}, Niklas\ \cite{niklas97}).

Unger et al.\ found no significant difference in the radio/far-infrared ratio
either between Hubble classes or between barred and non-barred galaxies. The average
value of $S_{6\rm cm}/S_{60\mu \rm m}$ in our sample also indicates no general excess
of radio emission from barred galaxies.

A close correlation between radio continuum emission and dust emission
in the far-infrared has been found within many galaxies
(Bicay \& Helou\ \cite{bicay+helou90}, Hoernes et al.\ \cite{hoernes+98}),
and between radio continuum and the mid-infrared emission ($\lambda15\,\mu$m)
in the spiral galaxy NGC~6946 at all spatial scales
(Frick et al.\ \cite{frick+01}).
ISOCAM images at $\lambda7\,\mu$m and $\lambda15\,\mu$m are available for several
galaxies in our sample: NGC~1097, 1365, 1433, 1672, 4535 and 7552
(Roussel et al.\ \cite{roussel+01a}). The similarity to our radio maps
is striking and shows that the relationship holds not only for the integrated
flux densities and luminosities, but also for spatial scales down to our
resolution.

As synchrotron emission dominates at radio wavelengths longer than about
$\lambda$3~cm, the radio--infrared correlation cannot be explained
solely by thermal processes. Various interpretations are dicussed by
Hoernes et al.\ (\cite{hoernes+98}). As suggested by Niklas \& Beck
(\cite{niklas+beck97}), the radio--[far-]infrared correlation for bright galaxies
holds if the magnetic field is connected to the star formation rate
where the gas clouds may serve as the physical link.
For radio-weak galaxies, however, the far-infrared emission is dominated by cold dust
heated by the general radiation field which is not related with recent
star formation (Hoernes et al.\ \cite{hoernes+98}). Most of the galaxies
in our sample are bright enough to ensure that their far-infrared emission
is indeed a measure of star formation intensity.

In normal spiral galaxies, cool gas and magnetic fields are compressed in
various shocks,
followed by an increase in star formation. However, large-scale shock fronts in a
galaxy do not always enhance star formation. For example, the non-barred spiral
galaxy NGC~2276 interacts with some external (intracluster) gas, so that
a large-scale shock front forms on the leading side producing a ridge of strong
total and regular magnetic field without significant effect
on star formation (Hummel \& Beck \cite{hummel+beck95}).
As a consequence, this galaxy deviates from the radio--far-infrared correlation.

Barred galaxies also host large-scale shock fronts, identified with dust lanes.
However, shock fronts in bars are non-standard shocks in that they have enhanced 
velocity shear across them, similar to bow shocks. If the shear rate
$\partial V_i/\partial x_j$
exceeds the inverse time for star formation,
the gas density enhancement in the shock does not trigger star formation.
If the magnetic field is compressed in the shock, the ratio of
radio/far-infrared flux densities would be higher than normal.
However, for our sample this ratio and the average total field strengths
(Sect.~5.3) are similar to those of non-barred galaxies.
This indicates that large-scale field compression in the bar is generally small
and that the magnetic field is not frozen into the flow in the regions of
strong compression and shear (the dust lanes).
Nevertheless, the average surface brightness in radio continuum
and far-infared increases with increasing bar length (see Sect.~5.2).

\subsection{Radio emission and bar strength}

\begin{figure}
\centering
\includegraphics[bb = 142 125 644 441,width=8.8cm,clip]{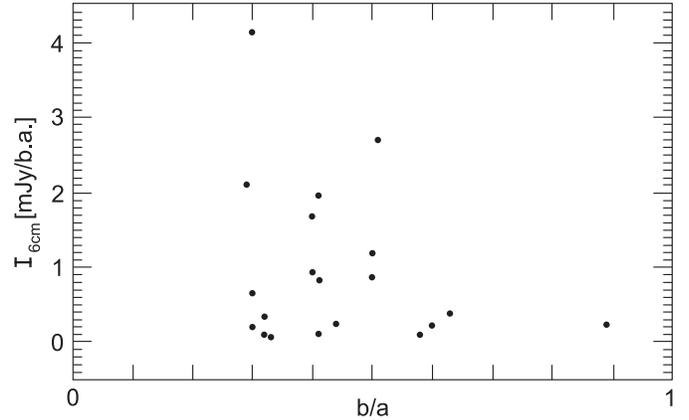}
\caption{Variation of the radio surface brightness $I_{6\rm cm}$
at $\lambda$6~cm with the deprojected aspect ratio $b/a$ of the bar
}
\label{fig:ba}
\end{figure}

Several quantitative measures of bar
strength have been suggested. Most of them are based
on purely geometric parameters such as the bar axial
ratio, where a smaller value of $b/a$ means a stronger bar
(Martin \cite{martin95}, Aguerri \cite{aguerri99},
Chapelon et al.\ \cite{chapelon+99},
Abraham \& Merrifield \cite{abraham+merrifield00}).
Smallest values of $b/a$ in our sample are
found in NGC~1300, 1433, 1493, 1559, 3059, 3359 and 7552
(see Tables~\ref{tab:vla-sample} and \ref{tab:atca-sample}),
but only NGC~1559
and NGC~7552 have a high surface brightness in radio continuum
(see Tables~\ref{tab:vla-flux} and \ref{tab:atca-flux}) and far-infrared.
NGC~1365 has the highest radio flux density $S_{6\rm cm}^*$
in our sample, although the aspect ratio of its bar is
relatively small ($b/a=0.51$). However, it hosts the longest bar
($\simeq29$~kpc)
in the sample. NGC~1559, 1672 and 7552 are fainter mainly because they are
smaller.
Their (distance-independent) radio surface bightness values $I_{6\rm cm}$
(measured in $\mu$Jy/beam area) are similar to or even larger than that of
NGC~1365
(see column 11 in Tables~\ref{tab:vla-flux} and \ref{tab:atca-flux}),
and the same is true for the typical
far-infrared surface brightness, which is
a measure of star formation rate per surface area.
Fig.~\ref{fig:ba} confirms that the radio surface brightness $I_{6\rm cm}$
is uncorrelated with the aspect ratio $b/a$ (correlation coefficient
of $0.58\pm0.15$).

\begin{figure}
\centering
\includegraphics[bb = 142 104 658 435,width=8.8cm,clip]{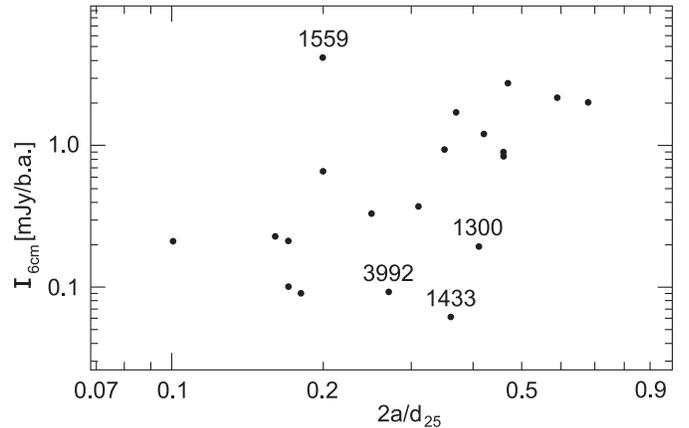}
\caption{Variation of the radio surface brightness $I_{6\rm cm}$
at $\lambda$6~cm with the relative bar length $2a/d_{25}$
(logarithmic scales). Deviating points are marked by their NGC name
}
\label{fig:ad}
\end{figure}


A physically motivated measure of the bar strength has been introduced by
Buta \& Block (\cite{buta+block01}) and Block et al. (\cite{block+01})
based, following Combes \& Sanders (\cite{combes+sanders81}),
on the maximum amplitude of
the tangential gravitational force relative to the mean axisymmetric radial
force. Thus defined, the strength parameter $Q_{\rm b}$ is sensitive
not only to the bar ellipticity, but also to its size and mass, and is
related to the quadrupole moment of the bar potential
(P.~Englmaier, priv. comm.).
A reliable estimation of $Q_{\rm b}$ involves careful analysis
of near-infrared galactic images (Quillen et al.\ \cite{quillen+94}).
Buta \& Block (\cite{buta+block01}) and Block et al. (\cite{block+01})
determine  $Q_{\rm b}$ for a selection of galaxies, but only six
of them belong to our sample. Therefore, we consider a simpler
(and admittedly incomplete) measure of the bar strength described
in what follows.

The quadrupole moment, with respect to the major axis,
of a homogeneous triaxial ellipsoid with semi-axes
$a,\ b$ and $c$ is given by
$\frac{1}{5}M(2a^2-b^2-c^2),$
where $M$ is the mass of the ellipsoid (Landau \& Lifshitz\ \cite{landau76}).
Assuming that the
vertical scale height of the bar is much smaller than its size
($c \ll a, b$) (e.g.,
Buta \& Block \cite{buta+block01}), the quadrupole moment normalized to
$M_{25}R_{25}^4$ (with $M_{25}$ the mass within the radius
$R_{25}$) is given by
\begin{equation}
\Lambda = \left(\frac{a}{R_{25}}\right)^4 \frac{b}{a}
 \left(1-{\textstyle\frac12}\frac{b^2}{a^2}\right)
 \frac{\sigma_{\rm b}}{\sigma}\;,
\label{qmoment}
\end{equation}
where $\sigma_{\rm b}$ and $\sigma$ are the average mass
surface densities of the bar and within $R_{25}$, respectively, and
we have omitted numerical factors of order unity.
The relative bar length $a/R_{25}=2a/d_{25}$, rather than the bar axial
ratio, is the dominant factor in $\Lambda$. Moreover,
70\% of galaxies in our sample have $b/a>0.4$, and $\Lambda$ varies
just by 30\% for $0.4<b/a<1$. Therefore, $b/a$ is a poor measure
of bar strength, especially for our sample, as it does
not discriminate well enough between galaxies with $b/a\ga0.5$.

Since $\Lambda$ depends strongly on the relative bar length $2a/d_{25}$, we
can reasonably expect that radio emission is correlated
with this parameter. This expectation is confirmed by
the high correlation between the radio surface brightness $I_{6\rm cm}$
at $\lambda6\cm$ and the relative bar length
(correlation coefficient of $0.86\pm0.06$, with NGC~1559 excluded
-- see below), confirmed by Student's $t$ test.
Fig.~\ref{fig:ad} shows this correlation in logarithmic scales.
Although the scatter is stronger than that in Fig.~\ref{fig:fir},
the correlation is not weaker than other correlations
discussed for barred galaxies in the current literature
(cf.\ Chapelon et al.\ \cite{chapelon+99}). We conclude that a stronger bar results
in an overall enhancement of the total radio emission in the bar region despite
a relatively weak compression of the regular magnetic field
near the dust lanes, as discussed in Section~\ref{RFIR}.
As noted by Block et al.\ (\cite{block+01}), longer bars can
produce more extensive deviations form axial symmetry in the gas velocity 
because the relative tangential force is stronger when the end of the bar is
farther from the (axisymmetric) bulge;
this may be the physical reason for the correlation
shown in Fig.~\ref{fig:ad}.

With the most strongly deviating galaxy excluded
(NGC~1559), the data shown in Fig.~\ref{fig:ad}
can be fitted with a power law
\[
I_{6cm}\propto (2a/d_{25})^{1.5\pm0.4}.
\]
A similar dependence is valid for the far-infrared surface brightness
$I_{60\mu m}$ (exponent $1.5\pm0.5$).

We have been unable to
include the dependence on the surface mass densities into our
measure of the bar strength, and this plausibly contributes
into the scatter of the data points around the fit. It
is difficult to say whether or not $\sigma_{\rm b}/\sigma$ is
correlated with $2a/d_{25}$. If $\sigma_{\rm b}/\sigma$ is
independent of $2a/d_{25}$, the above fit implies an
approximate scaling
\[
I_{6cm}\propto \Lambda^{0.4\pm0.1}.
\]

There are a few deviations from the above correlation
(see Fig.~\ref{fig:ad}).
NGC~1559 has the largest ratio of radio to far-infrared flux
densities and the highest radio surface brightness,
and so deviates strongly from the radio--infrared
correlation as well (see Fig.~\ref{fig:fir}).
NGC~1559 is not a member of any group or cluster of galaxies and
has no nearby companion (Zaritsky et al.\ \cite{zaritsky+97}).
High-resolution radio and optical observations are required.

On the other hand, NGC~1300, 1433 and 3992 are radio-weak in spite
of their relatively long bars (Fig.~\ref{fig:ad}). Their far-infrared flux
density and thus their star formation rate is low.
Apart from an usually small value of $\sigma_b/\sigma$
for these galaxies, other reasons for these deviations are concievable.
Martinet \& Friedli (\cite{martinet+friedli97}) argue that some galaxies
with strong bars have settled into a quiescent state after an episode of
vigorous star formation which has transformed most of the gas into stars.
Alternatively, Tubbs (\cite{tubbs82}) and Reynaud \& Downes
(\cite{reynaud+downes98})
found indications for suppression of star formation in fast flows of the
gas along the bar.
The field strength should be low in the first case, because there is not enough
gas to hold the field or the dynamo is not able to maintain a strong
magnetic field. In the second case the field should be strong, but the galaxy
does not host enough cosmic-ray electrons to generate strong synchrotron radiation.
This can be verified by comparing regular magnetic field strengths
deduced from Faraday rotation and polarized intensity 
from further radio observations with higher sensitivity.

Measurements of the star formation efficiency {\rm SFE} may also help:
In the first case, the content of molecular gas should be low, with a {\rm SFE}
similar to that in spiral galaxies, while in the second case the {\rm SFE}
should be exceptionally small. 
Existing data seem to favour a {\it higher} {\rm SFE}
in barred galaxies compared to non-barred ones (Young \cite{young93}), but
the infrared luminosity is dominated by the central region
where star formation is triggered by gas inflow (see Roussel et al.\ \cite{roussel+01b}).
The {\rm SFE} in the bar itself (and its possible suppression by a fast gas flow)
should be subject to future investigations.

\subsection{Magnetic field strength}

The estimates of the total magnetic field strength in our Galaxy, derived from
$\gamma$-ray data and the local cosmic-ray energy density (Strong et
al.\ \cite{strong+00}), agree well with equipartition
values from radio continuum data (Berkhuijsen, in Beck \cite{beck01}),
so that the equipartition assumption is a useful estimate, at least on scales
of more than a few kpc.

From the integrated flux density $S_{6\rm cm}$ at $\lambda$6~cm and the
solid angle
of the integration area, the surface brightness $I_{6\rm cm}$
and the corresponding
equipartition strength of the {\it total} magnetic field $B_{\rm tot}$ were computed
(Tables~\ref{tab:vla-flux} and \ref{tab:atca-flux}), assuming for all
galaxies a thermal contribution to the surface brightness
at $\lambda6$~cm of 20\% and a spectral index
$\alpha_n$ of the nonthermal emission of 0.85, which is the mean value
for spiral galaxies of type Sb and later (Niklas et al.\ \cite{niklas+97}).

Spectral indices $\alpha$ between $\lambda22$~cm and $\lambda6$~cm
($S_{\nu}\propto \nu^{-\alpha}$)
can be computed from our VLA data (given in Table~\ref{tab:vla-flux}).
The values lie in the range 0.71 and 0.97 which is in the range typical 
of normal spiral galaxies (Niklas\ \cite{niklas95}). 
Nonthermal spectral indices $\alpha_n$ cannot be
determined with data at only two frequencies.

We adopted the standard cosmic-ray proton-to-electron ratio $K$ of 100, a pathlength
through the disc of 1~kpc/$|\cos i|$, and assumed that
the regular field is in the galaxy's plane and
the random field is statistically isotropic.  Uncertainties in any of these
parameters of $\le50\%$ lead to an error of $\le13\%$ in $B_{\rm tot}$.
We estimate the total error in $B_{\rm tot}$ to be about 30\%. The relative errors
between galaxies are smaller because some of the input parameters
(e.g.\ the proton-to-electron ratio) are not expected to vary strongly
from one galaxy to another.

With the above assumptions, $B_{\rm tot}$ is related to the
average synchrotron volume emissivity $\epsilon$
and surface brightness $I$ (neglecting a term weakly varying with inclination $i$) by
\[
B_{\rm tot}\propto\epsilon^{1/(\alpha_n +3)}\quad (\mbox{where }\epsilon \propto I |\cos i|).
\]
Note that the equipartition field strengths are about 10\% larger than
the field strengths derived from the standard minimum-energy formula
(which should be used with caution, see Beck\ \cite{beck00}).

The average total magnetic field strength $B_{\rm tot}$, according to
Tables~\ref{tab:vla-flux} and \ref{tab:atca-flux}
(representing the average synchrotron emissivity), is
a function of neither Hubble type (SBb--SBc) nor luminosity class (I--III),
which has also been found for a much larger sample of barred and non-barred
spiral galaxies (Hummel\ \cite{hummel81}). The average total field strength
$B_{\rm tot}$ is $10\pm3\mkG$ for our sample,
similar to the average minimum-energy field strength of $\simeq8\mkG$
for the large galaxy sample (Hummel et al.\ \cite{hummel+88}) and
to the mean equipartition value of $11\pm4\mkG$ of the sample of 146 late-type
galaxies calculated by Fitt \& Alexander (\cite{fitt+alexander93}),
corrected to $K=100$.
Niklas (\cite{niklas95}) derived a mean equipartition value of $9\pm3\mkG$
for his sample of 74 spiral galaxies.
Hummel (\cite{hummel81}) also found no significant emissivity difference
between barred and non-barred galaxies.

The following galaxies have the strongest total magnetic field
in our sample, as evidenced by their high radio surface brightness:
NGC~1365, 1559, 1672 and 7552 (see Tables~\ref{tab:vla-flux} and \ref{tab:atca-flux}).
This indicates that the total field strength is highest for galaxies with the
(relatively) longest bars, with the exception of NGC~1559 that
has a short bar (see Fig.~\ref{fig:ad}).

The last column in Tables~\ref{tab:vla-flux} and \ref{tab:atca-flux} gives
the average equipartition strength
$B_{\rm reg}$ of the {\it resolved regular} magnetic field, derived from the
polarized surface brightness averaged over the galaxy.
$B_{\rm reg}$, in contrast to $B_{\rm tot}$, depends on
the linear resolution within a galaxy and thus on its physical size, its
distance and its inclination.
However, $p_{\lambda}$ and $B_{\rm reg}$ in Tables~\ref{tab:vla-flux} and
\ref{tab:atca-flux} do not correlate with distance of the galaxy.
As a test, we smoothed the $\lambda$6~cm map of NGC~1097 by enlarging the beam size
from 30\arcsec\ to 60\arcsec\ and to 90\arcsec\, which corresponds to
increasing the galactic distance by factors 2 and 3. The
degree of polarization decreased from 8.5\% to 7\% and 6\%,
respectively, and the strength of the resolved regular magnetic field decreased
from 4.3 to 4.0 and $3.6\mkG$, respectively,
remaining above the sample average.
Hence the values of $p_{\lambda}$ and $B_{\rm reg}$
in Tables~\ref{tab:vla-flux} and \ref{tab:atca-flux}
seem to depend only weakly on distance to the galaxies,
implying that our observations generally resolve most of the structure
in the regular magnetic field, at least for large galaxies and
at distances of up to about 40\,Mpc.

Average polarized surface brightness (and thus $B_{\rm reg}$) values
are similar for the galaxies of our sample. The exceptions are NGC~1097
and NGC~1559 with $B_{\rm reg}\simeq 4\mkG$, above the average of $2.5\pm0.8\mkG$.
NGC~1097 probably drives a strong dynamo where field amplification is supported
by shear in the velocity field (Moss et al.\ \cite{moss+01}, Paper~II). The
degree of polarization at $\lambda$6~cm, signature of the
degree of uniformity of the resolved field, is also high in NGC~1097
(see Fig.~25). NGC~1559, 1672 and 7552 are similar
candidates for a strong dynamo, but the present radio observations (Fig.~26) have
insufficient linear resolution at the relatively large distances of these galaxies
to reveal the true strength of the regular fields and their detailed structure.

NGC~1300, NGC~3992 and NGC~4535 have the highest degrees of
polarization but only low total surface brightness. They host weak
but ordered magnetic fields with spiral patterns, similar
to those in non-barred galaxies (Beck \cite{beck00}).

\subsection{Bars and global magnetic field structure}

A classification system of barred galaxies was
introduced by Martinet \& Friedli (\cite{martinet+friedli97}), based on the
axis ratio $b/a$ (see Tables~\ref{tab:vla-sample} and \ref{tab:atca-sample}) and on the
star formation rate (SFR) measured by the far-infrared luminosity.
Galaxies with large $b/a$ are generally weak in star formation (class I),
but some have a high SFR (class II). Galaxies with small $b/a$ have a large
spread in SFR: from high (class III) to weak (class IV).
Galaxies of class IV in Martinet \& Friedli
have strong bars, but low SFR (see Sect.~5.2).

Here we propose that there are basic differences among barred galaxies
concerning their magnetic field structure and strength which may reflect
physical properties of barred galaxies like the gas flow, the shock strength
in the bar and the presence of a circumnuclear ring.

Firstly, barred galaxies can have low radio luminosity because they are small
(NGC~1313, 1493 and 5068), or because their gas content and star formation
activity is small in spite of their large bars (NGC~1300 and 1433).
Little or no polarization is detected in these galaxies.
In galaxies with small bars the radio
continuum morphology is formed as a result of star formation
in the spiral arms, as in NGC~2336, 3359, 3953, 3992, 4535, 5643, and also M83
observed previously by Beck (\cite{beck00}). The bar is
of little importance for the overall radio properties of these galaxies.
The average degree of radio polarization (i.e., the degree of field regularity)
seems to be controlled by the spiral structure rather than the bar, being
low in flocculent spirals and high when massive spiral arms are present.
Regular fields are often enhanced in {\it interarm} regions
between optical spiral arms, e.g. in NGC~3359, NGC~4535 and M83, similar to
non-barred galaxies.

Secondly, galaxies with long bars and strong star formation
have a high radio luminosity and a strong total magnetic field
($B_{\rm tot}\ge10\mkG$) (NGC~1097, 1365, 1672, 2442 and 7552, and
also NGC~3627 observed previously by Soida et al. \cite{soida+01}).
NGC~1097, 1365, 1672 and 7552 have a
high polarization surface brightness and a strong regular field which is
enhanced {\it upstream} of the shock fronts in the bar.
The magnetic field lines upstream of the dust
lanes are oriented at large angles with respect to the bar and turn
smoothly towards the dust lanes along the major axis of the bar.
This is accompanied by large-scale field enhancements associated with, e.g.,
strong shear in the velocity field and/or strong dynamo action rather than
enhanced gas density.
Gas inflow along the bar may lead to circumnuclear rings which have been
detected already in NGC~1097 (Hummel et al.\ \cite{hummel+87},
Gerin et al.\ \cite{gerin+88}), NGC~2442 (Mihos \& Bothun \cite{mihos+bothun97}),
NGC~7552 (Forbes et al.\ \cite{forbes+94a}, \cite{forbes+94b})
and possibly in NGC~1365 (Sandqvist et al.\ \cite{sandqvist+95}),
and should be searched for in the other
radio-bright galaxies.

NGC~7479 is anomalous in the radio range as it possesses
a nuclear `jet' (Laine \& Gottesman \cite{laine+gottesman98}).
Indications of a weaker nuclear jet have been found
in NGC~1365 by Sandqvist et al.\ (\cite{sandqvist+95}).

For NGC~986, 1559 and 3059
the resolution and sensitivity of the present observations
are insufficient to reveal their detailed field structure.

NGC~7552 is a special case. Its radio surface brightness
is high (i.e., the total magnetic field is strong, see Table~\ref{tab:atca-flux}),
but still too low to be consistent with its far-infrared flux density
(see Fig.~\ref{fig:fir}). NGC~7552 hosts a starburst ring and may drive
a `galactic superwind' (Forbes et al.\ \cite{forbes+94a}).
As a member of a galaxy group, it may be subject to tidal interactions.
It seems possible that the magnetic
field is still not strong enough to hold the large number of cosmic-ray electrons
produced due to the high star formation activity. However, major distortions
of our radio map by instrumental effects cannot be excluded. Further
radio observations are required.

\section{Conclusions}

We observed a sample of 20 barred galaxies with the VLA and ATCA radio
telescopes. Polarized radio emission was detected in 17 galaxies.

The flux densities in the radio continuum and the far-infrared spectral ranges
are closely correlated in our sample. The average radio/far-infrared
flux density ratio and equipartition strength of the total magnetic
field are similar to
those in non-barred galaxies. These properties are
apparently connected to the star formation rate and possibly
controlled by the density of cool gas.
Radio surface brightness and present star formation activity are
highest for galaxies with a high content of molecular gas
and long bars where the velocity field is distorted over a large volume.
The radio surface brightness is correlated with a newly
introduced measure of bar strength proportional to the quadrupole
moment of the gravitational potential.
However, a few galaxies with strong bars are
not radio bright, possibly because their molecular gas has been
depleted in a star formation burst.

In barred galaxies with low or moderate radio surface brightness, the regular field
(traced by the polarized radio emission) is strongest between the optical
spiral arms (e.g.\ NGC~3359 and 4535) or has a diffuse distribution
(e.g.\ NGC~3059 and 3953).
In radio-bright galaxies, the pattern of the regular field can,
however, be significantly
different: the regular magnetic field may have a broad local maximum
in the bar region upstream of the dust lanes, and the field lines are
oriented at large angles with respect to the bar (NGC~1097, 1365, 1672 and 7552).
We propose that shear in the velocity field around a large bar may enhance
dynamo action and explain the observed strong regular fields.
Strong bar forcing induces shear in the velocity field and enhancements in the
regular magnetic field, and polarized emission traces such shear motions.

The southern galaxies NGC~986, 1559, 1672 and 7552 show strong polarization and
are promising candidates for further studies with high resolution.
Circumnuclear rings are already known to exist in NGC~1097, NGC~1365,
NGC~2442 and NGC~7552 and should be searched for in NGC~986, 1559 and 1672.

\begin{acknowledgements}

The authors would like to express special thanks to Dr.~Elly
Berkhuijsen for critical reading of the manuscript and many useful
suggestions. Useful discussions with Drs.~Peter Englmaier and
David Moss are gratefully acknowledged.
ME is grateful to the ATNF for providing support and facilities.
His work in Australia was funded through grant No.\ Eh 154/1-1 from the
Deutsche Forschungsgemeinschaft.
VS acknowledges financial support from the RFBR/DFG programme
no.~96-02-00094G and from the Royal Society.
DDS acknowledges financial support from the RFBR programme
no.~01-02-16158.
JIH acknowledges the Australian Academy of Science for
financial support under their Scientific Visits to Europe Program
and the Alexander von Humboldt Foundation for its support.
VS, JIH, AS and DS are grateful to the MPIfR for support and hospitality.
This work was supported by the PPARC Grant PPA/G/S/1997/00284,
the NATO Collaborative Linkage
Grant PST.CLG~974737 and the University of Newcastle (Small Grants Panel).
This work benefited from the use of NASA's Astrophysics Data System
Abstract Service.

\end{acknowledgements}

\begin{figure*}
\hbox to \textwidth{
\includegraphics[bb = 39 173 567 657,width=8.8cm,clip]{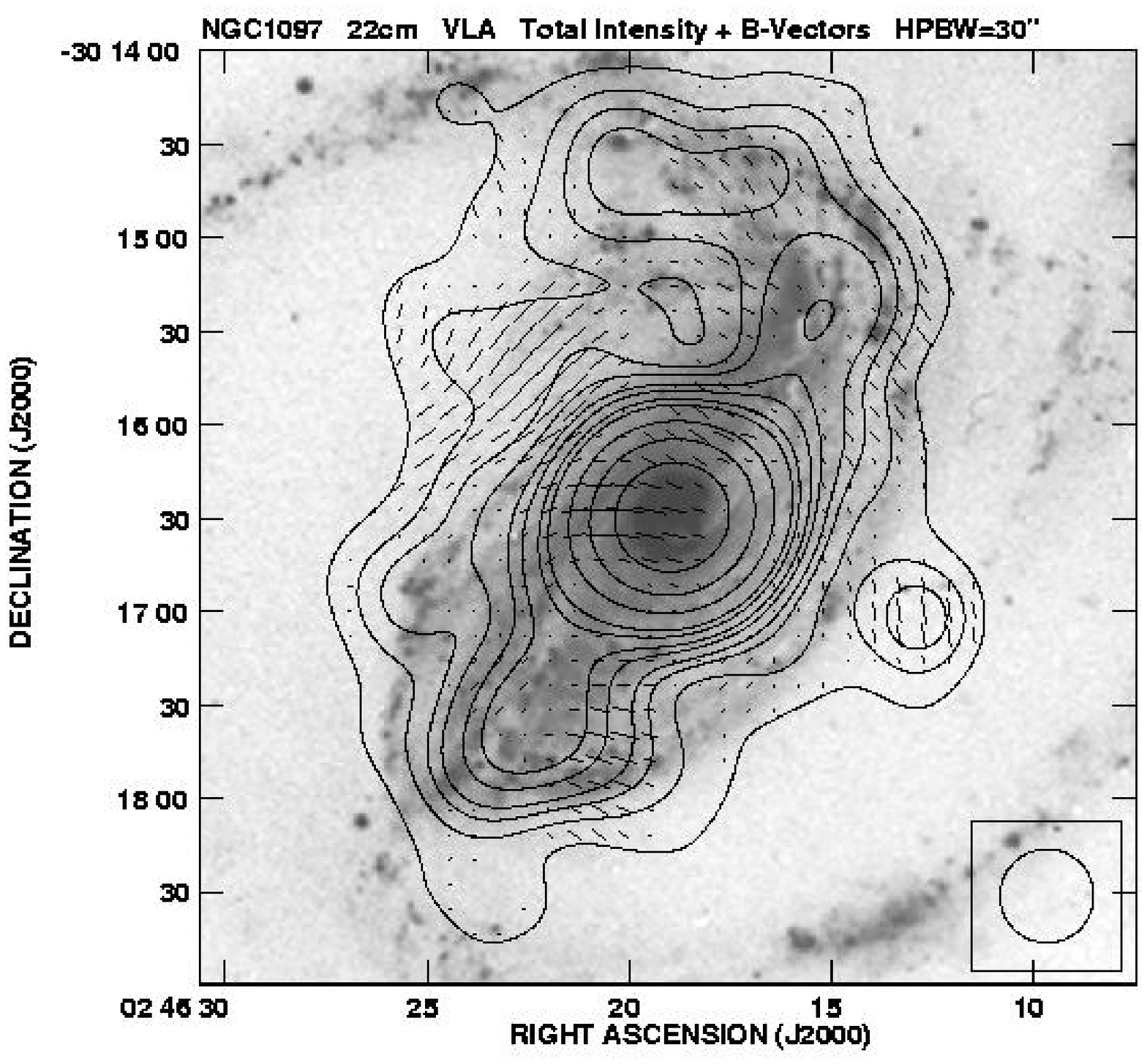}
\hfill
\includegraphics[bb = 39 173 567 657,width=8.8cm,clip]{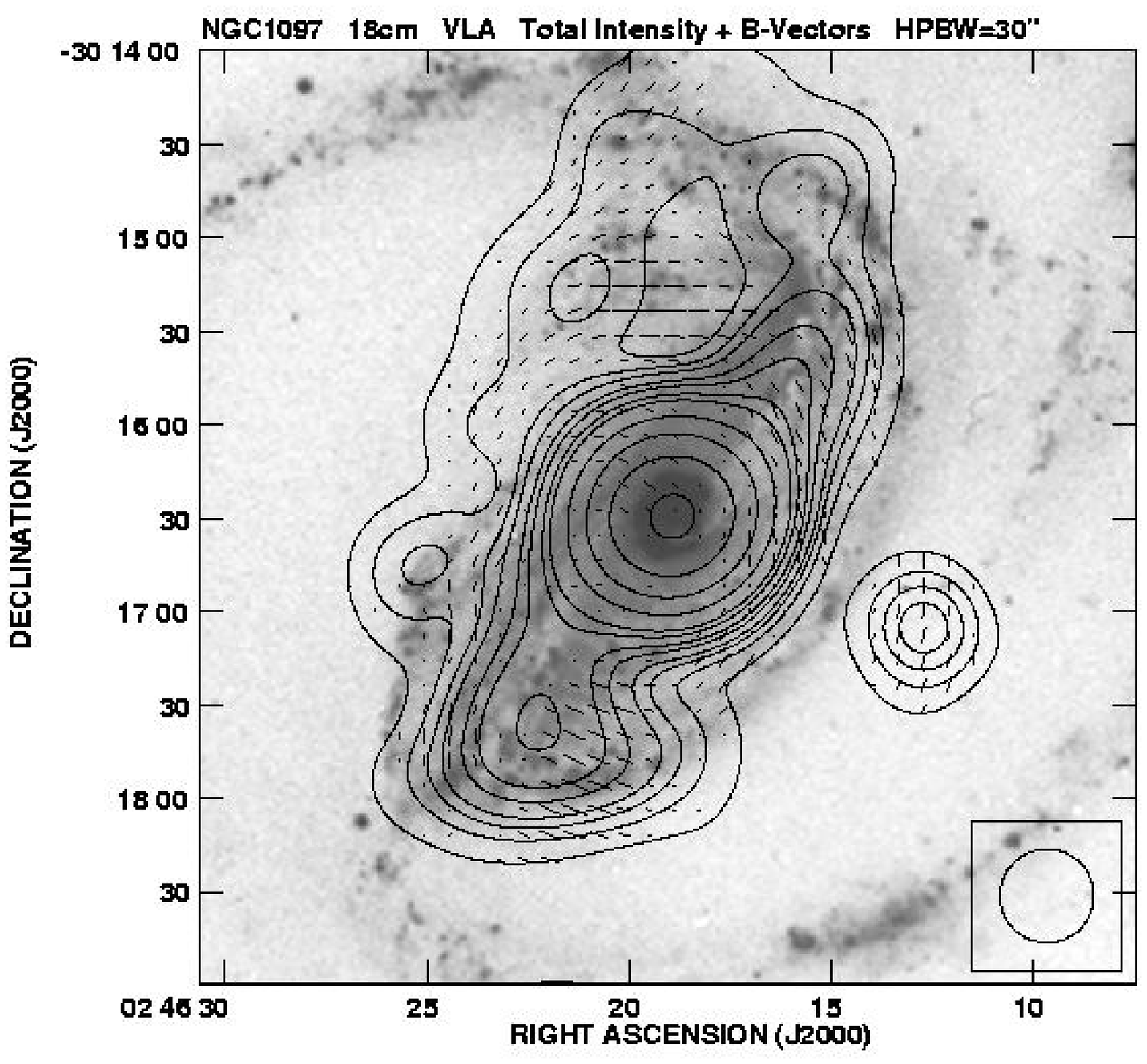}
}
\vspace{1.5cm}
\hbox to\textwidth{
\includegraphics[bb = 39 173 567 657,width=8.8cm,clip]{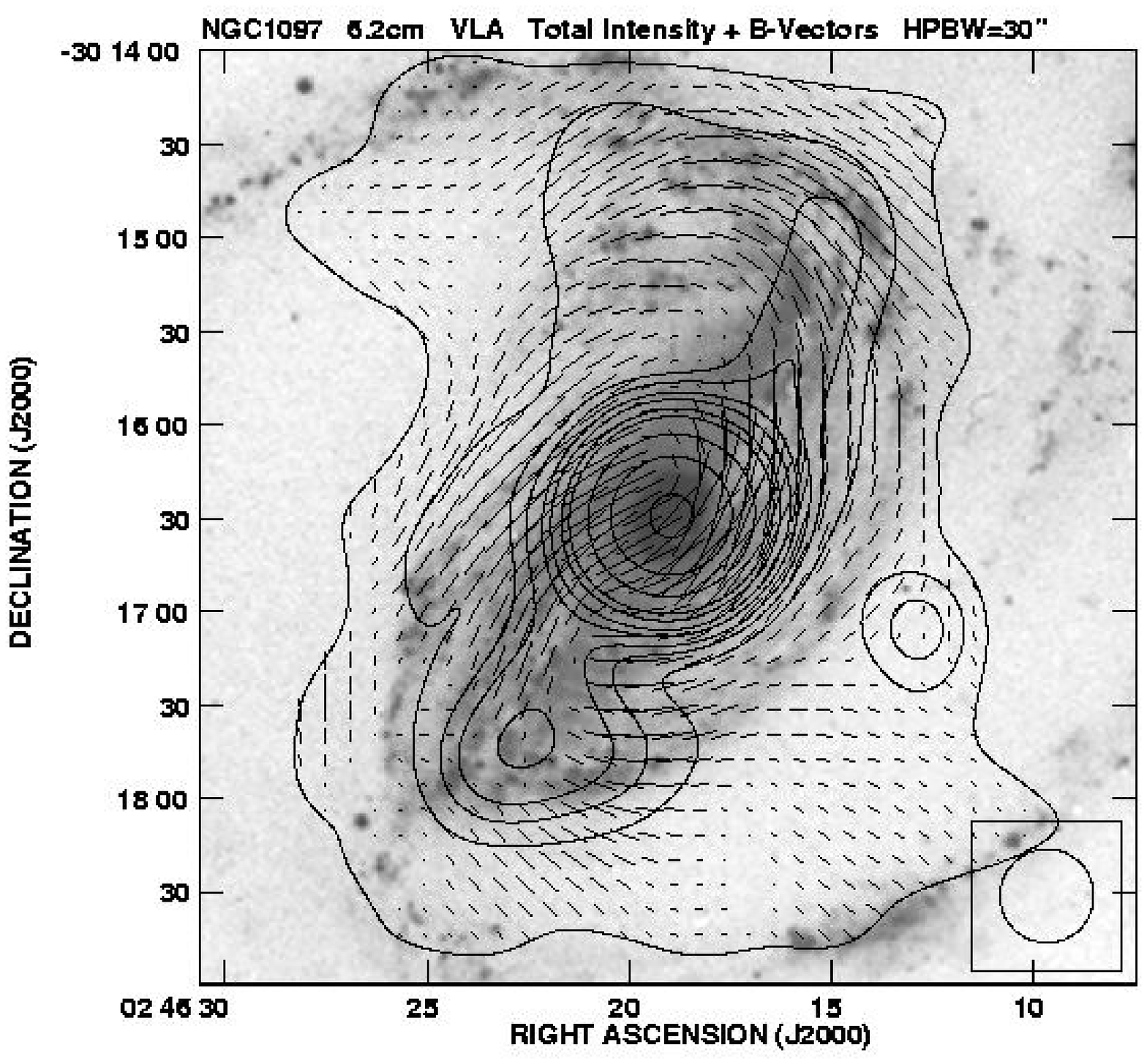}
\hfill
\includegraphics[bb = 39 173 567 657,width=8.8cm,clip]{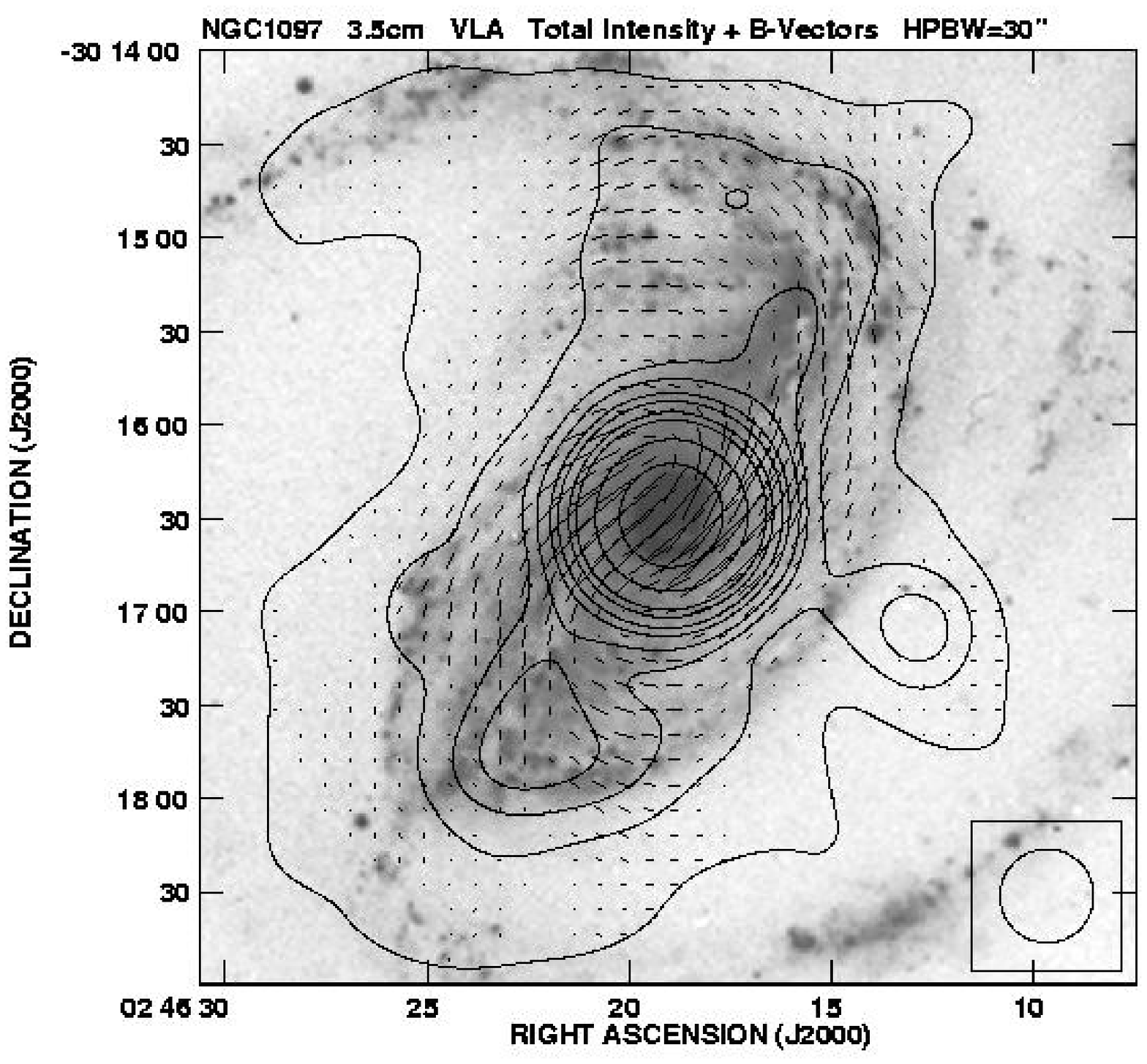}
}
\caption{Total intensity
contours and the observed $B$-vectors of polarized intensity
($E$-vectors turned by 90\degr, uncorrected
for Faraday rotation) of NGC~1097, overlayed onto an optical image
kindly provided by H.~Arp.
The contour intervals are at 1, 2, 3, 4, 6, 8, 12, 16, 32, 64, 128, 256 $\times$
the basic contour level which is 700, 500, 500 and 400 $\mu$Jy/beam area
at $\lambda$22, 18, 6.2 and 3.5~cm, respectively.
A vector of 1\arcsec\ length corresponds to a polarized intensity
of 20~$\mu$Jy/beam area.
The half-power width of the synthesized beam is shown in the corner
of each panel.}
\end{figure*}


\begin{figure*}
\hbox to \textwidth{
\includegraphics[bb = 39 163 567 667,width=8.8cm,clip]{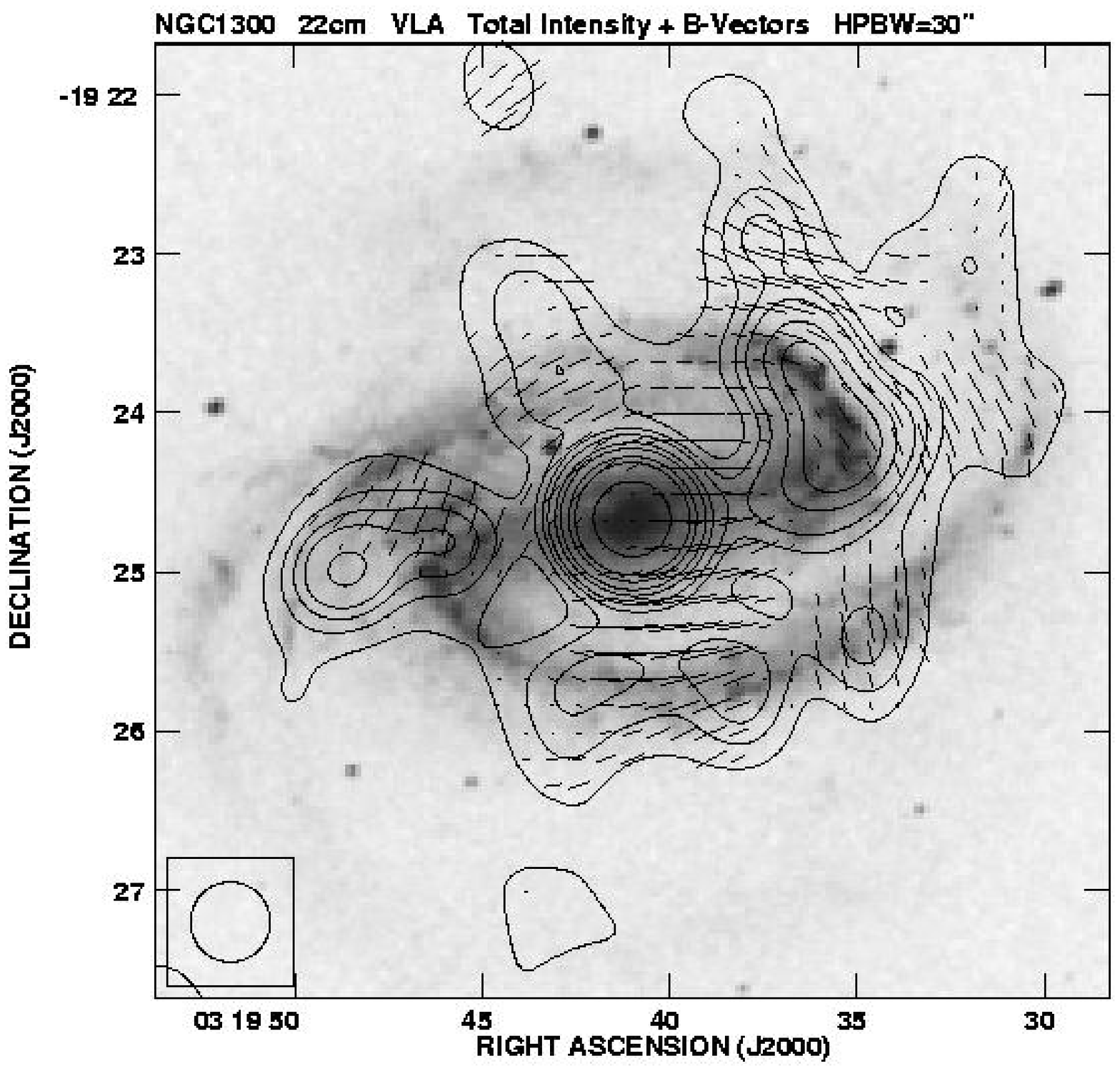}
\hfill
\includegraphics[bb = 39 163 567 667,width=8.8cm,clip]{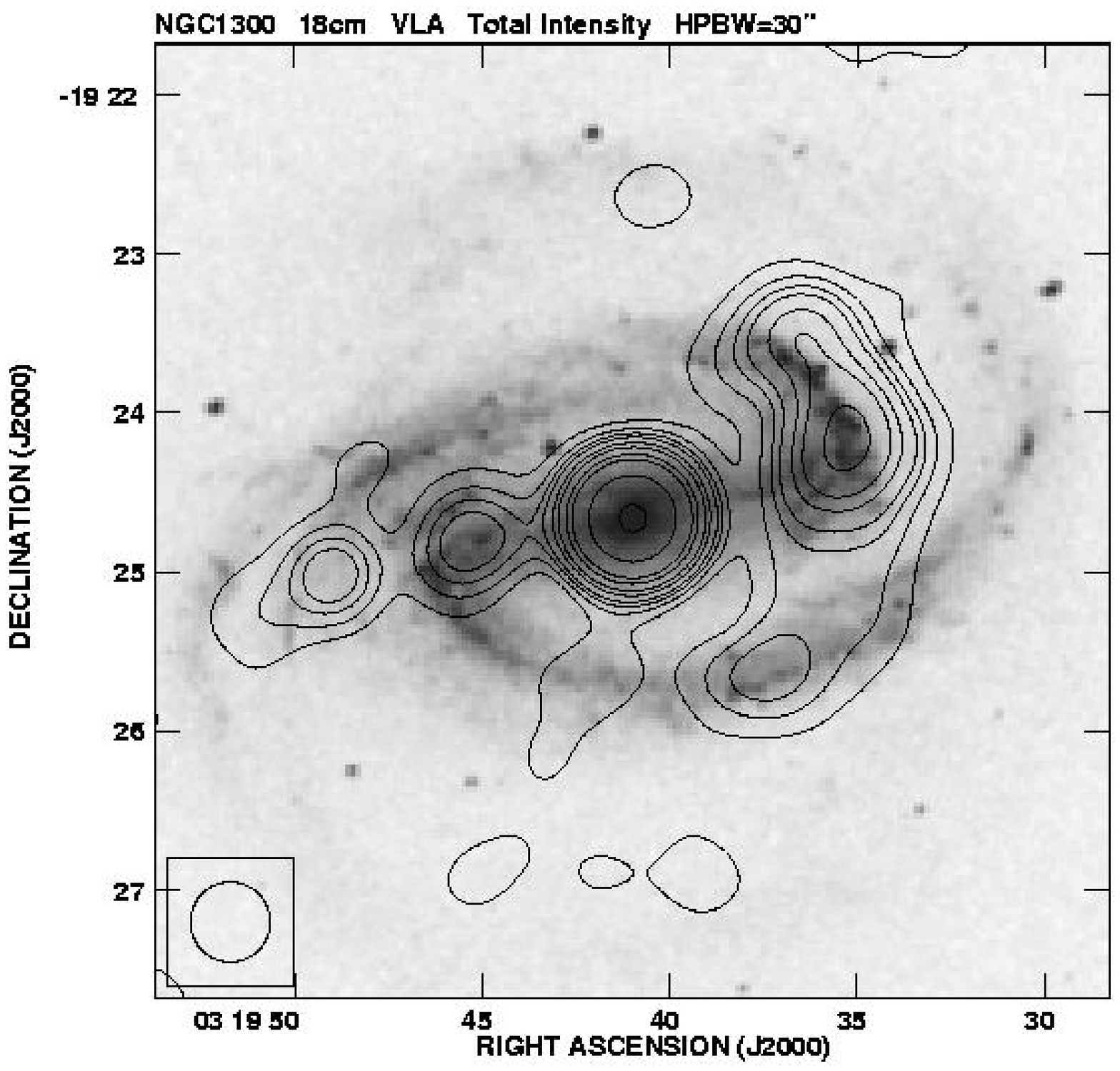}
}
\vspace{1.5cm}
\hbox to\textwidth{
\includegraphics[bb = 39 163 567 667,width=8.8cm,clip]{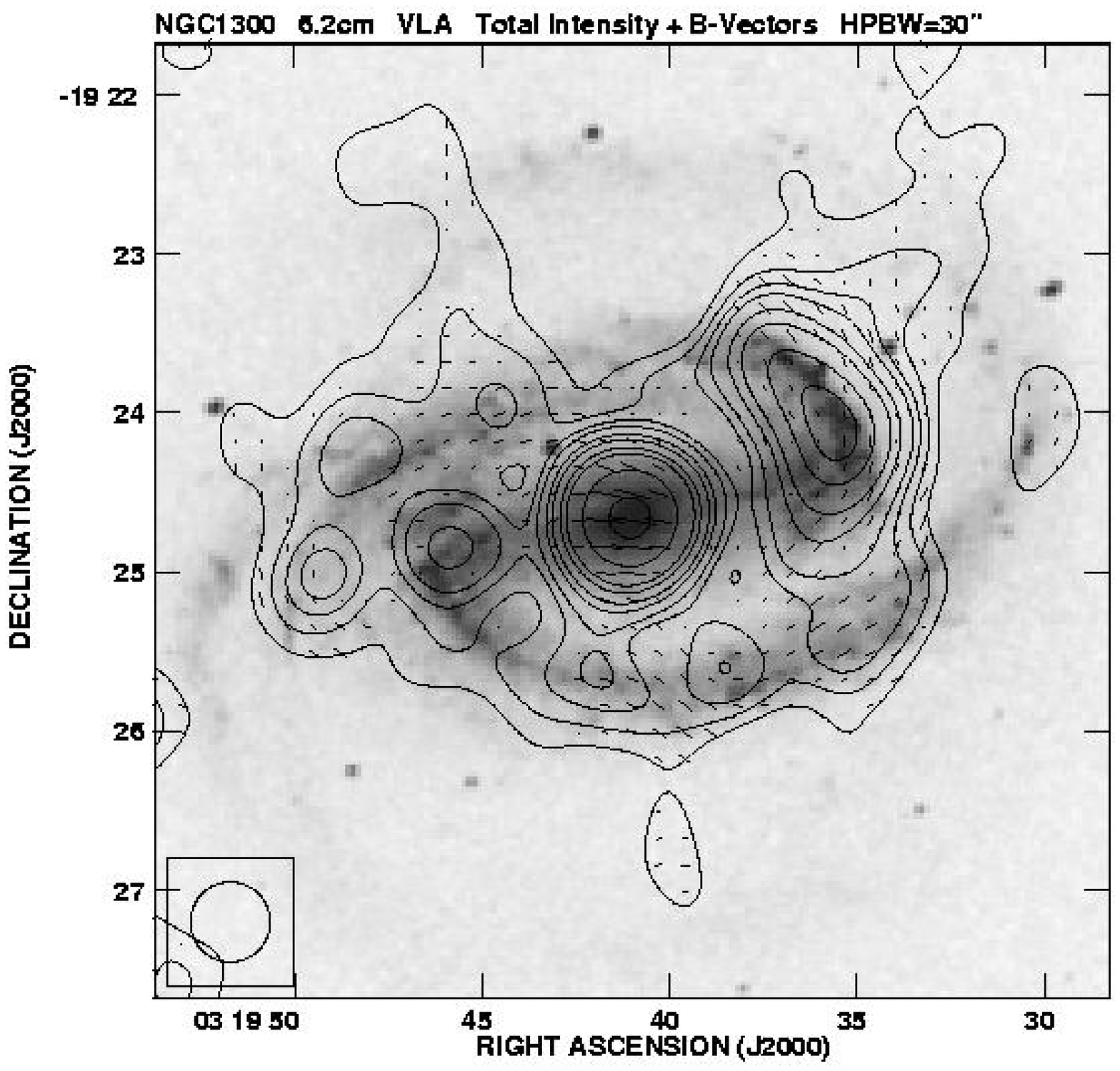}
\hfill
\includegraphics[bb = 39 163 568 667,width=8.8cm,clip]{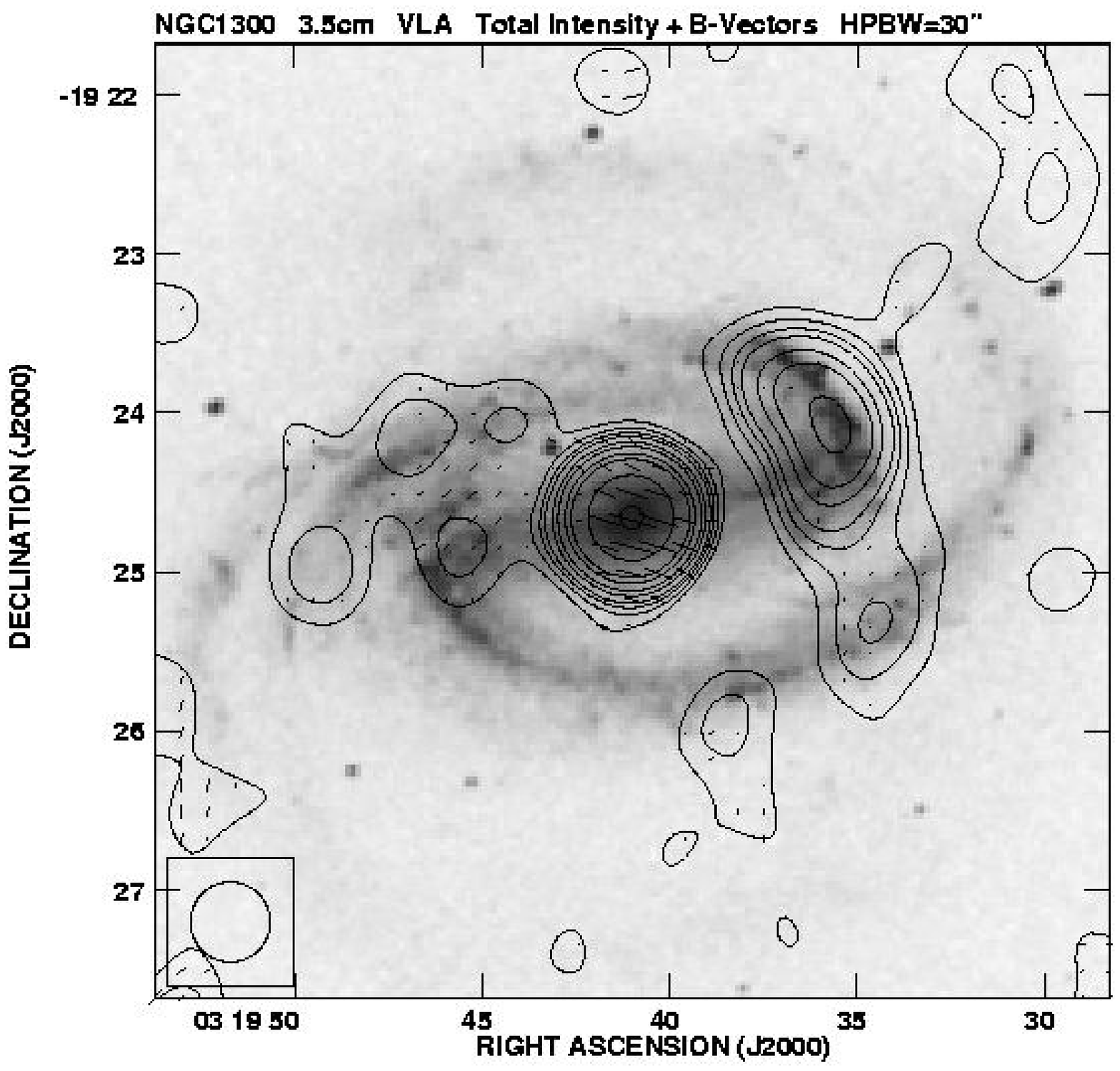}
}
\caption{Total intensity contours and the observed $B$-vectors of polarized emission
of NGC~1300, overlayed onto an optical image from the Digitized
Sky Surveys. The contour intervals are 1, 2, 3, 4, 6, 8, 12,
16, 32, 64, 128, 256 $\times$ the basic contour level, which is
150, 100, 50 and 40 $\mu$Jy/beam area
at $\lambda$22, 18, 6.2 and 3.5~cm, respectively.
A vector of 1\arcsec\ length corresponds to a polarized intensity
of 10~$\mu$Jy/beam area.}
\end{figure*}


\begin{figure*}
\hbox to \textwidth{
\includegraphics[bb = 39 173 567 657,width=8.8cm,clip]{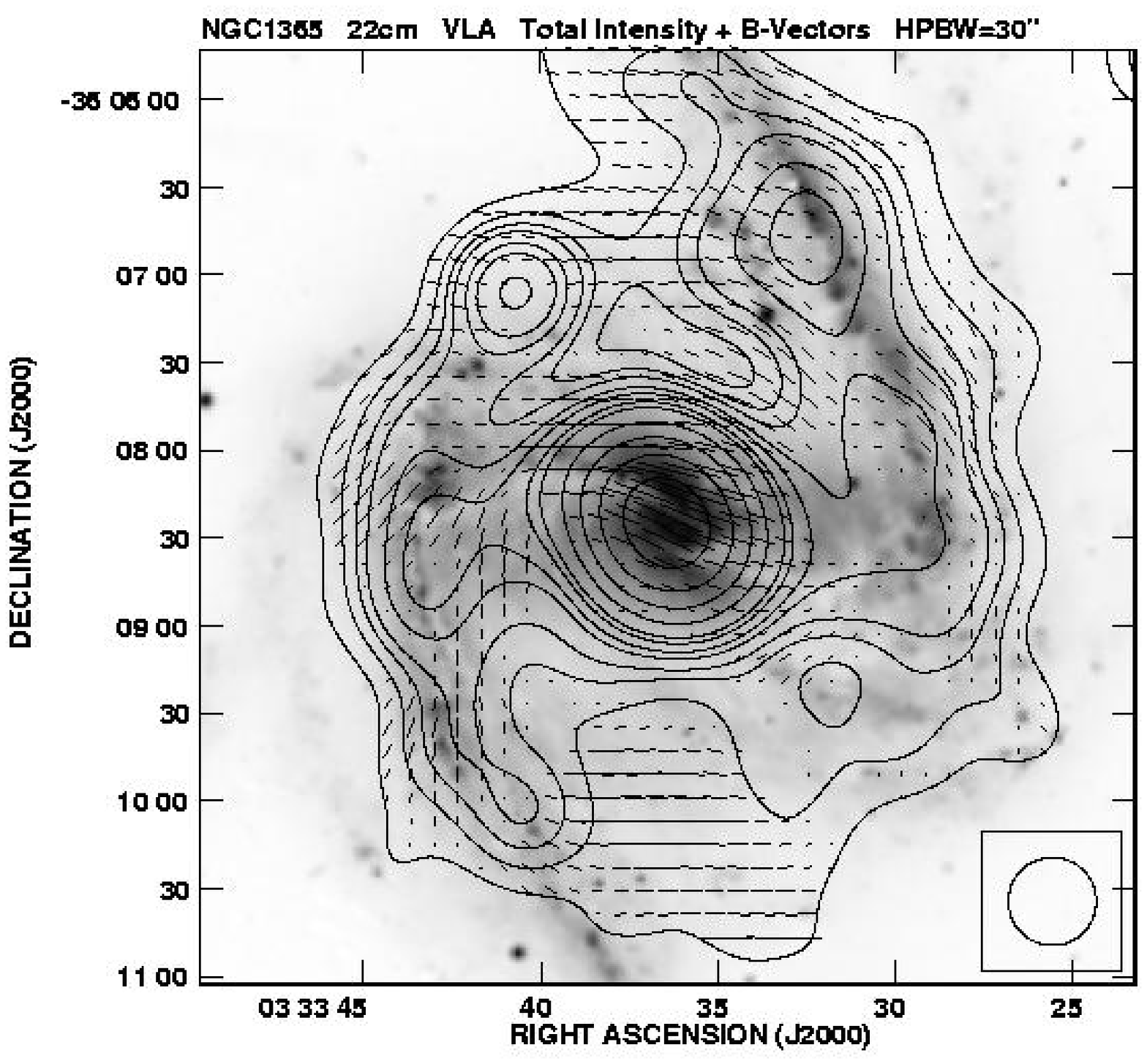}
\hfill
\includegraphics[bb = 39 173 567 657,width=8.8cm,clip]{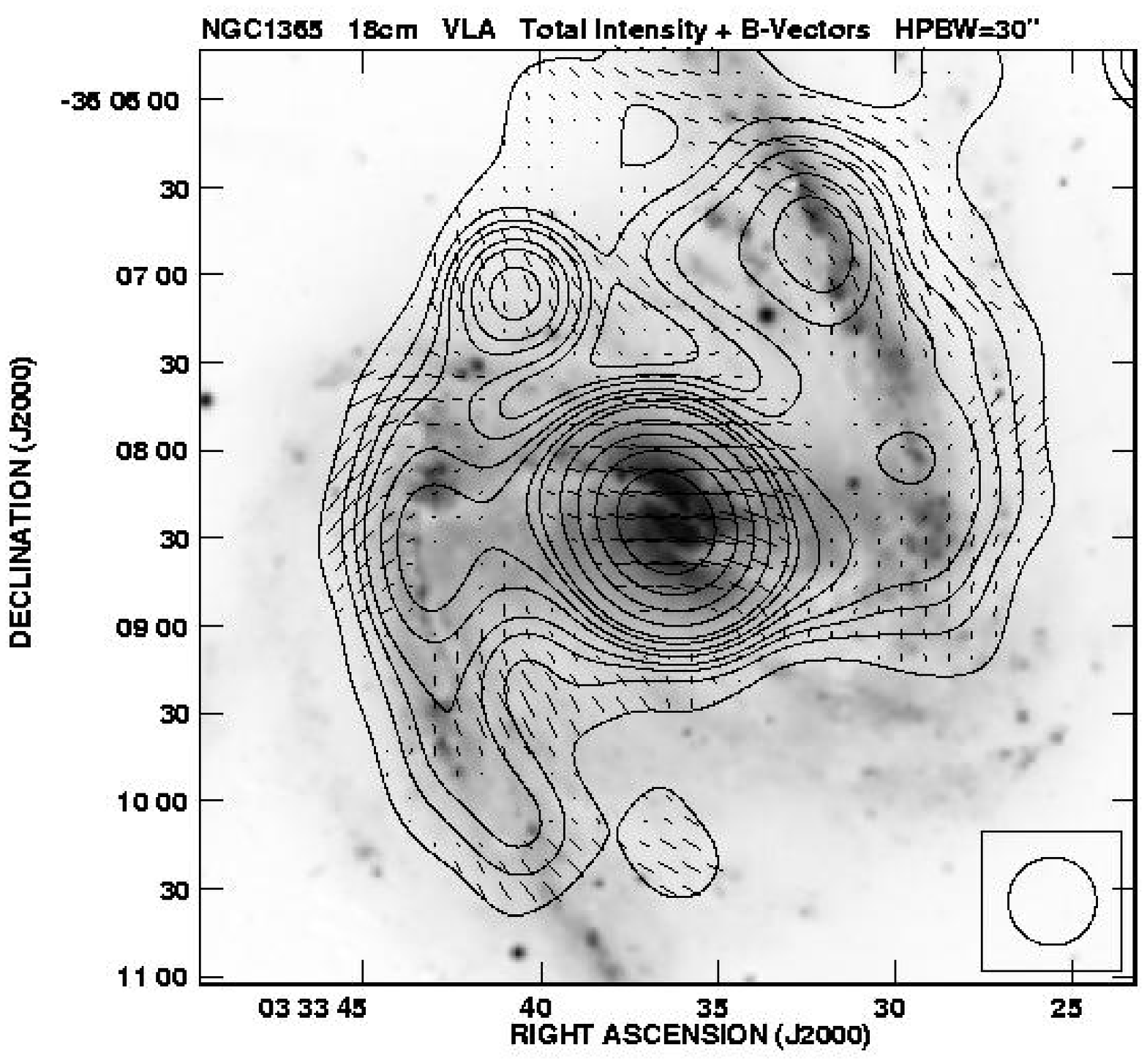}
}
\vspace{1.5cm}
\hbox to\textwidth{
\includegraphics[bb = 39 173 567 657,width=8.8cm,clip]{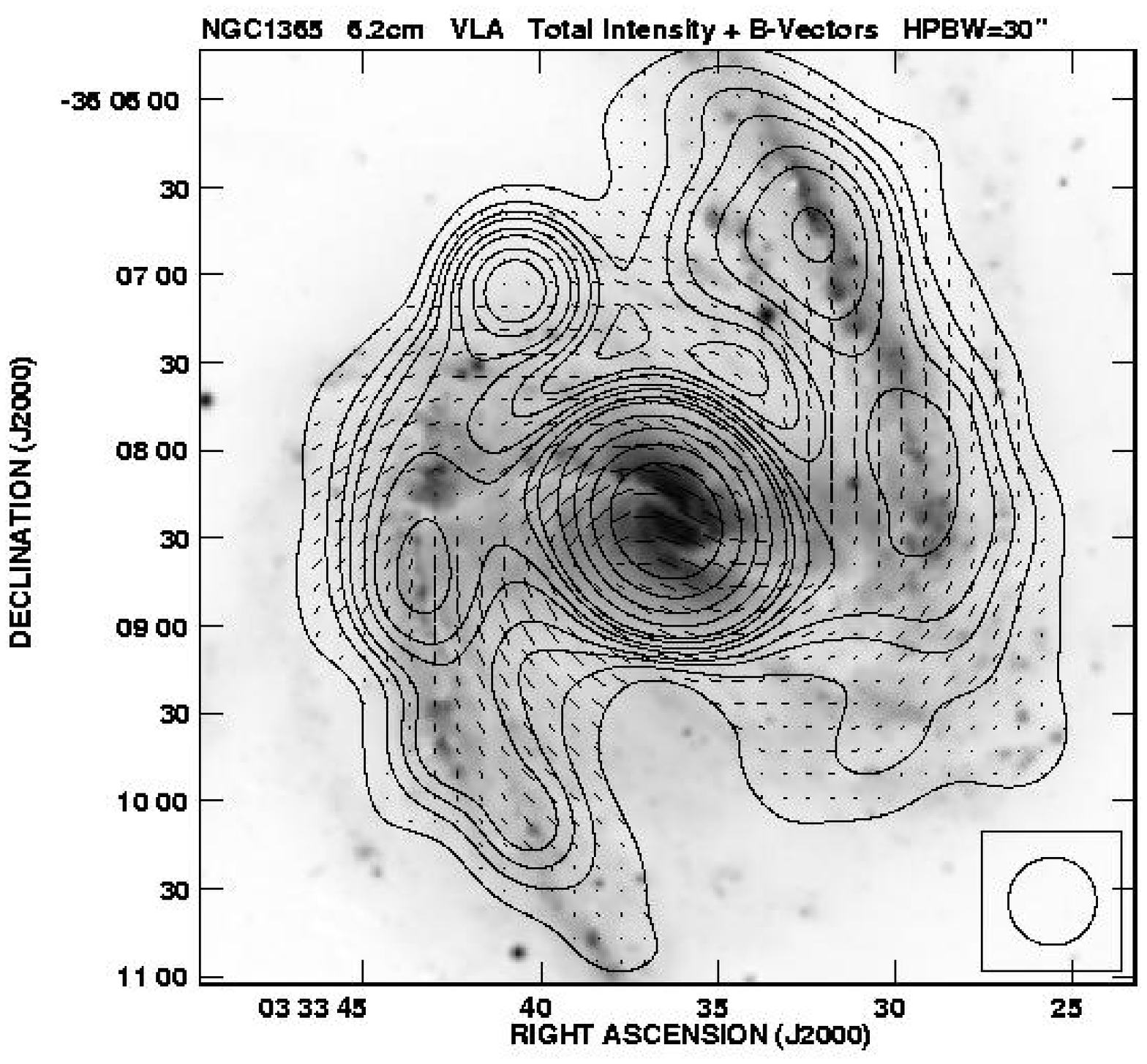}
\hfill
\includegraphics[bb = 39 173 567 657,width=8.8cm,clip]{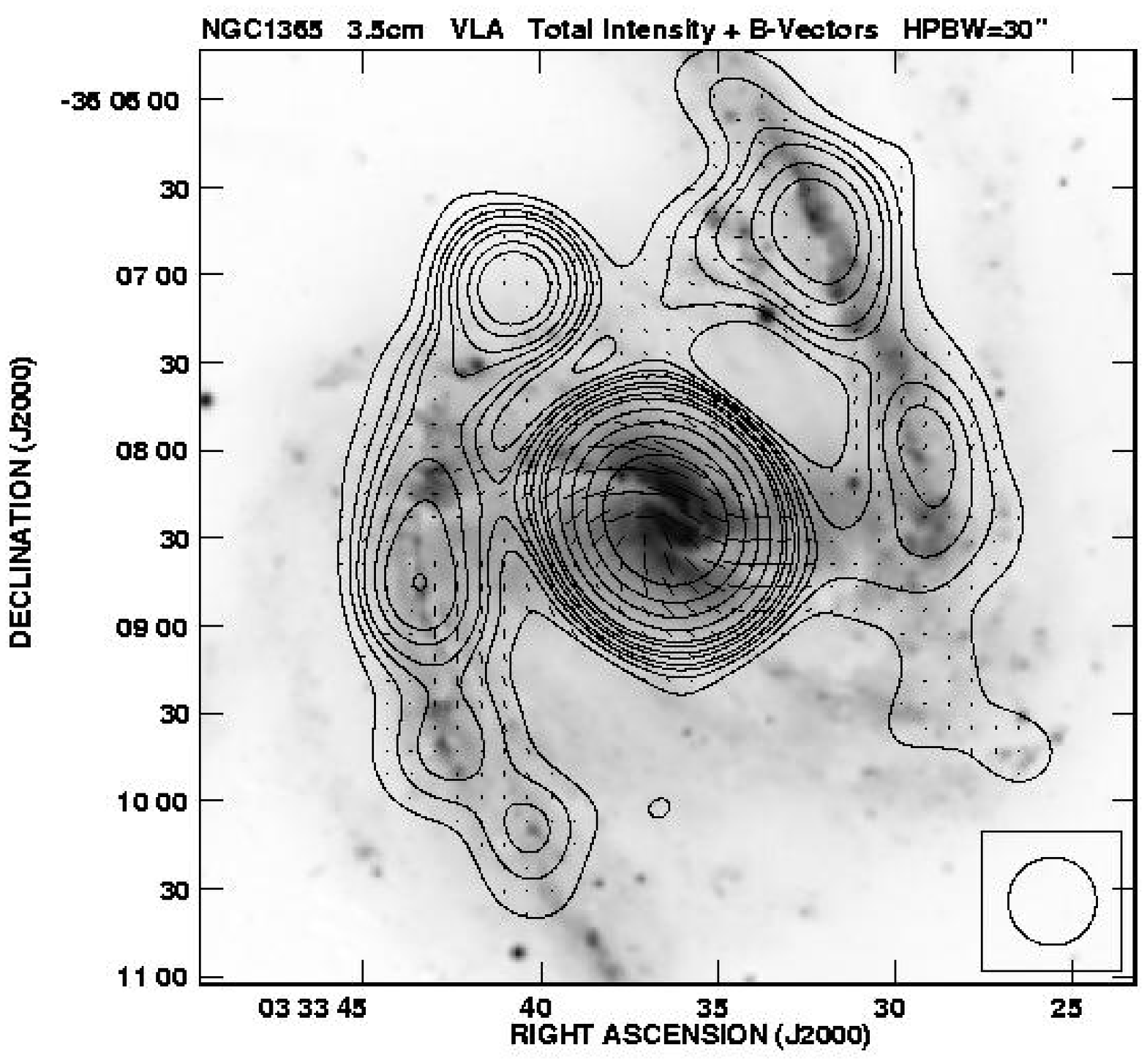}
}
\caption{Total intensity contours and the observed $B$-vectors of polarized emission
of NGC~1365, overlayed onto an optical image taken at the ESO
3.6m telescope by P.O.~Lindblad\ (\cite{lindblad99}).
The basic contour levels are 700, 500, 200 and 100 $\mu$Jy/beam area
in decreasing wavelength order, the contour intervals
are as in Fig.~5. A vector of 1\arcsec\ length corresponds to a
polarized intensity of 20~$\mu$Jy/beam area.}
\end{figure*}


\begin{figure*}
\hbox to \textwidth{
\includegraphics[bb = 39 155 568 665,width=8.8cm,clip]{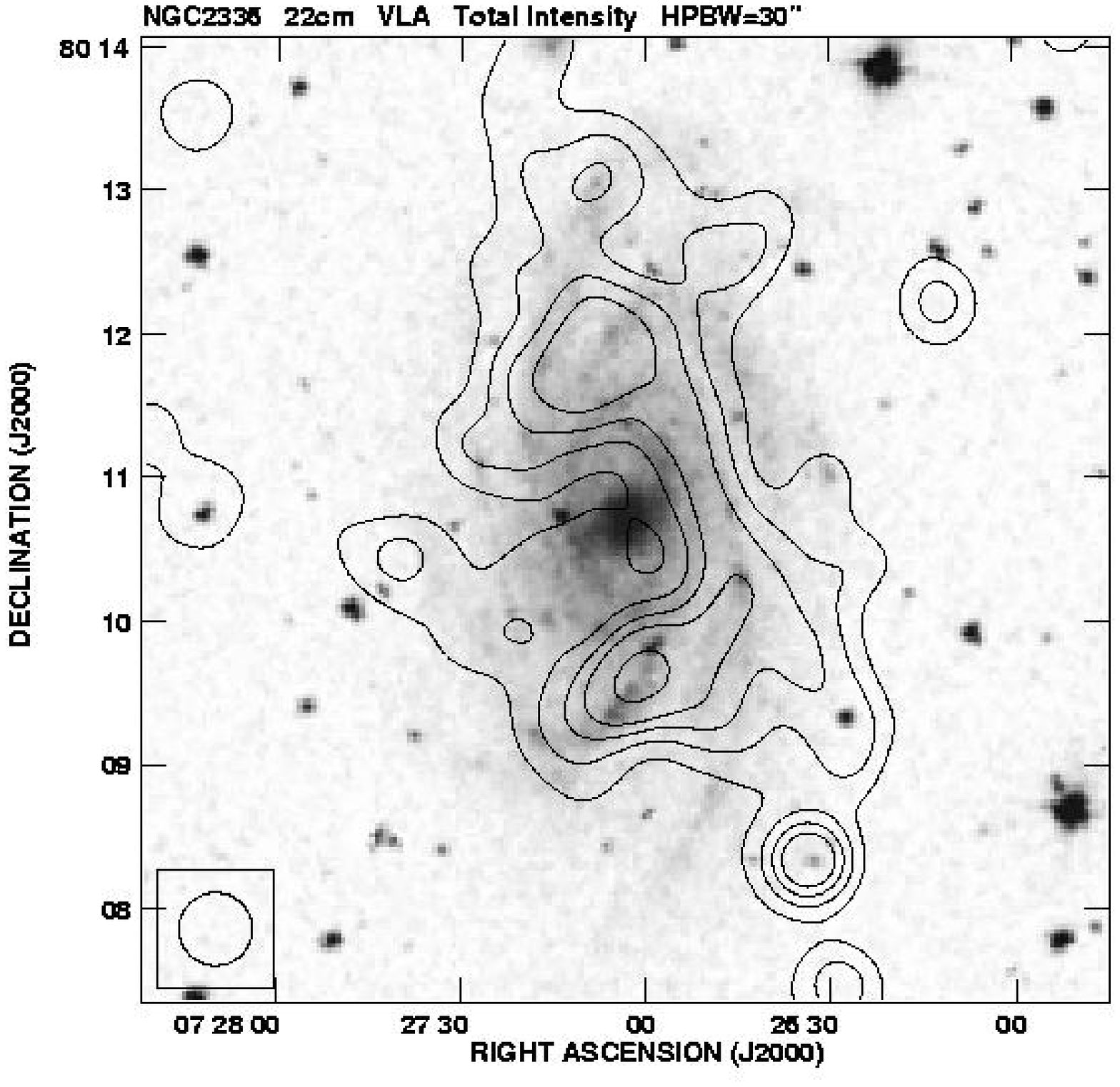}
\hfill
\includegraphics[bb = 39 155 568 665,width=8.8cm,clip]{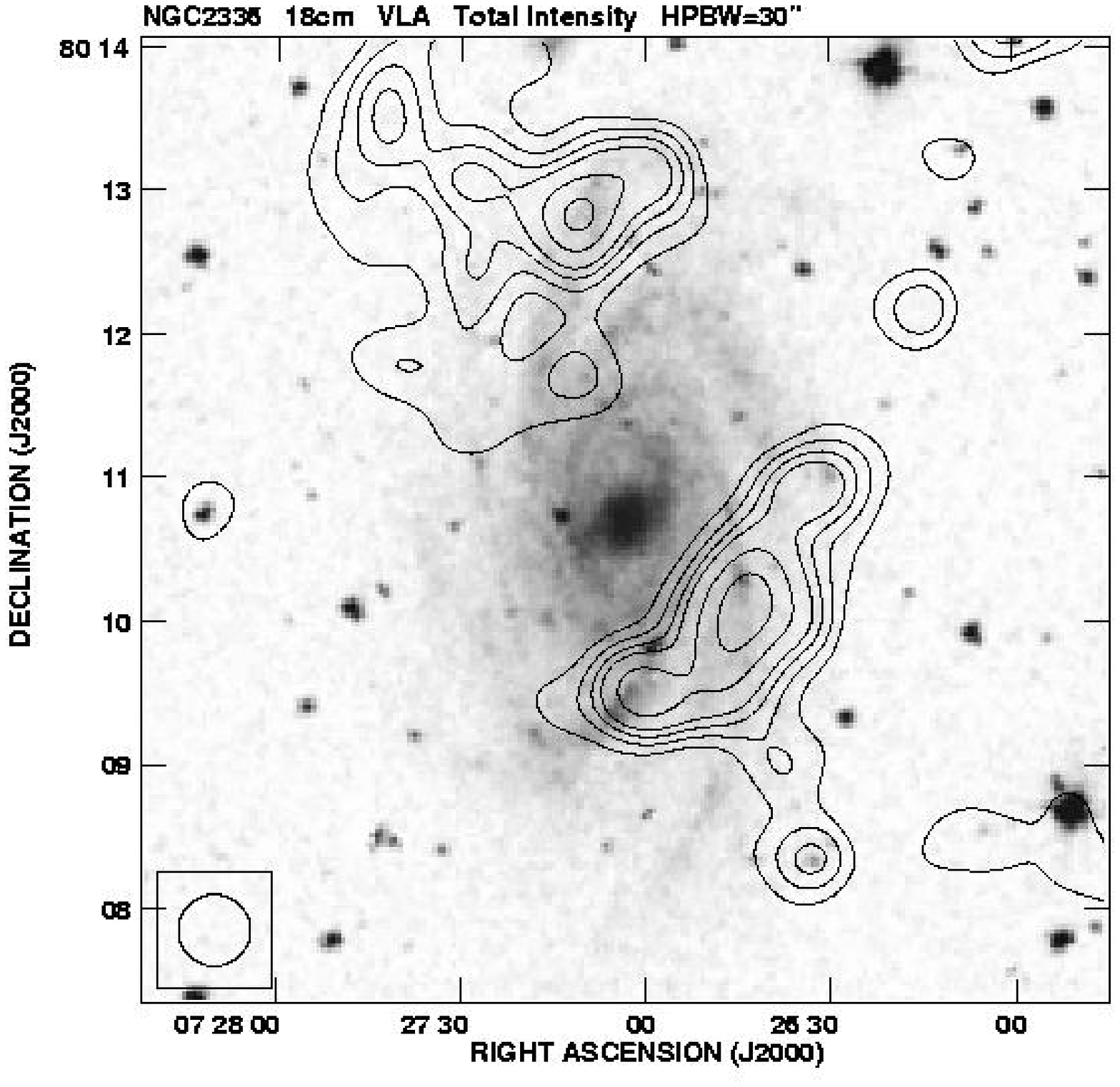}
}
\vspace{1.5cm}
\hbox to\textwidth{
\includegraphics[bb = 39 155 568 665,width=8.8cm,clip]{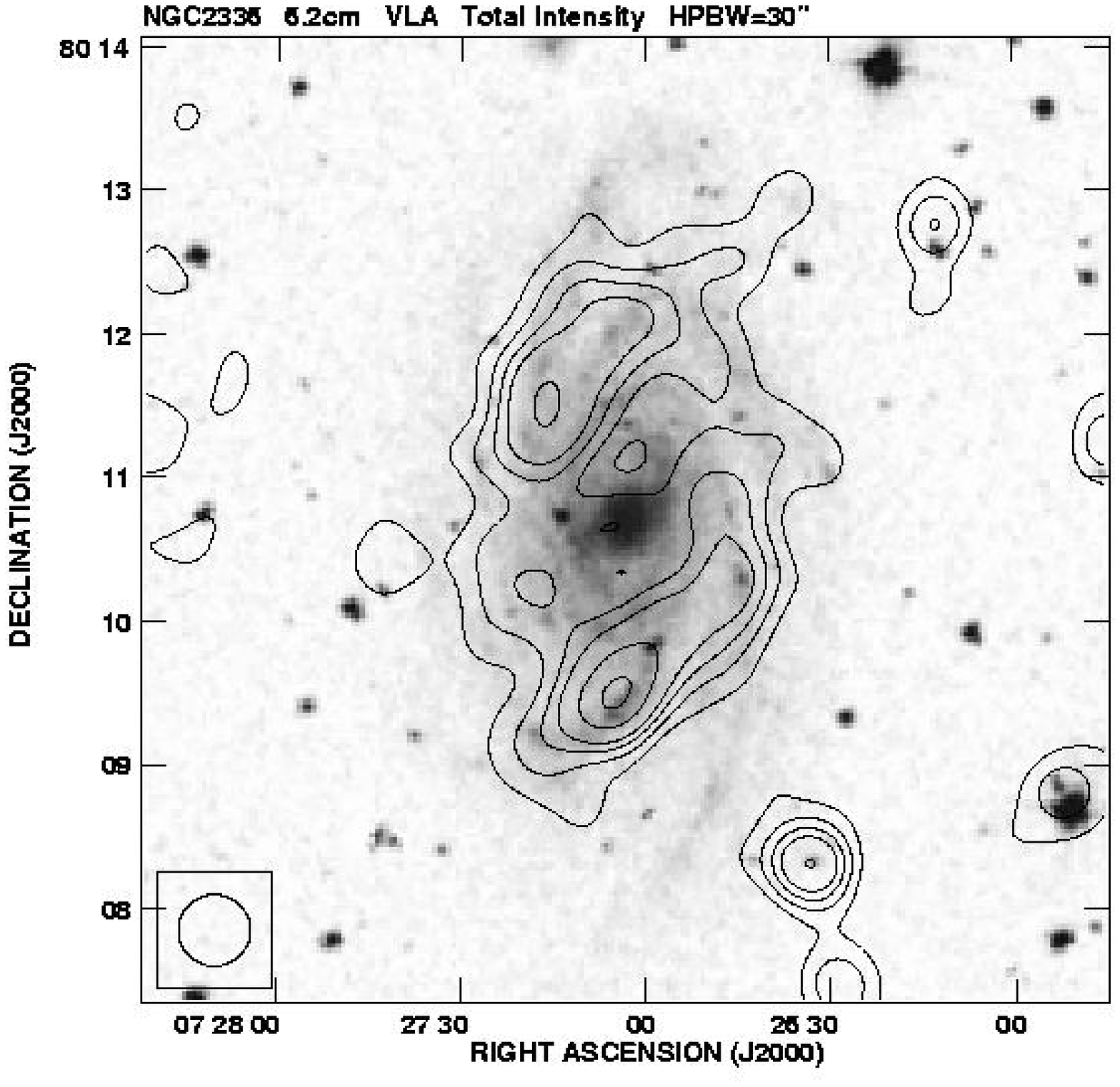}
\hfill
\includegraphics[bb = 39 155 568 665,width=8.8cm,clip]{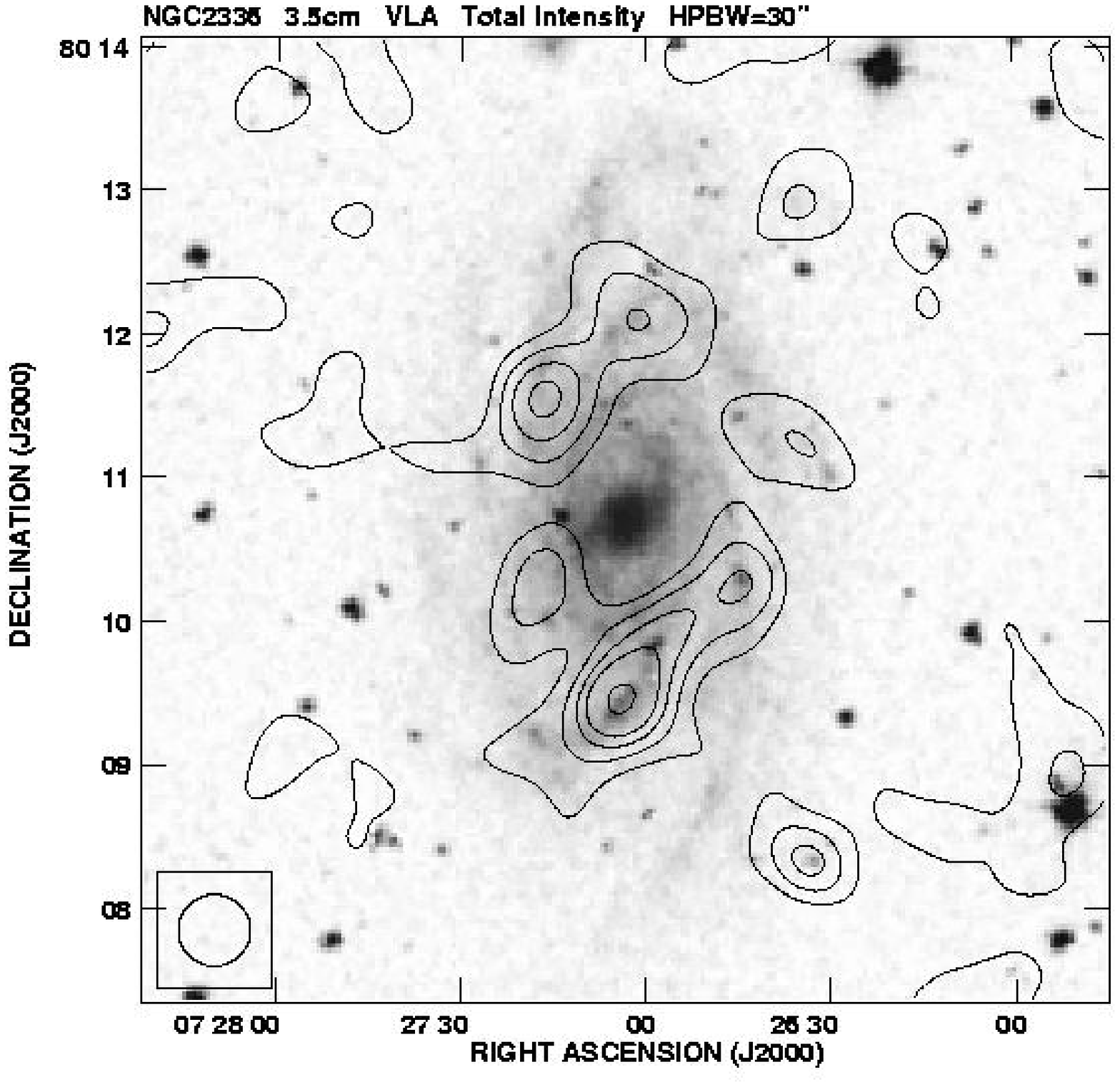}
}
\caption{Total intensity contours
of NGC~2336, overlayed onto an optical image from the Digitized
Sky Surveys. The contours are as in Fig.~6.}
\end{figure*}


\begin{figure*}
\hbox to \textwidth{
\includegraphics[bb = 39 109 567 721,width=8.8cm,clip]{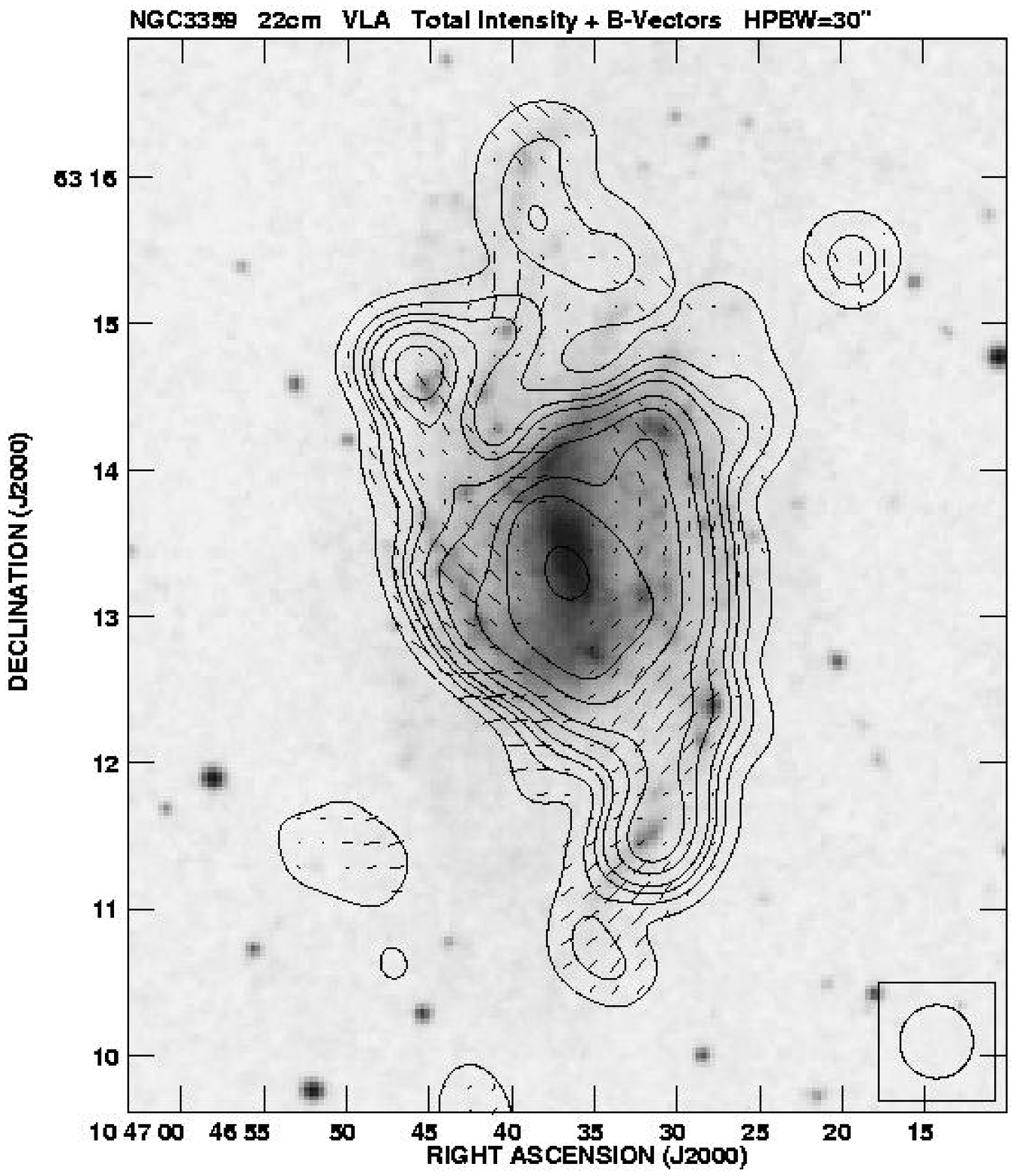}
\hfill
\includegraphics[bb = 36 102 569 714,width=8.8cm,clip]{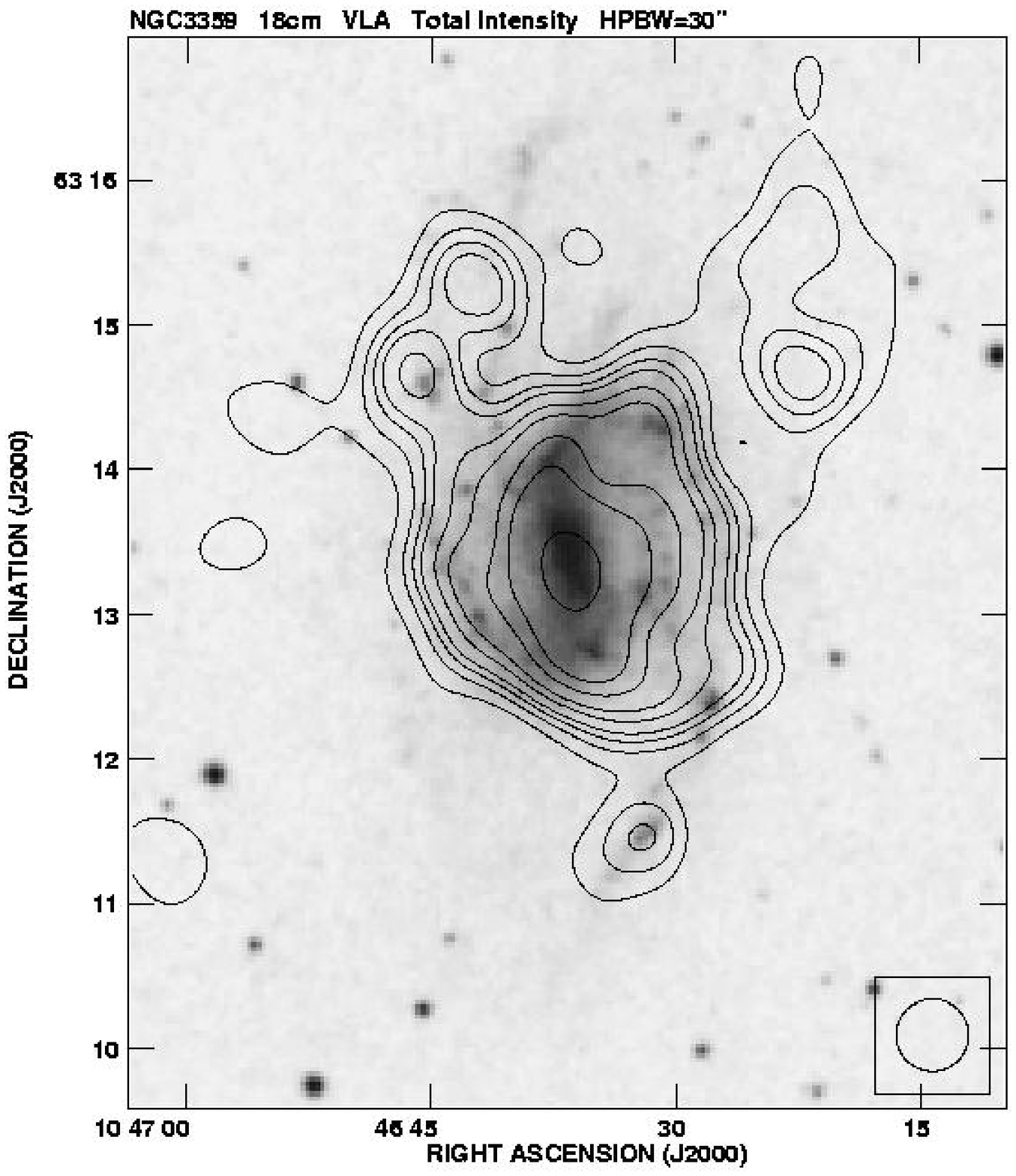}
}
\vspace{1.5cm}
\hbox to\textwidth{
\includegraphics[bb = 39 109 567 721,width=8.8cm,clip]{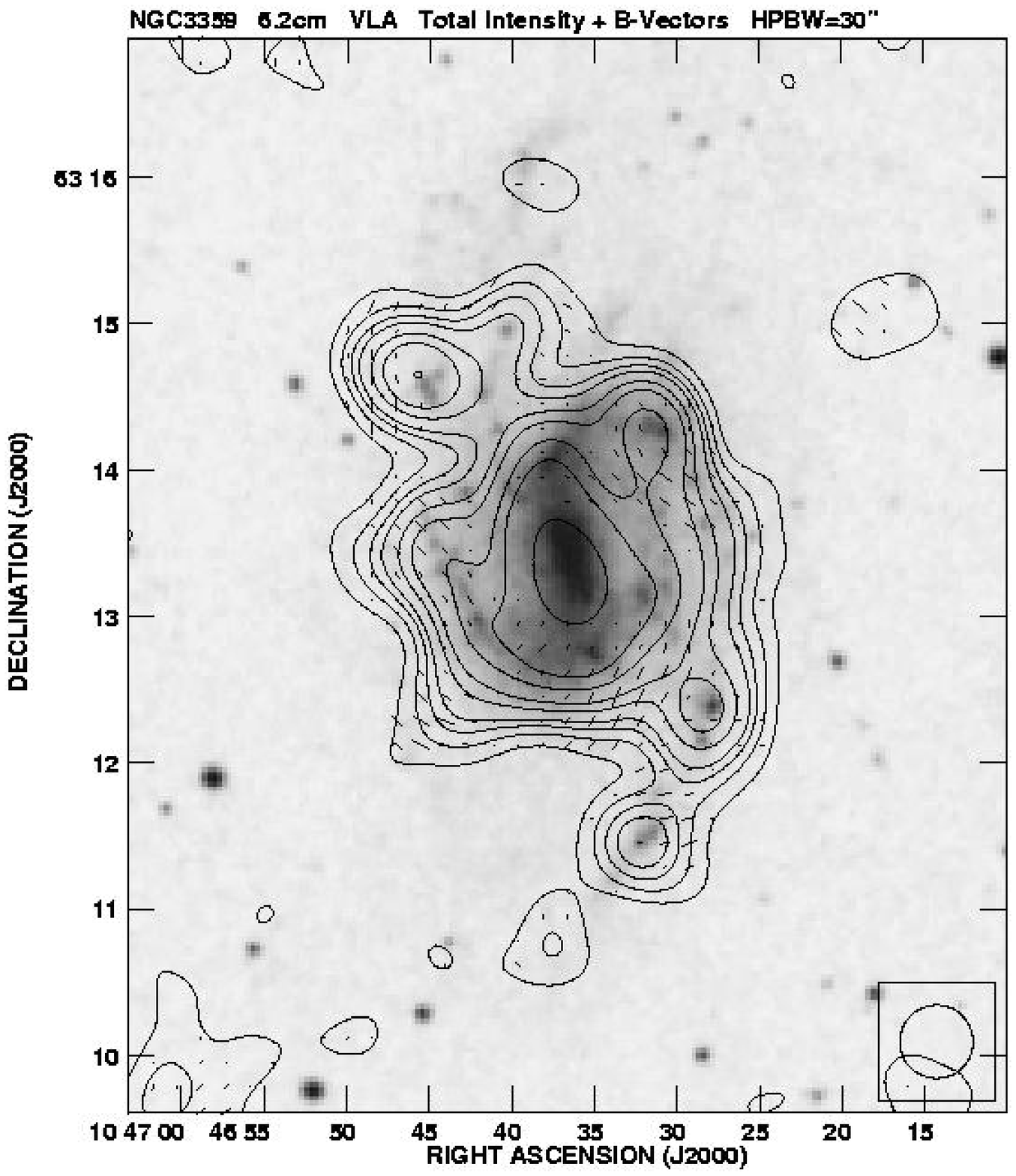}
\hfill
\includegraphics[bb = 36 102 569 714,width=8.8cm,clip]{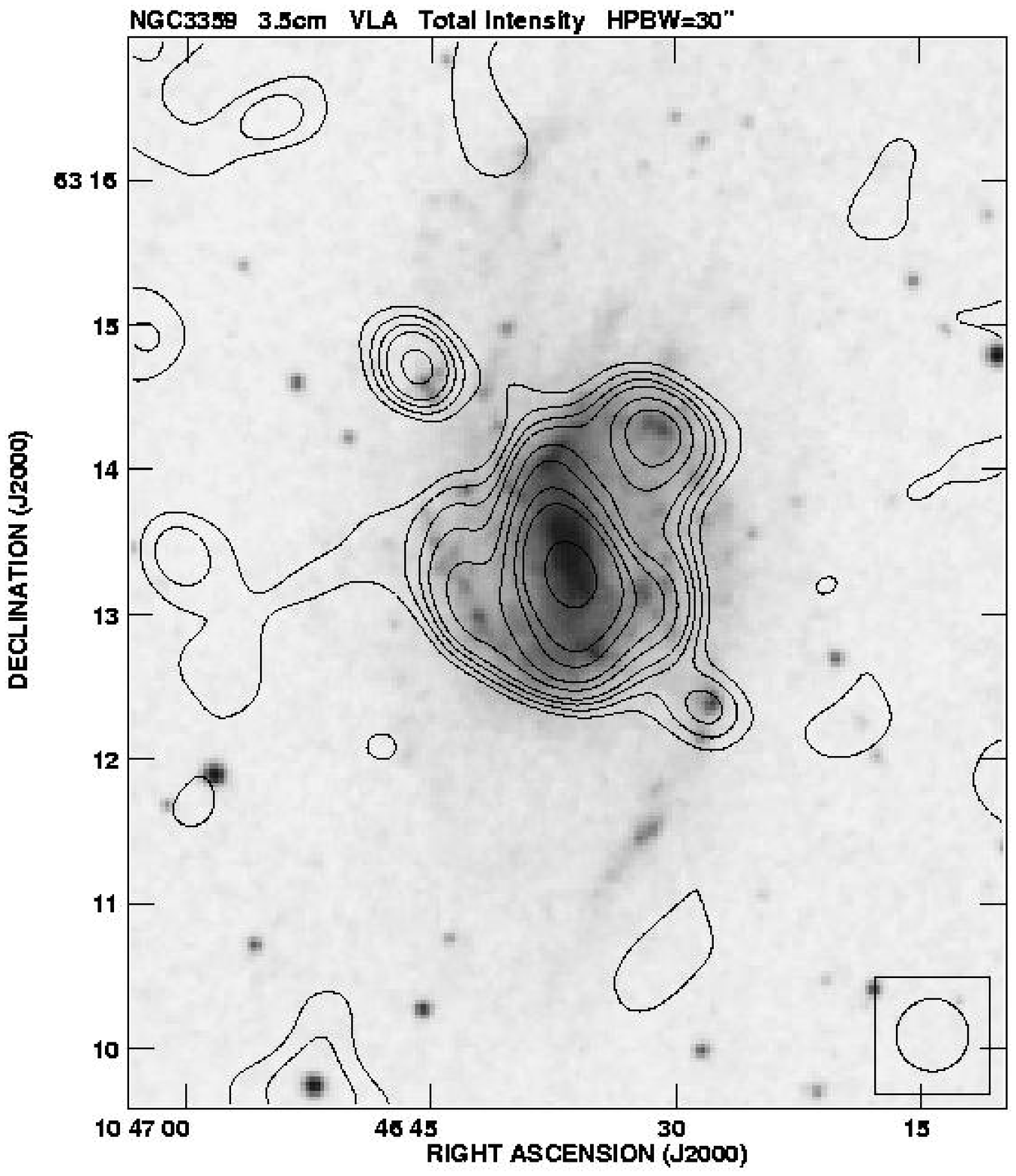}
}
\caption{Total intensity contours and the observed $B$-vectors of polarized emission
of NGC~3359, overlayed onto an optical image from the Digitized
Sky Surveys. The contours and the vector scale are as in Fig.~6.
}
\end{figure*}


\begin{figure*}
\hbox to \textwidth{
\includegraphics[bb = 39 109 567 721,width=8.8cm,clip]{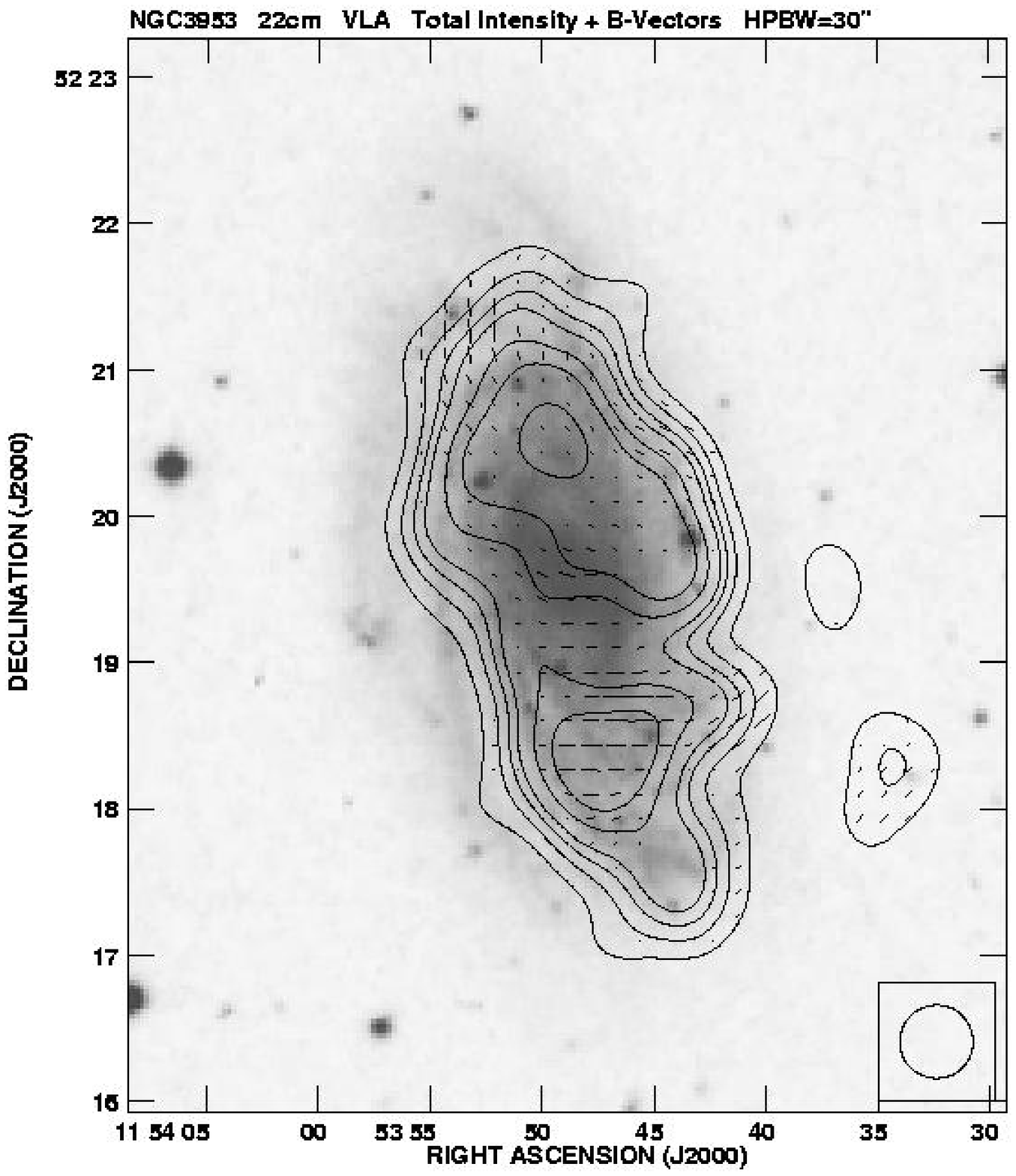}
\hfill
\includegraphics[bb = 39 102 567 714,width=8.8cm,clip]{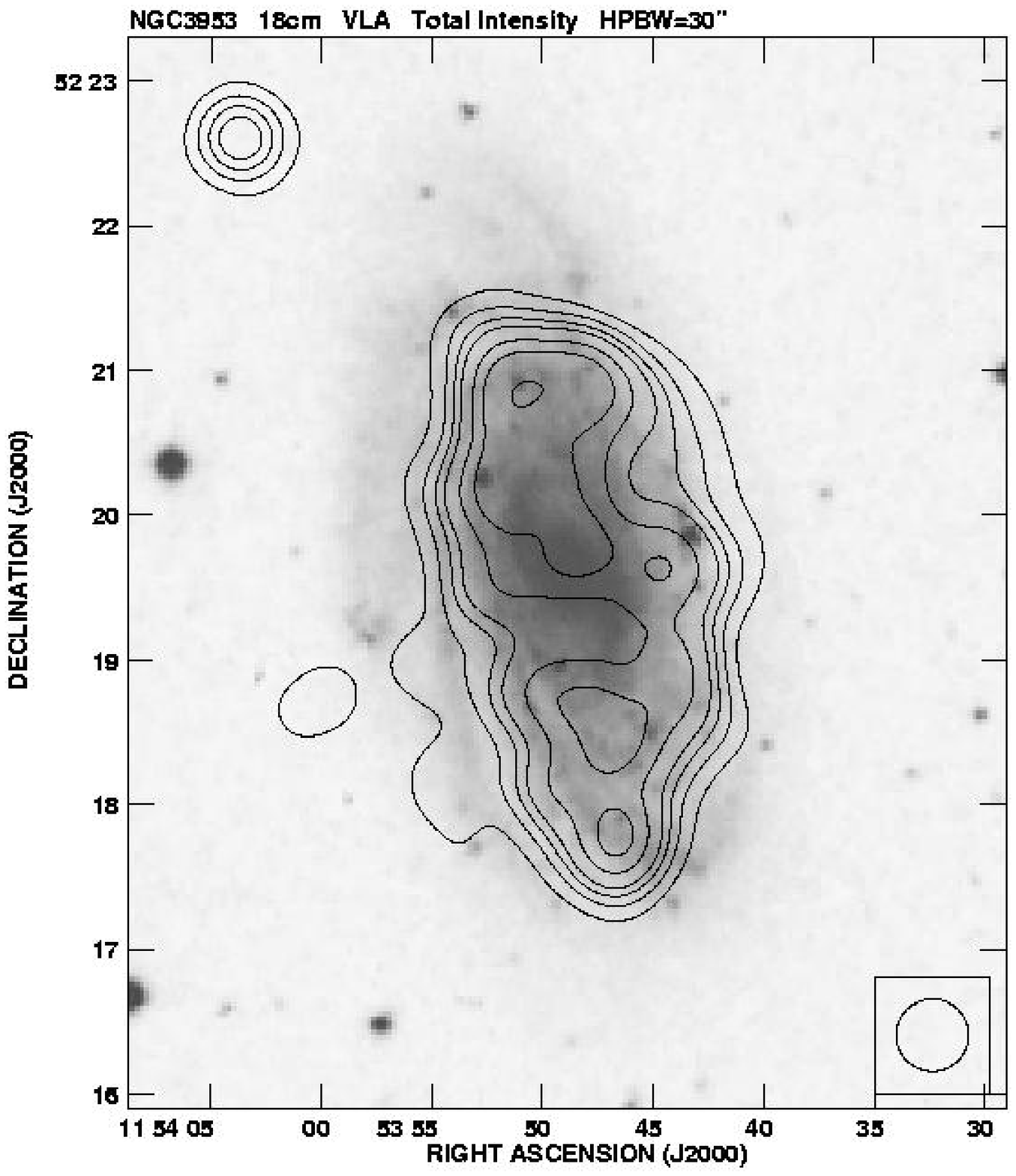}
}
\vspace{1.5cm}
\hbox to\textwidth{
\includegraphics[bb = 39 109 567 721,width=8.8cm,clip]{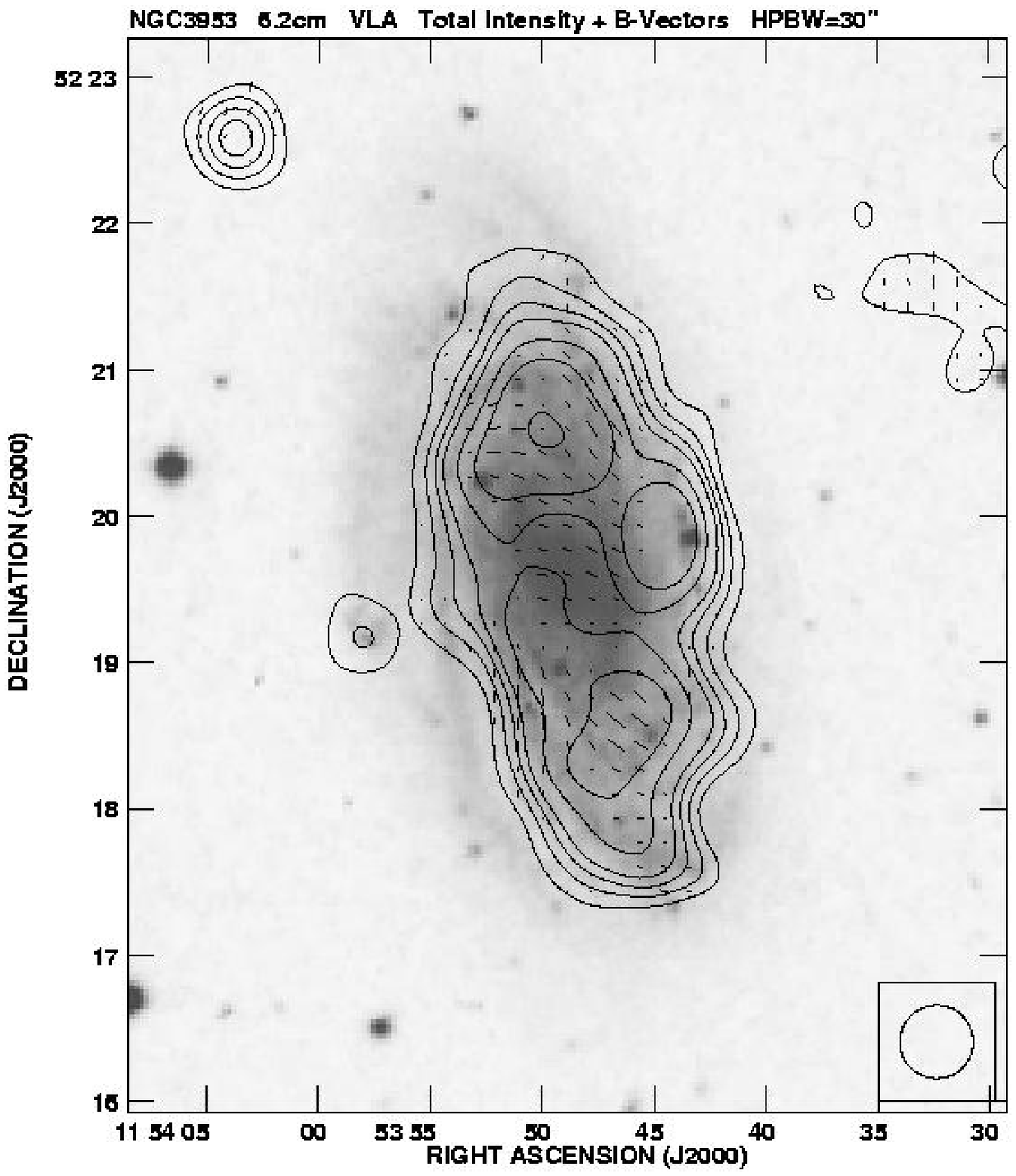}
\hfill
\includegraphics[bb = 39 102 567 714,width=8.8cm,clip]{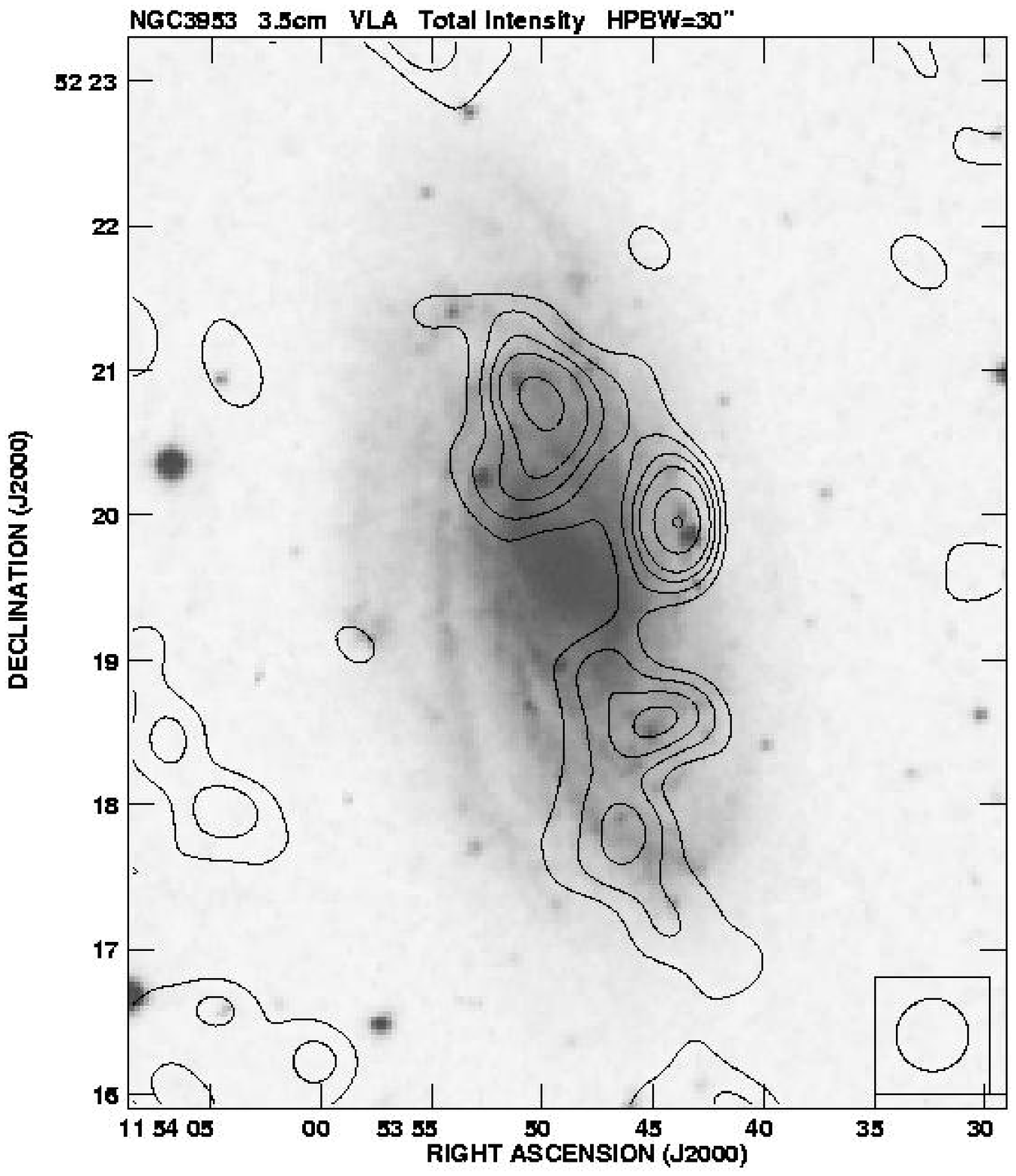}
}
\caption{Total intensity contours and the observed $B$-vectors of polarized emission
of NGC~3953, overlayed onto an optical image from the Digitized
Sky Surveys. The contours and the vector scale are as in Fig.~6.}
\end{figure*}


\begin{figure*}
\hbox to \textwidth{
\includegraphics[bb = 58 109 517 678,angle=270,width=8.8cm,clip]{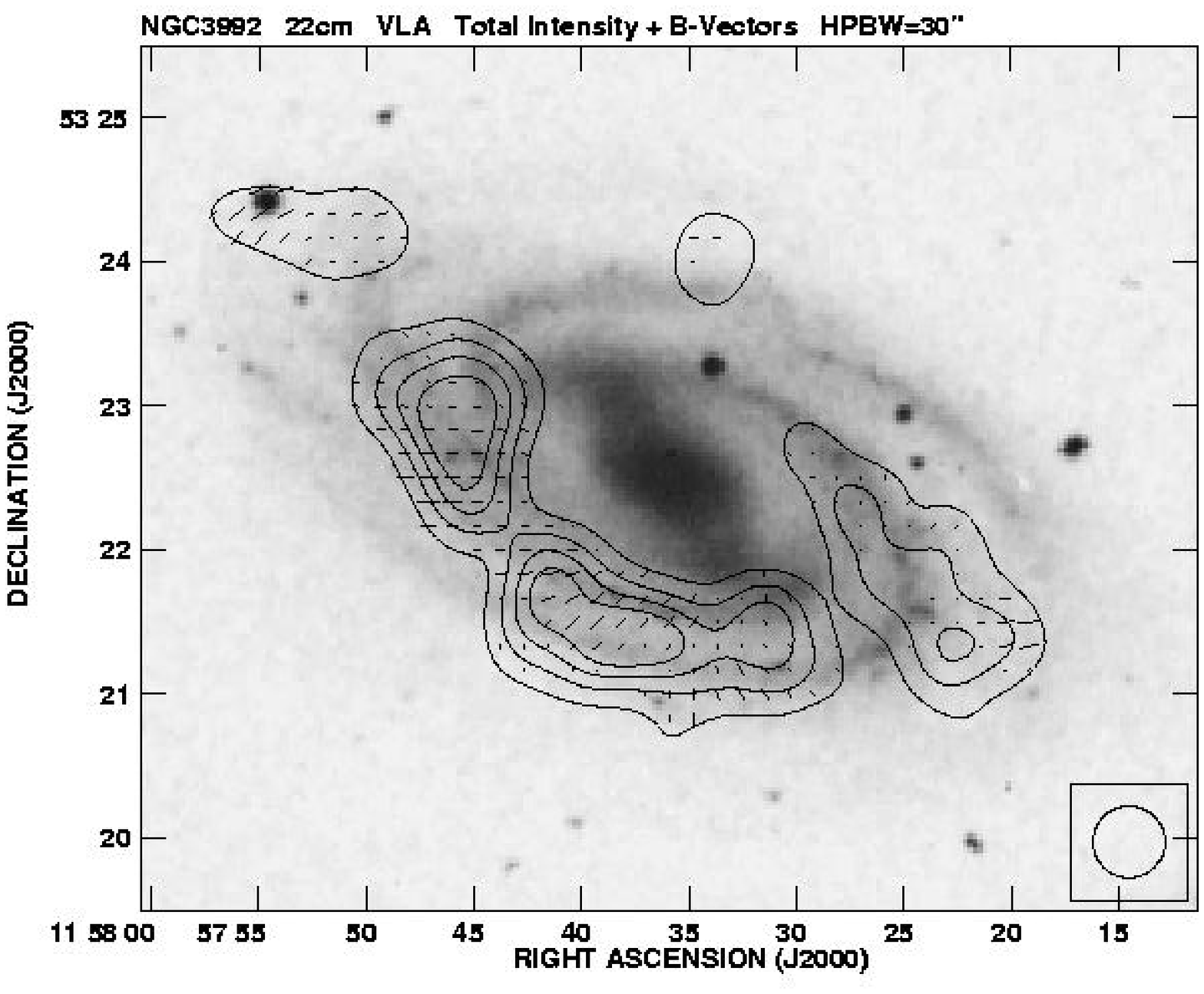}
\hfill
\includegraphics[bb = 58 100 531 686,angle=270,width=8.8cm,clip]{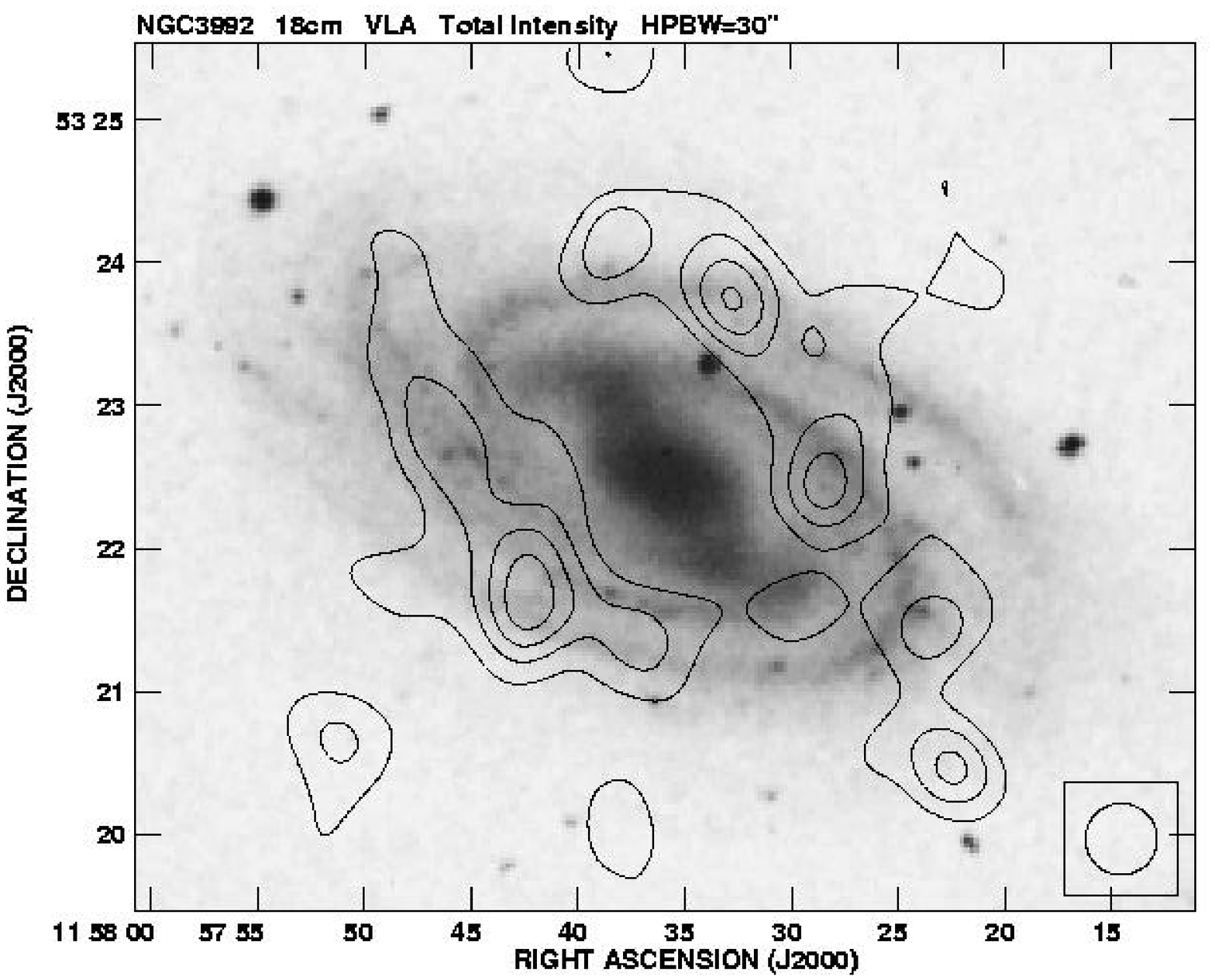}
}
\vspace{1.5cm}
\hbox to\textwidth{
\includegraphics[bb = 58 109 517 678,angle=270,width=8.8cm,clip]{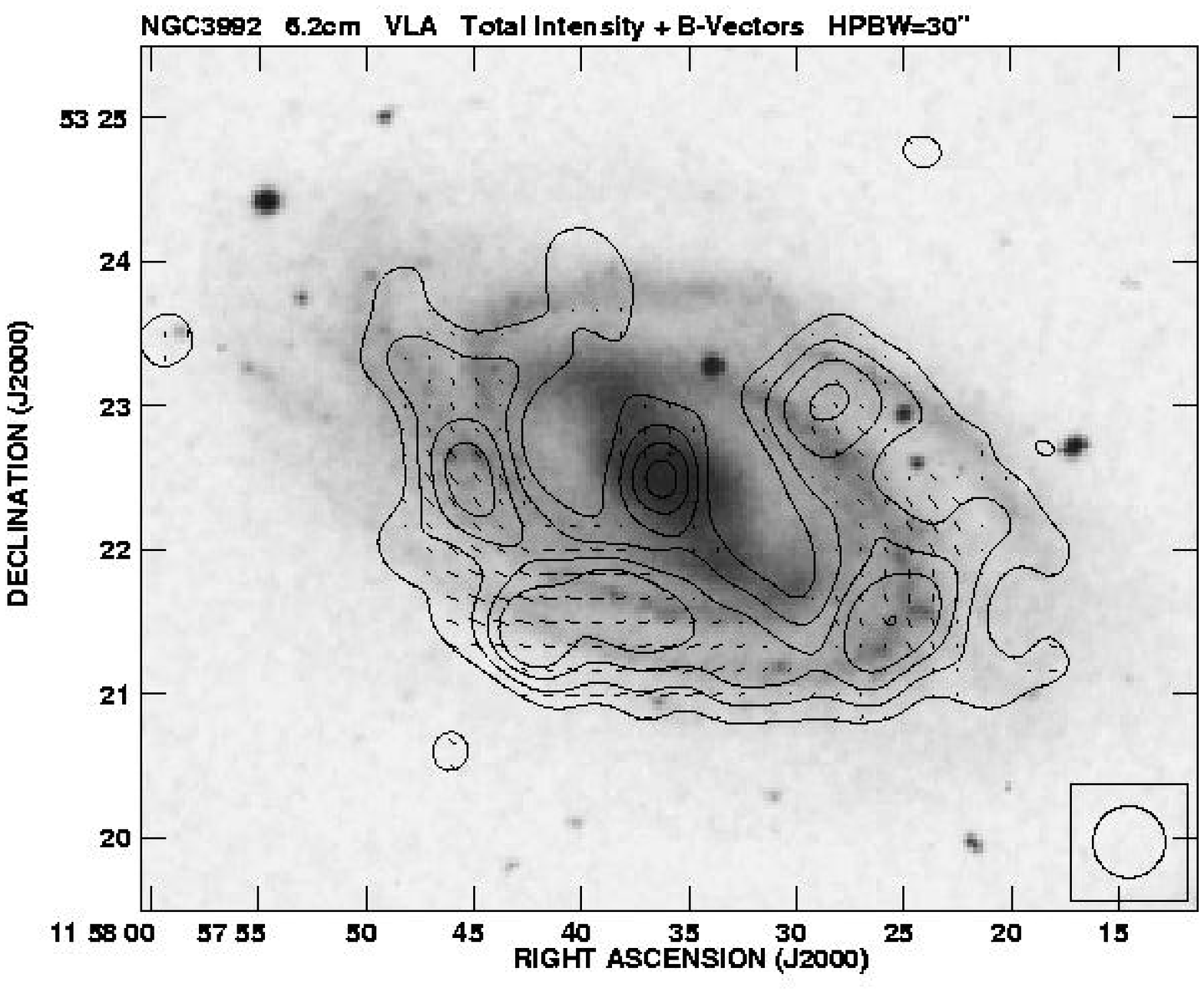}
\hfill
\includegraphics[bb = 58 100 531 686,angle=270,width=8.8cm,clip]{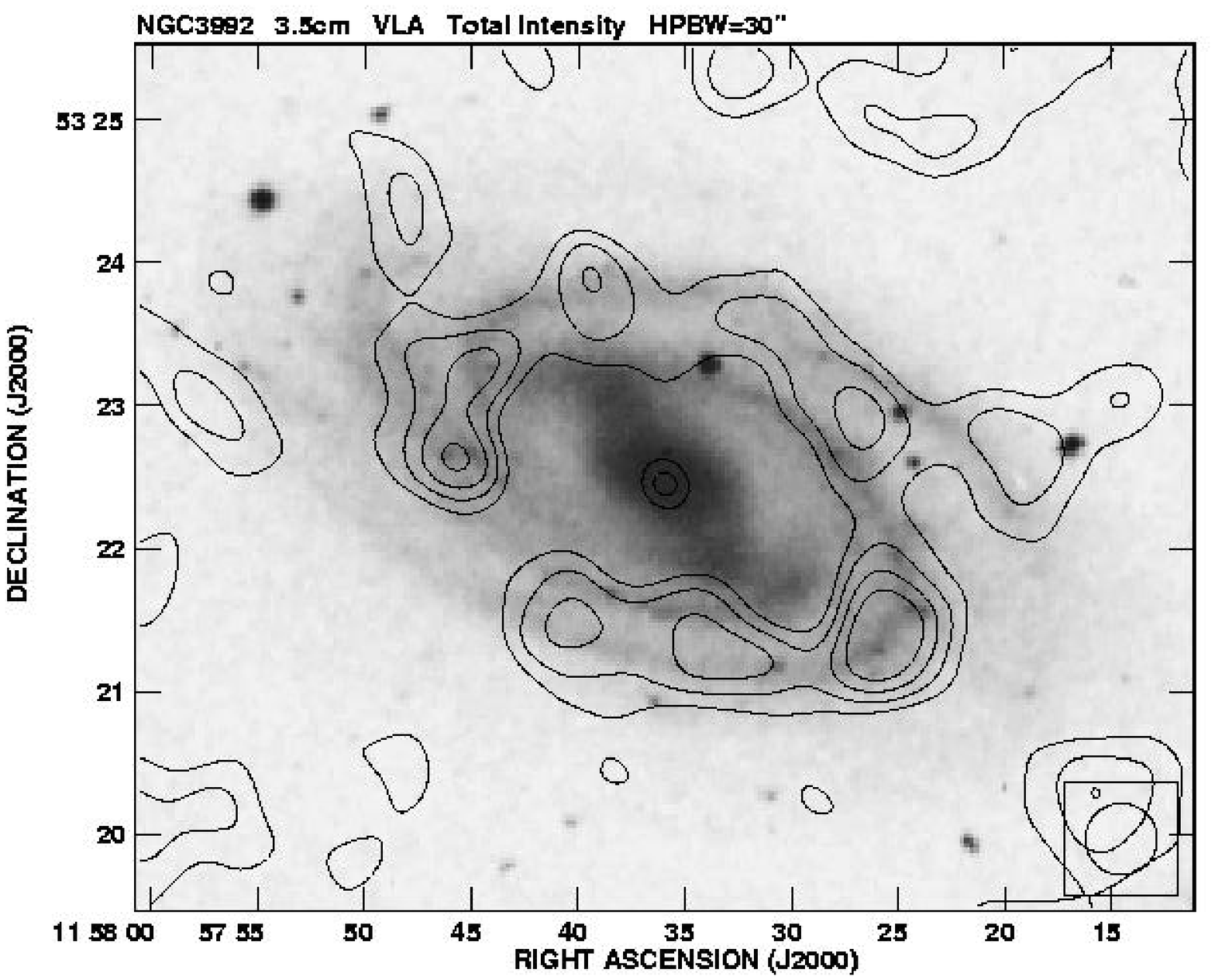}
}
\caption{Total intensity contours and the observed $B$-vectors of polarized emission
of NGC~3992, overlayed onto an optical image from the Digitized
Sky Surveys. The contours and the vector scale are as in Fig.~6.}
\end{figure*}


\begin{figure*}
\hbox to \textwidth{
\includegraphics[bb = 39 169 567 660,width=8.8cm,clip]{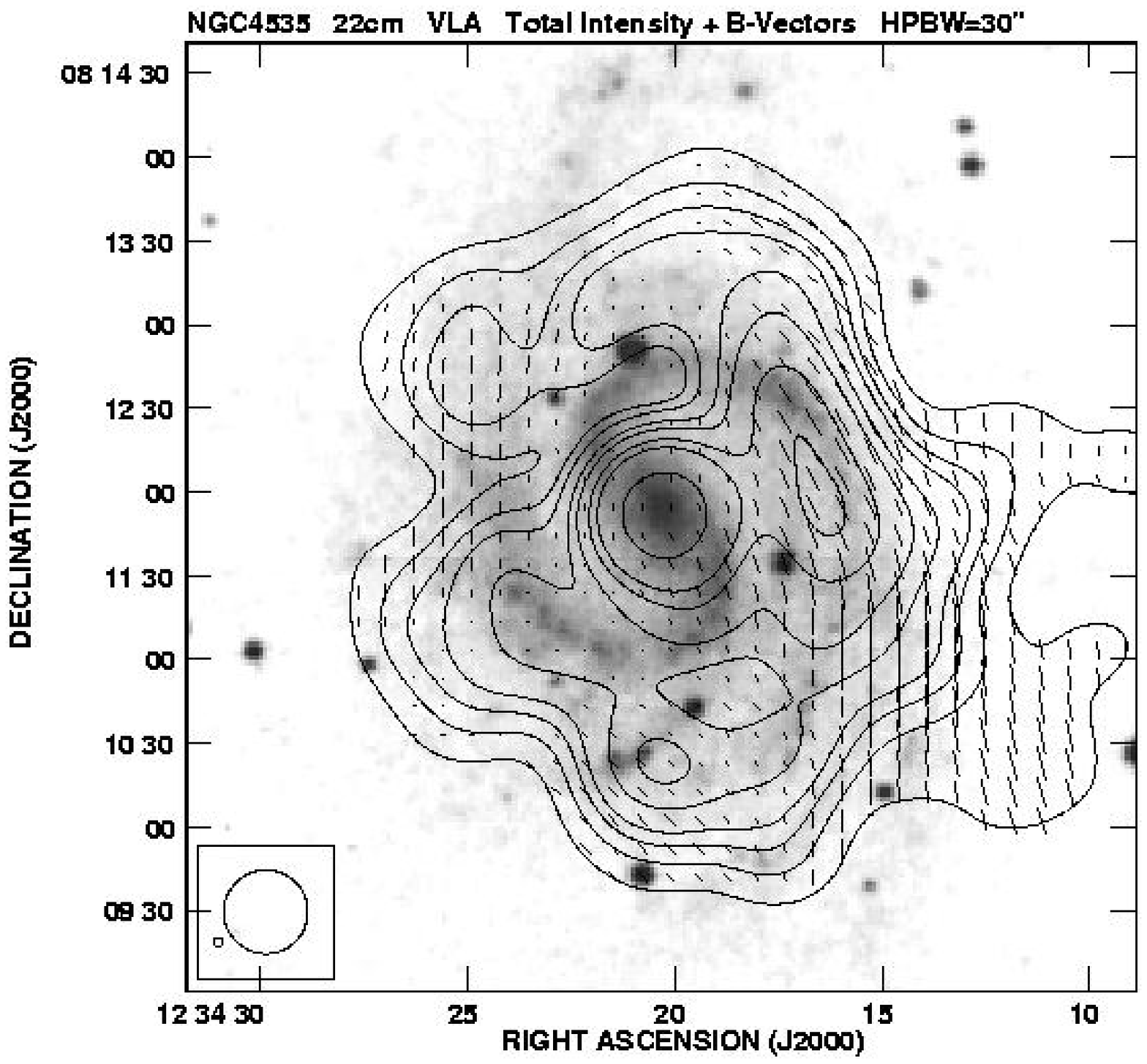}
\hfill
\includegraphics[bb = 39 169 567 660,width=8.8cm,clip]{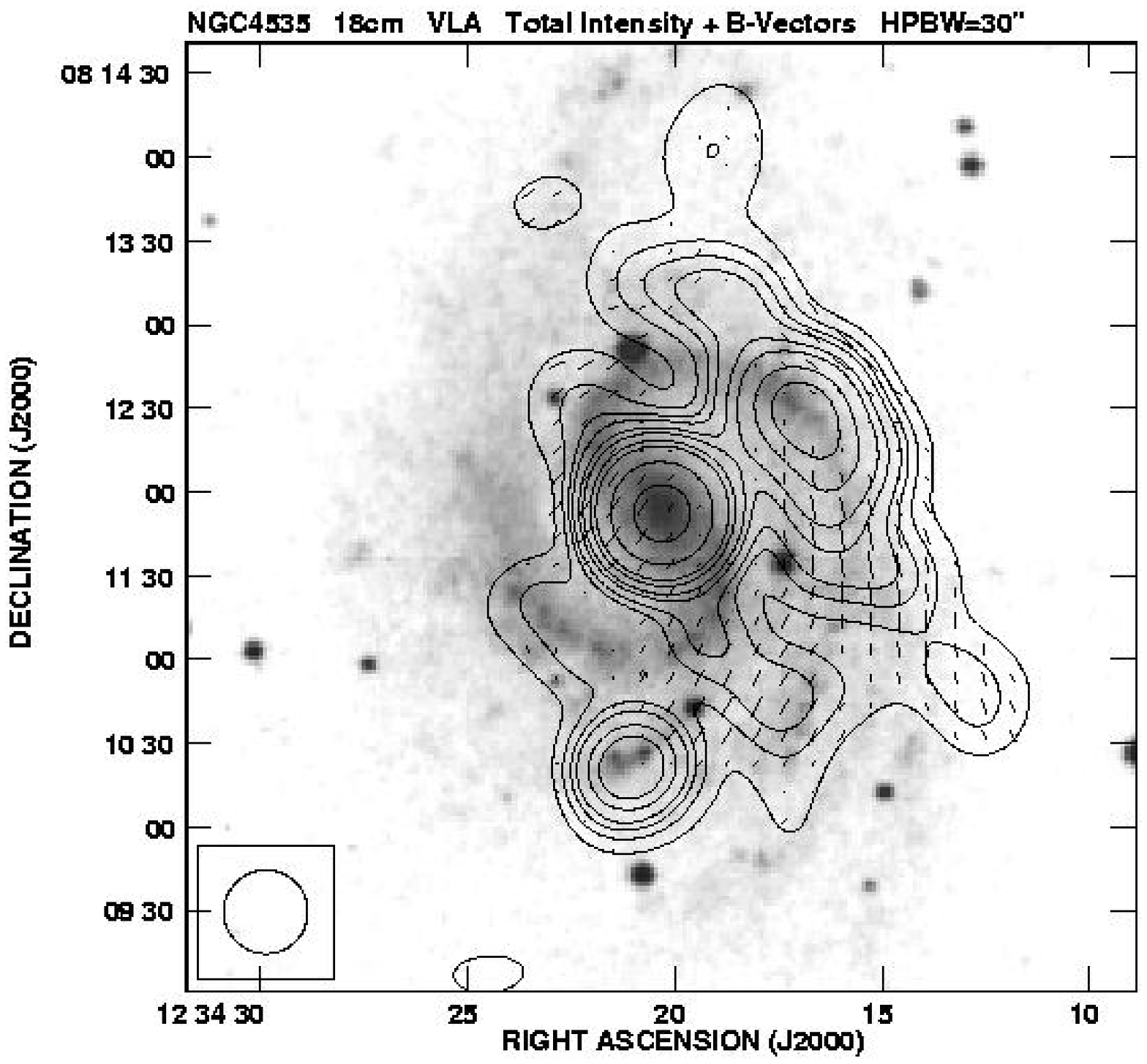}
}
\vspace{1.5cm}
\hbox to\textwidth{
\includegraphics[bb = 39 169 567 660,width=8.8cm,clip]{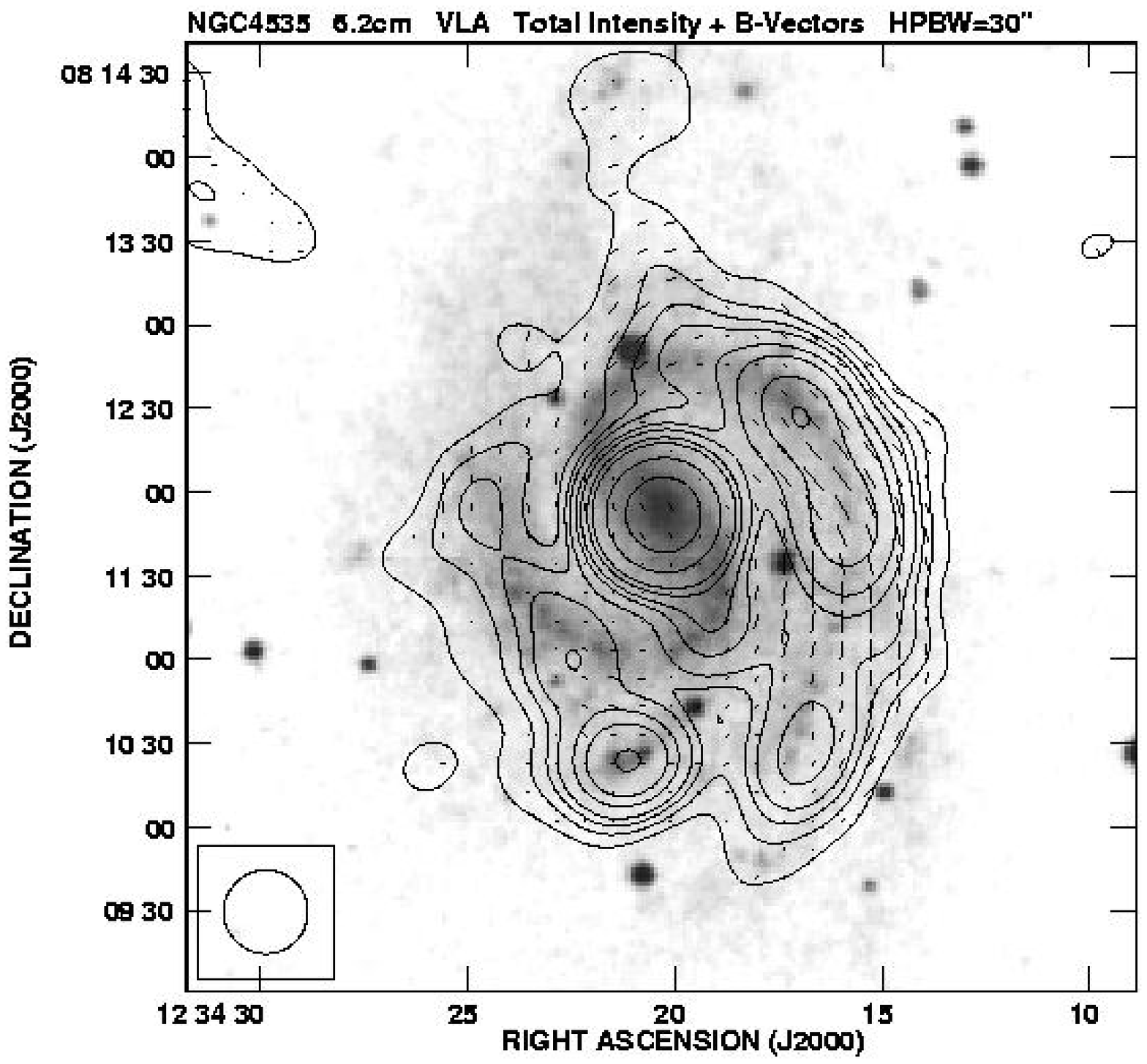}
\hfill
\includegraphics[bb = 39 169 567 660,width=8.8cm,clip]{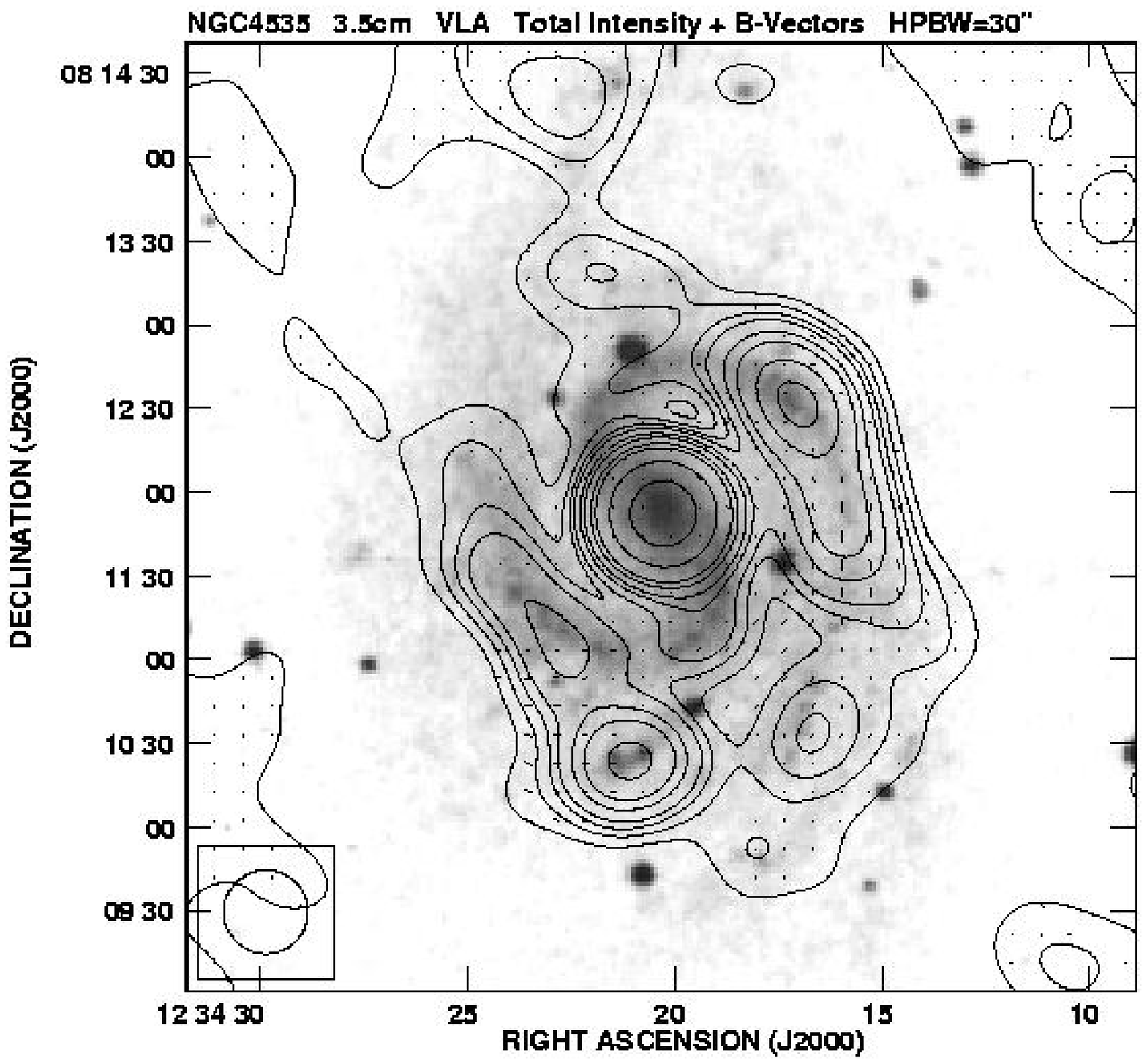}
}
\caption{Total intensity contours and the observed $B$-vectors of polarized emission
of NGC~4535, overlayed onto an optical image from the Digitized
Sky Surveys. The contours are as in Fig.~6. A vector of 1\arcsec\ length
corresponds to a polarized intensity of 20~$\mu$Jy/beam area.}
\end{figure*}


\begin{figure*}
\hbox to \textwidth{
\includegraphics[bb = 39 156 567 660,width=8.8cm,clip]{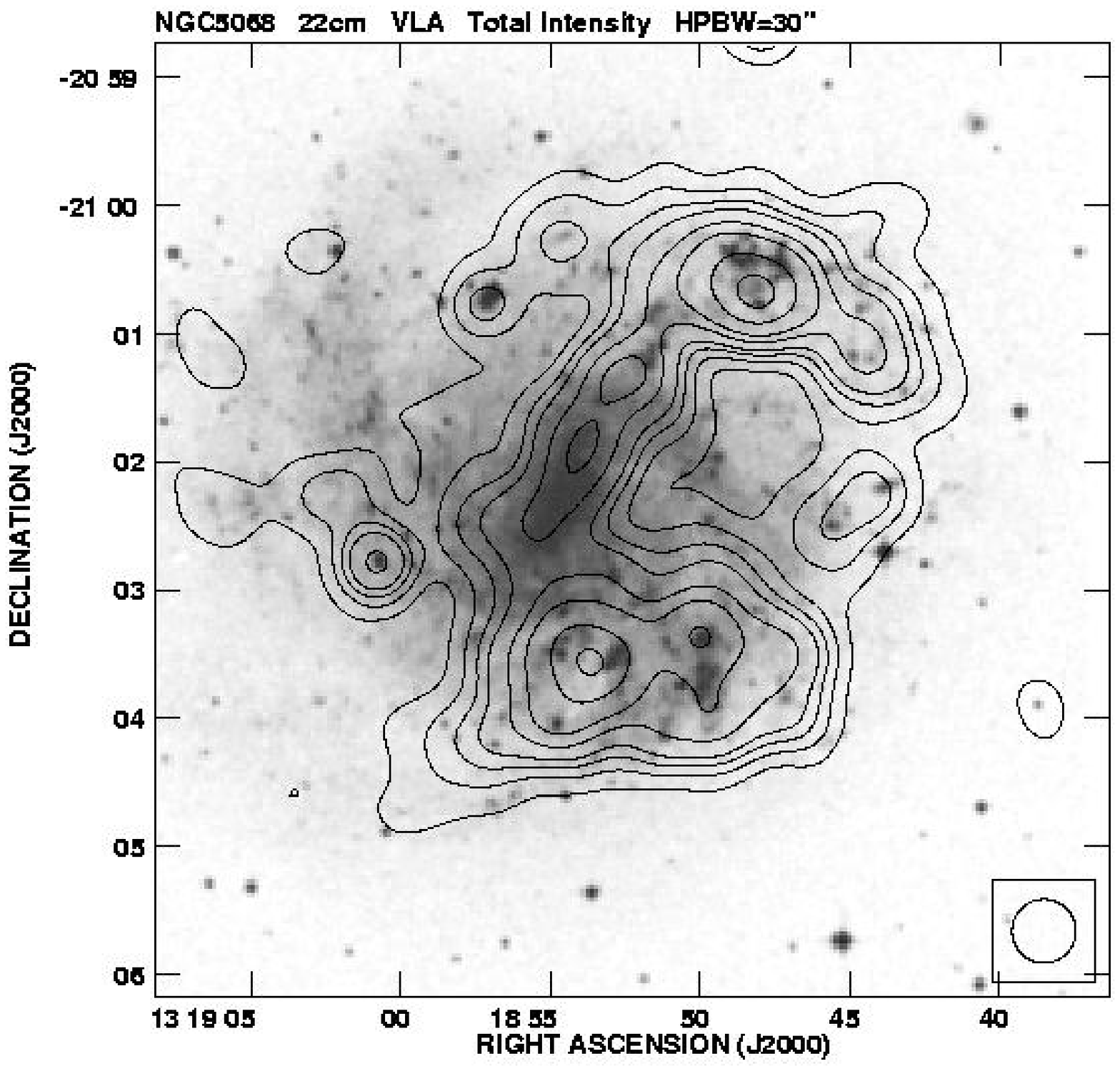}
\hfill
\includegraphics[bb = 36 156 567 660,width=8.8cm,clip]{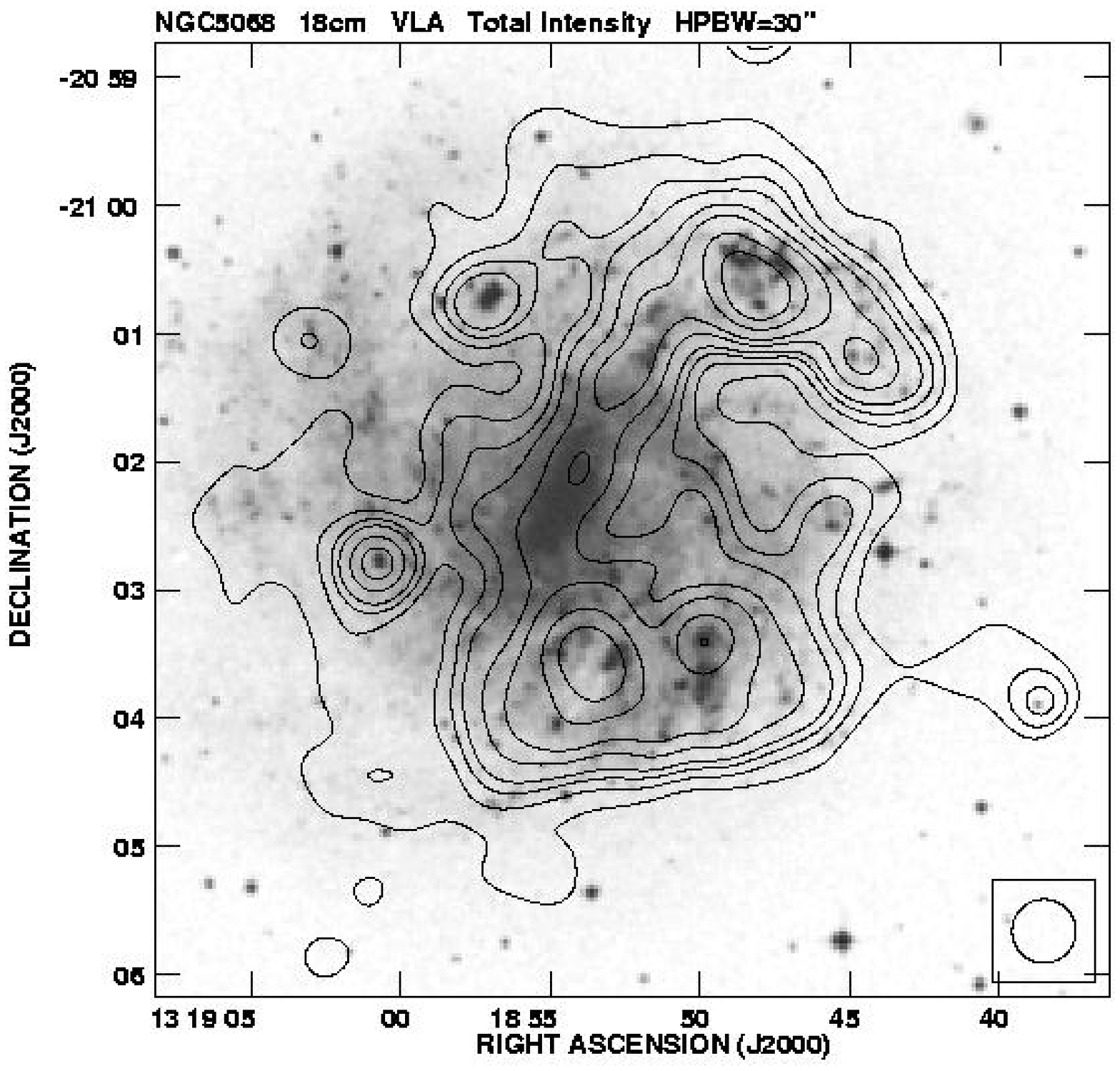}
}
\vspace{1.5cm}
\hbox to\textwidth{
\includegraphics[bb = 40 164 567 666,width=8.8cm,clip]{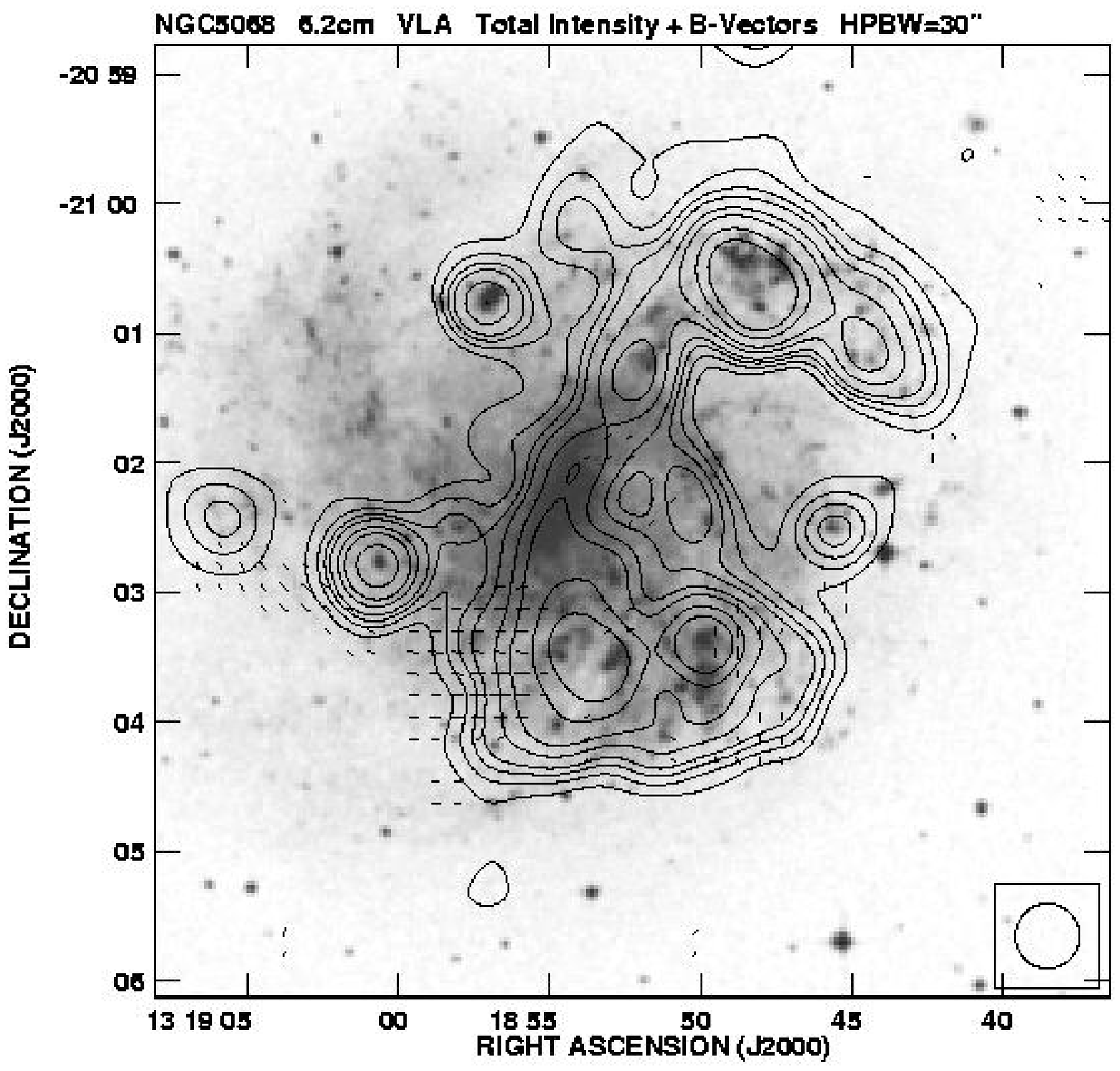}
\hfill
\includegraphics[bb = 36 156 567 660,width=8.8cm,clip]{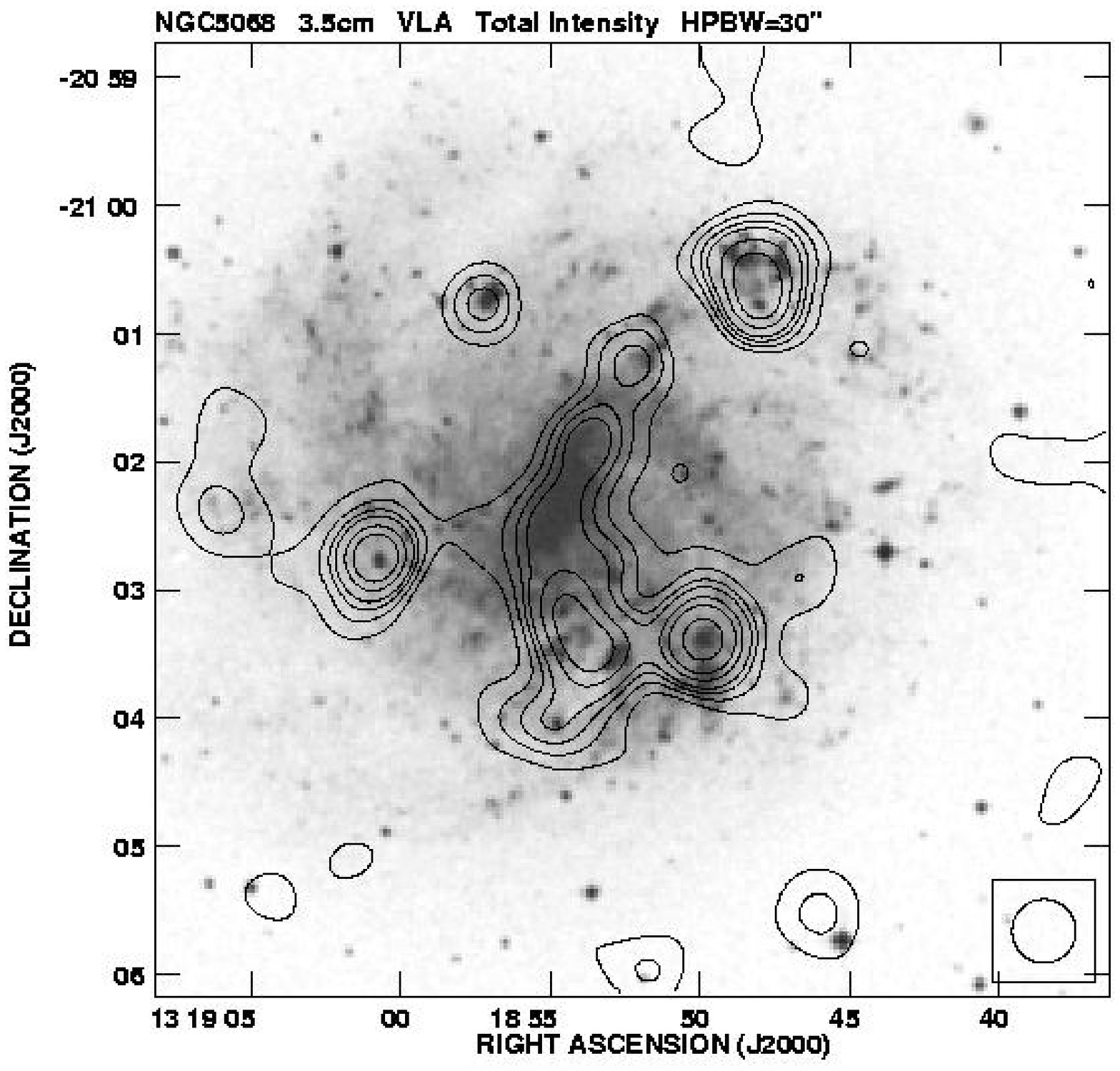}
}
\caption{Total intensity contours and the observed $B$-vectors of polarized emission
of NGC~5068, overlayed onto an optical image from the Digitized
Sky Surveys. The contours and the vector scale are as in Fig.~6.}
\end{figure*}


\begin{figure*}
\hbox to \textwidth{
\includegraphics[bb = 39 138 568 692,width=8.8cm,clip]{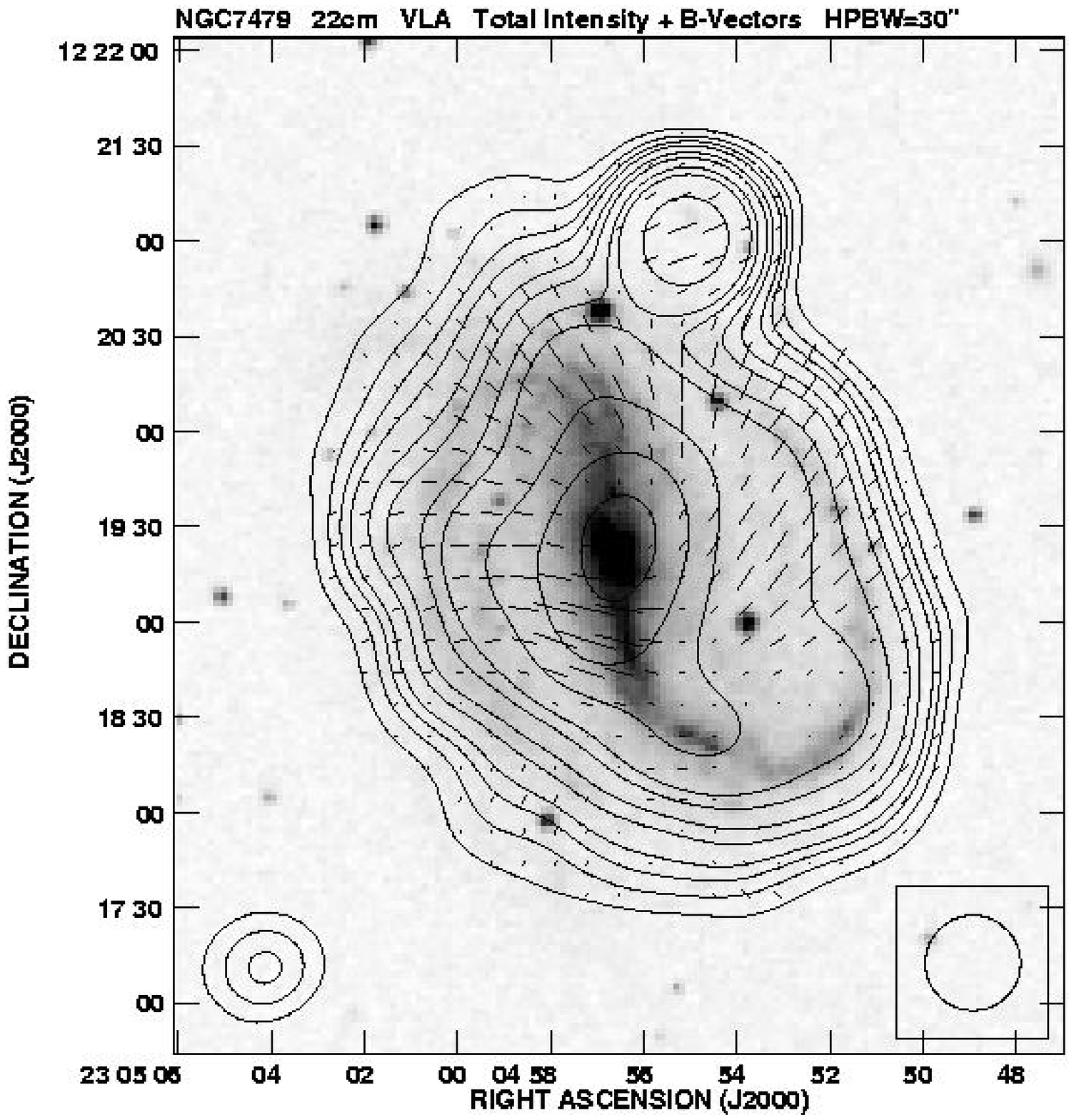}
\hfill
\includegraphics[bb = 36 138 568 692,width=8.8cm,clip]{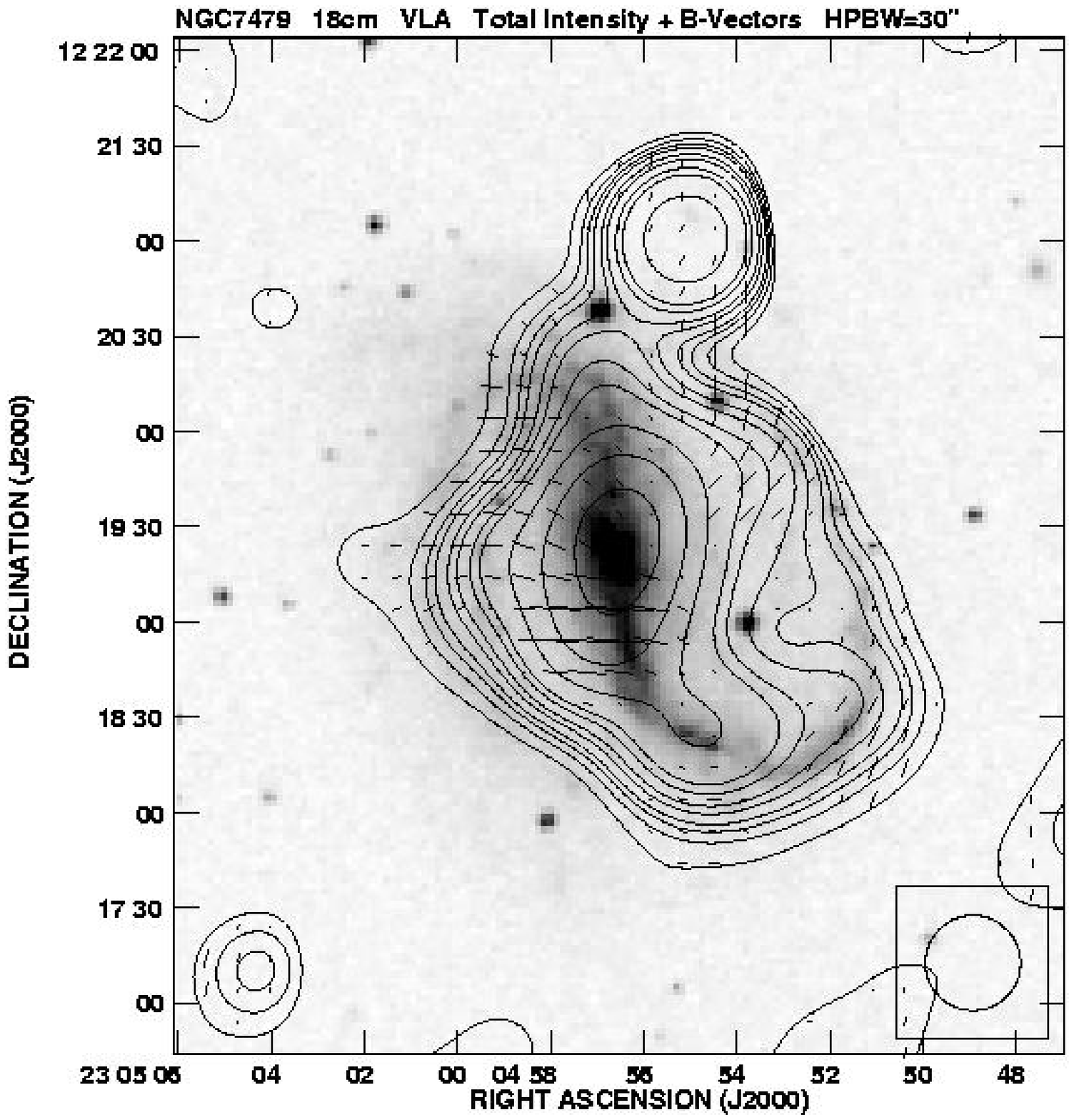}
}
\vspace{1.5cm}
\hbox to\textwidth{
\includegraphics[bb = 39 138 568 692,width=8.8cm,clip]{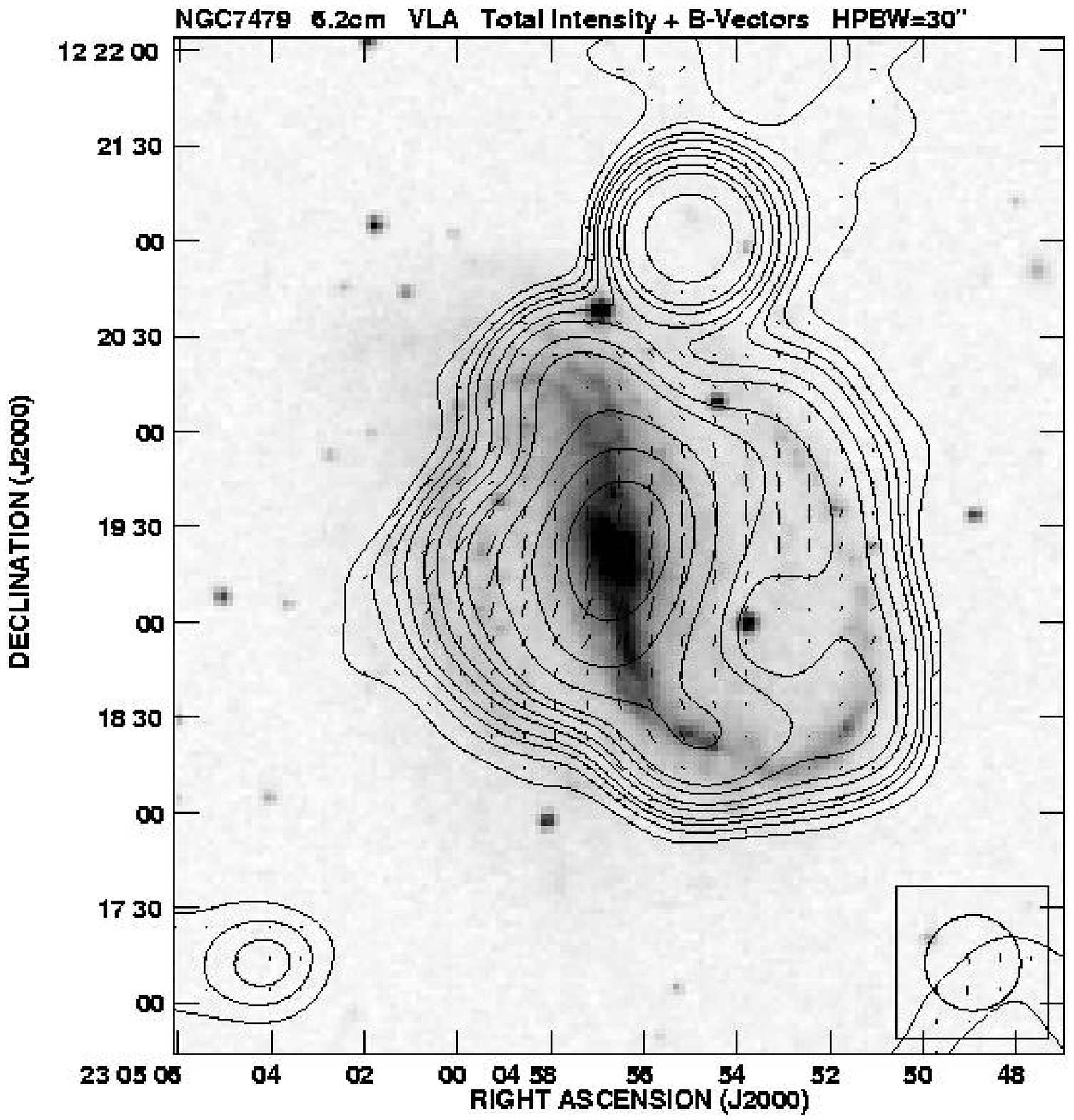}
\hfill
\includegraphics[bb = 36 138 568 692,width=8.8cm,clip]{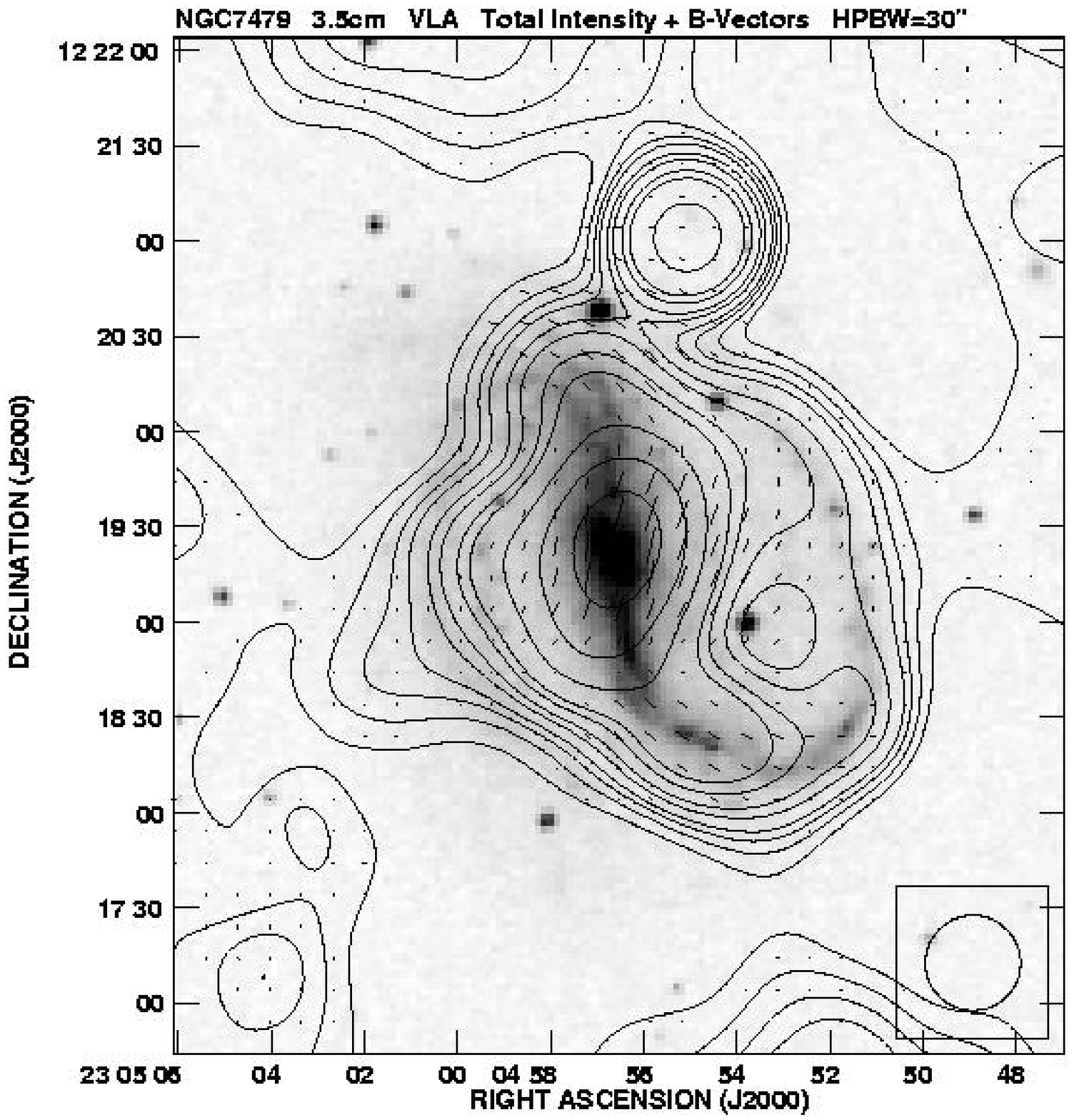}
}
\caption{Total intensity contours and the observed $B$-vectors of polarized emission
of NGC~7479, overlayed onto an optical image from the Digitized
Sky Surveys. The contours are as in Fig.~6. A vector of 1\arcsec\
length corresponds to a polarized intensity of 20~$\mu$Jy/beam area.}
\end{figure*}


\begin{figure*}
\begin{minipage}[t]{8.8cm}
\includegraphics[bb = 39 173 568 657,width=8.8cm,clip]{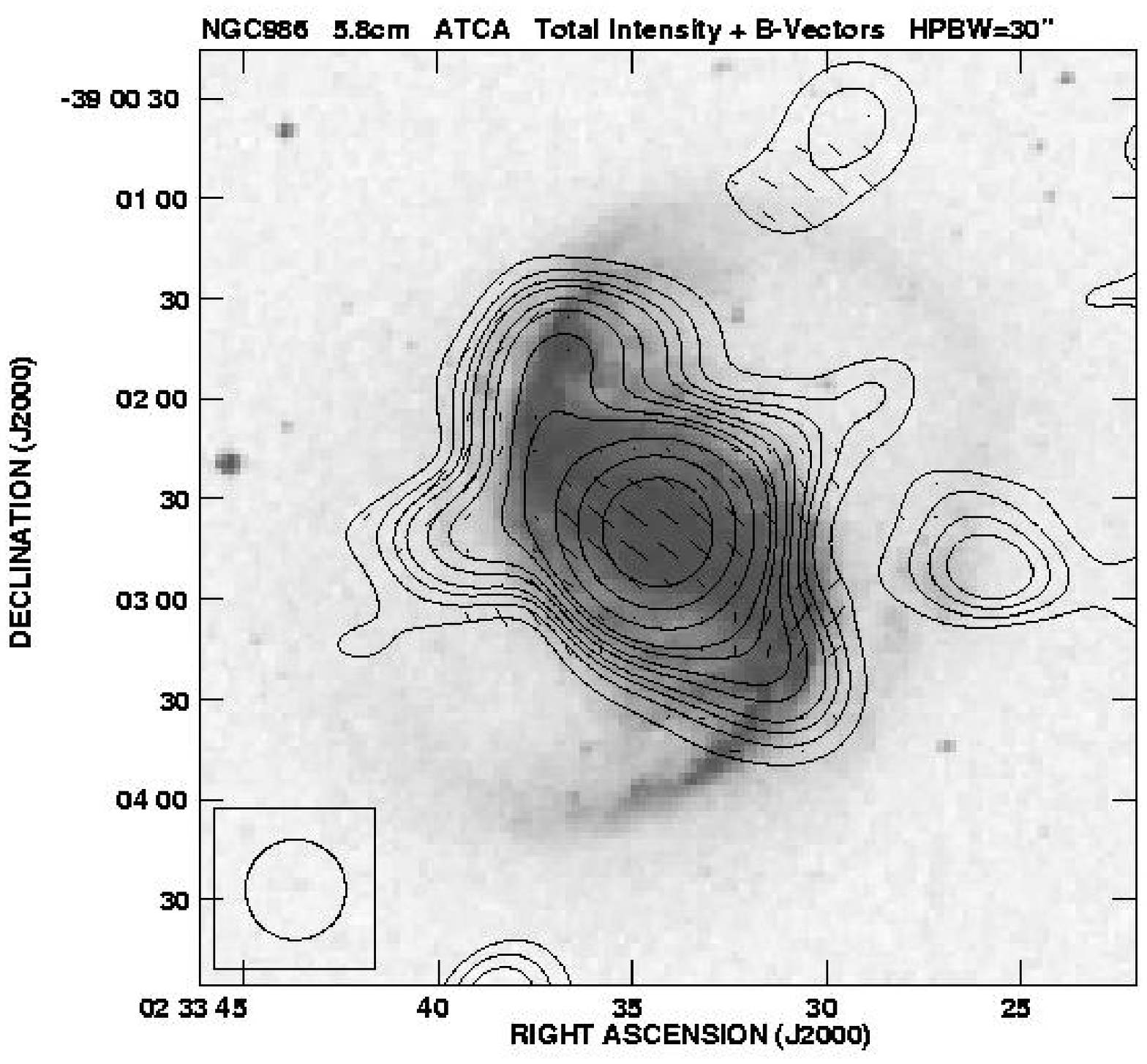}
\caption{Total intensity contours and the observed $B$-vectors of polarized emission
of NGC~986, overlayed onto an optical image from the Digitized
Sky Surveys. The contour levels are 1, 2, 3, 4, 6, 8, 12, 16, 32,
64, 128 $\times$ the basic contour level of 100~$\mu$Jy/beam.
A vector of 1\arcsec\ length corresponds to a polarized intensity
of 10~$\mu$Jy/beam area.
}
\end{minipage}\hfill
\begin{minipage}[t]{8.8cm}
\includegraphics[bb = 36 155 568 660,width=8.8cm,clip]{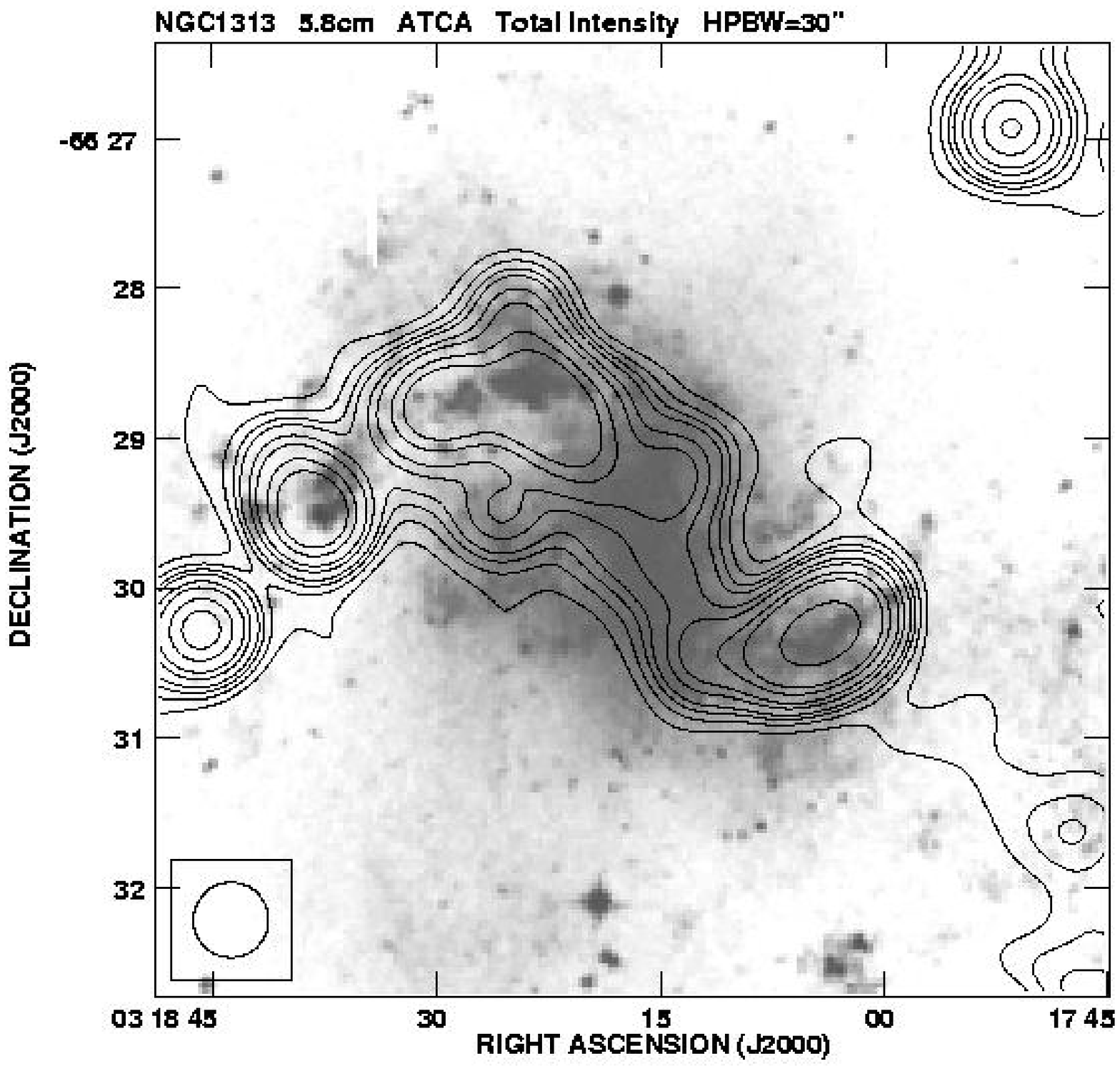}
\caption{Total intensity contours
of NGC~1313, overlayed onto an optical image from the Digitized
Sky Surveys. The contours are as in Fig.~15.}
\end{minipage}
\vspace{1cm}
\end{figure*}


\begin{figure*}
\begin{minipage}[t]{8.8cm}
\includegraphics[bb = 36 165 568 650,width=8.8cm,clip]{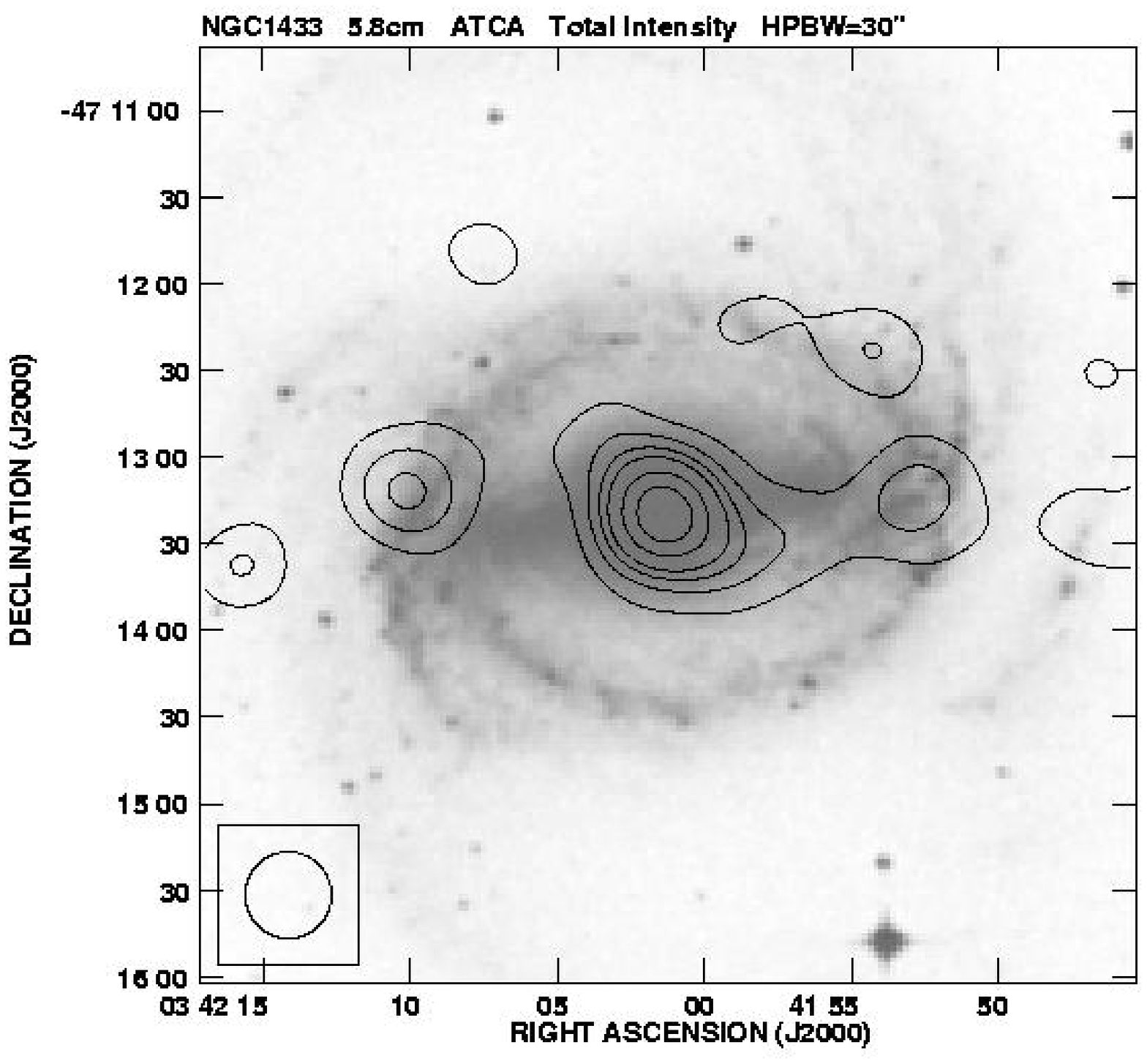}
\caption{Total intensity contours
of NGC~1433, overlayed onto an optical image from the Digitized
Sky Surveys. The contours are as in Fig.~15.}
\end{minipage}\hfill
\begin{minipage}[t]{8.8cm}
\includegraphics[bb = 39 173 568 657,width=8.8cm,clip]{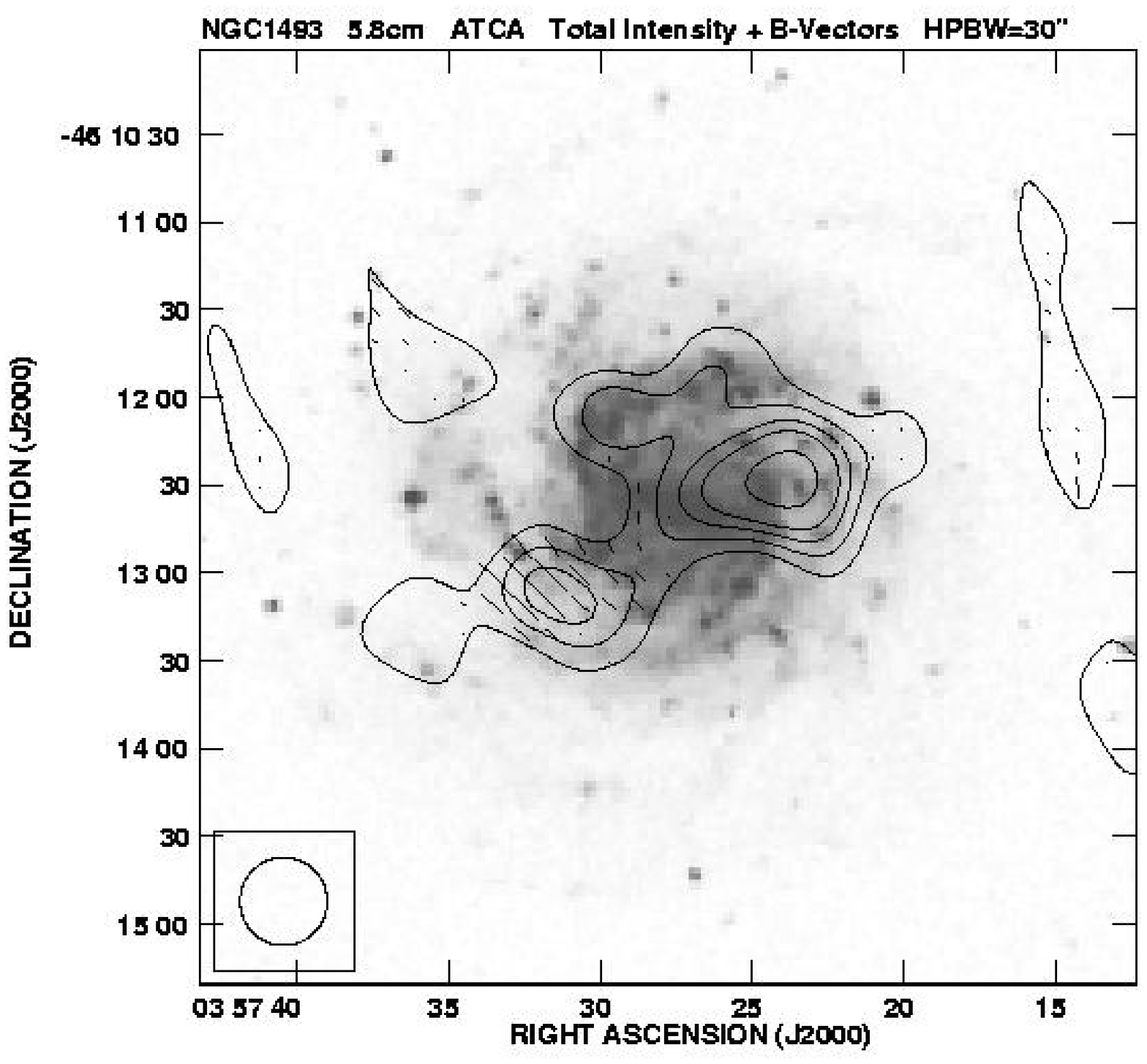}
\caption{Total intensity contours and the observed $B$-vectors of polarized emission
of NGC~1493, overlayed onto an optical image from the Digitized
Sky Surveys. The contours and the vector scale are as in Fig.~15.}
\end{minipage}
\end{figure*}


\begin{figure*}
\hbox to \textwidth{
\includegraphics[bb = 40 173 569 656,width=8.8cm,clip]{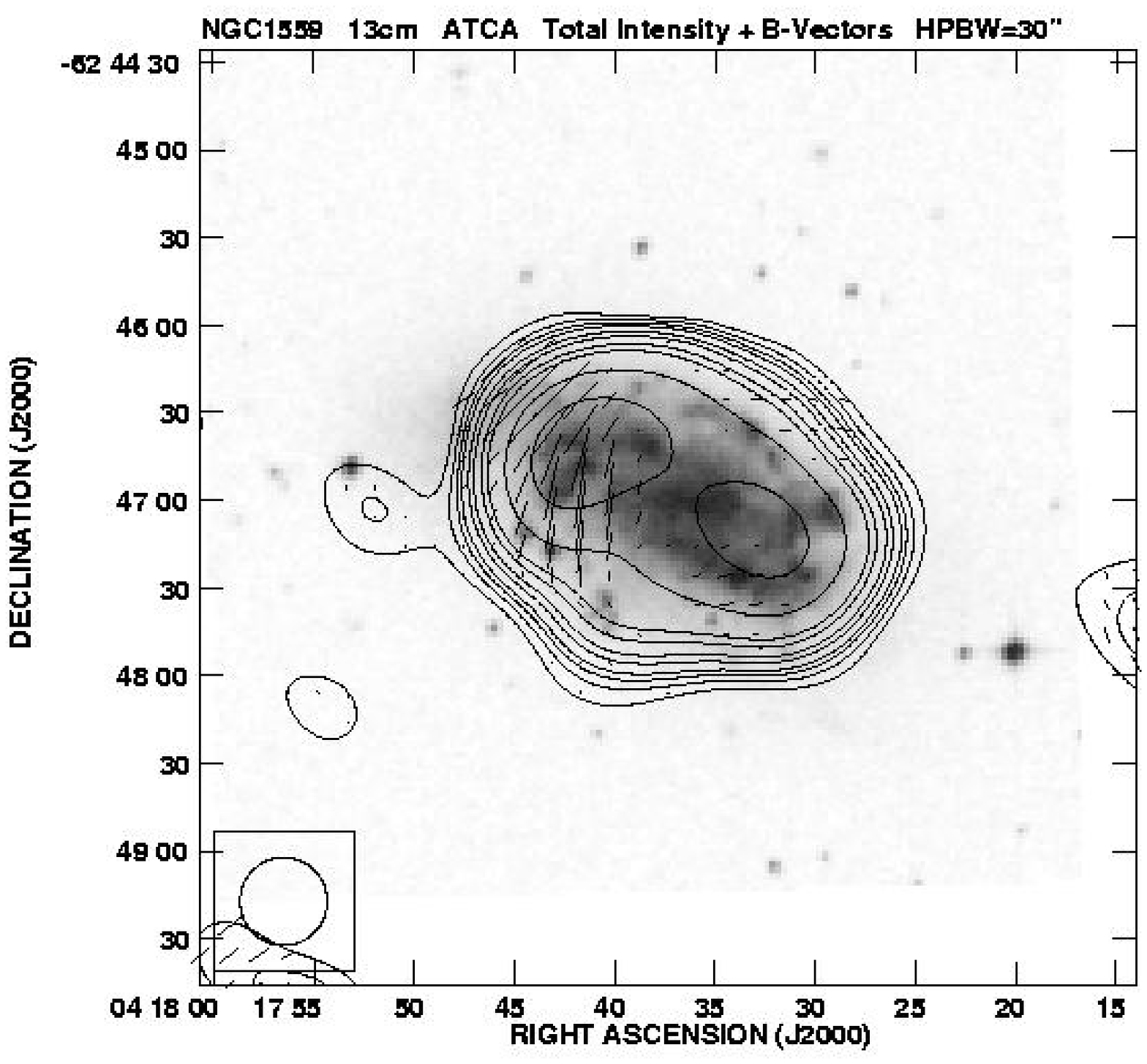}
\hfill
\includegraphics[bb = 40 173 567 656,width=8.8cm,clip]{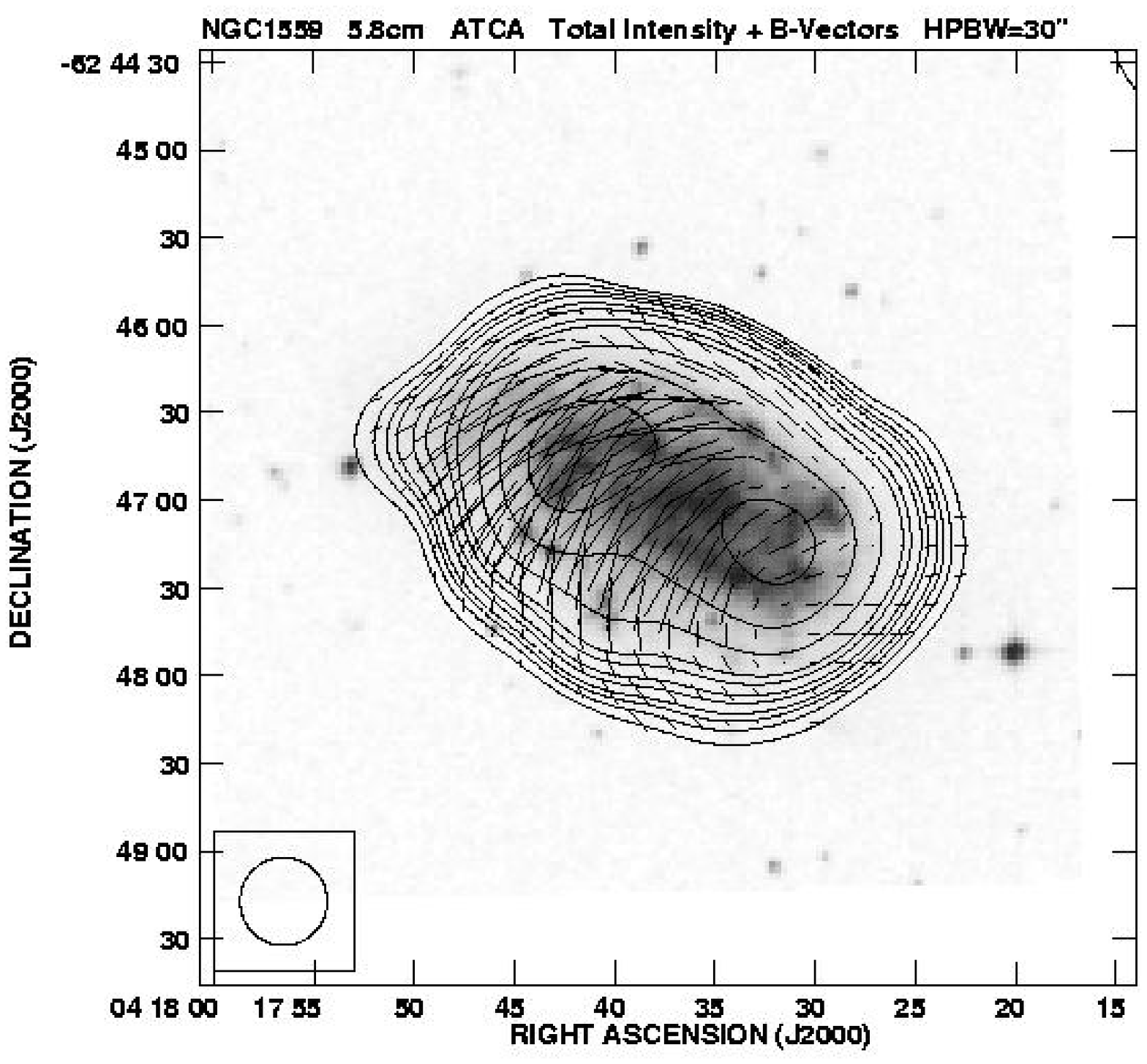}
}
\caption{Total intensity contours and the observed $B$-vectors of polarized emission
of NGC~1559, overlayed onto an optical image from the Digitized
Sky Surveys. The basic contour levels are 300 and 100~$\mu$Jy/beam, the
contour intervals and the vector scale are as in Fig.~15.}
\end{figure*}


\begin{figure*}
\hbox to \textwidth{
\includegraphics[bb = 40 173 567 656,width=8.8cm,clip]{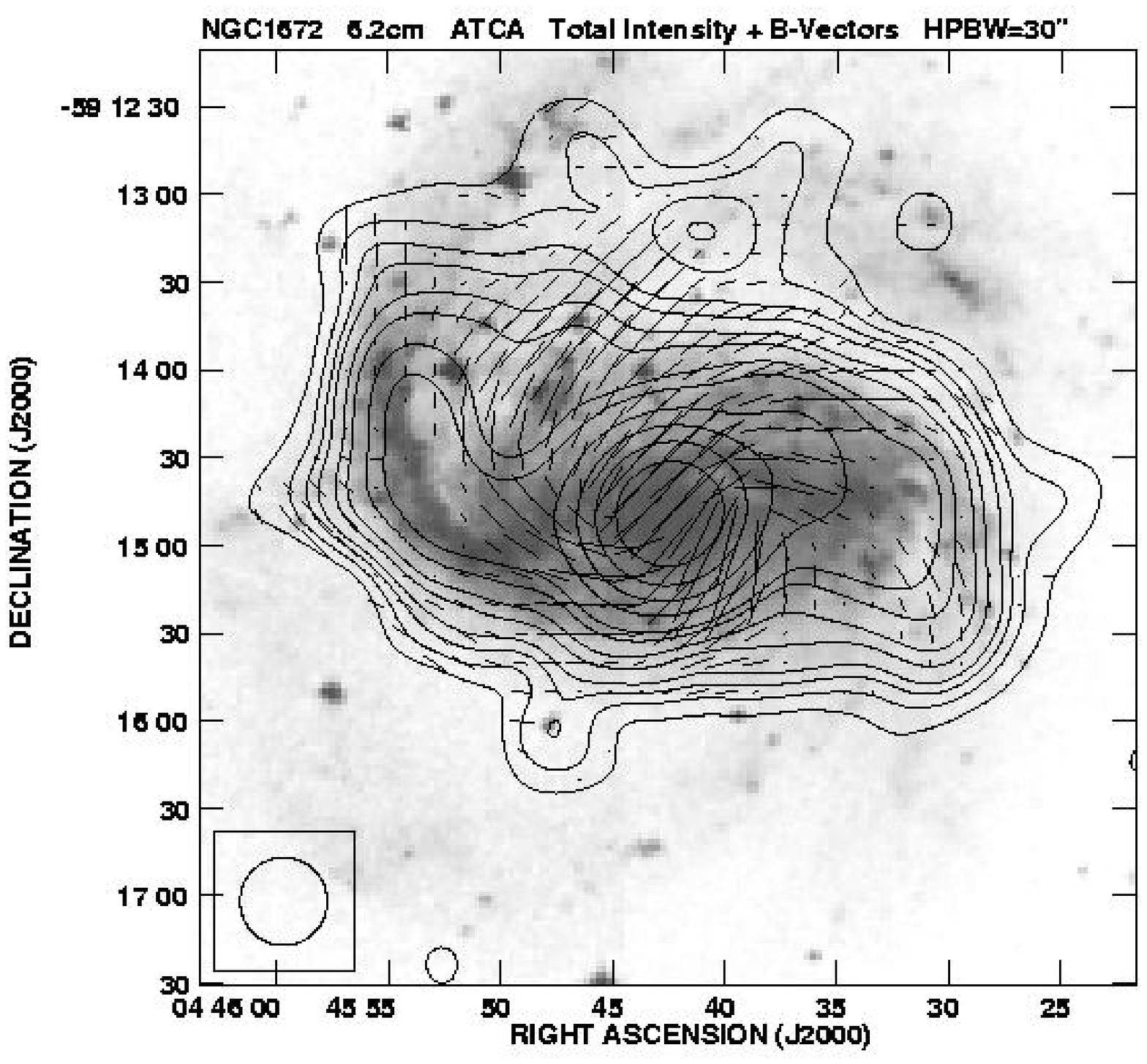}
\hfill
\includegraphics[bb = 40 173 568 656,width=8.8cm,clip]{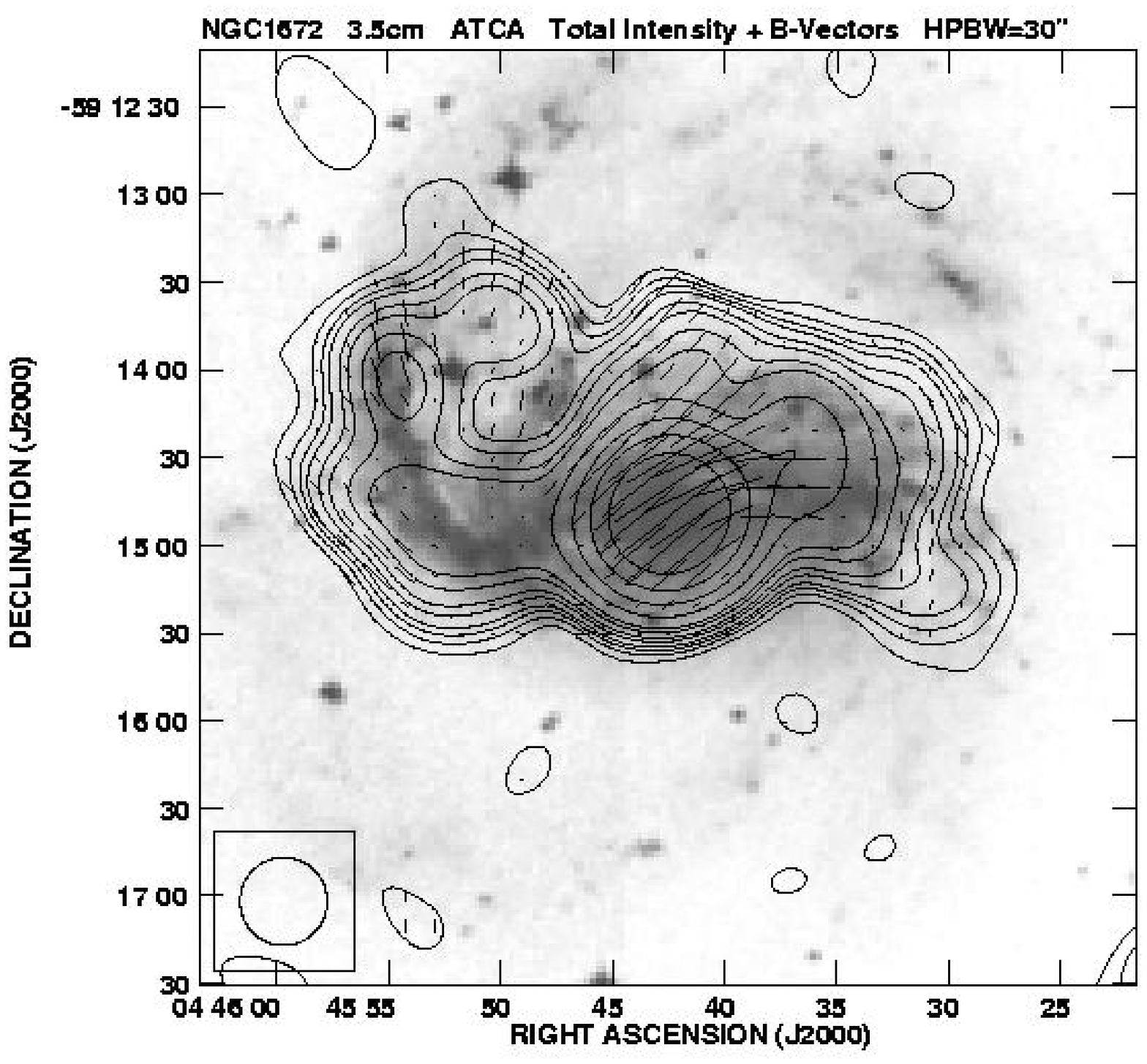}
}
\caption{Total intensity contours and the observed $B$-vectors of polarized emission
of NGC~1672, overlayed onto an optical image from the Digitized
Sky Surveys. The basic contour levels are 100 and 40~$\mu$Jy/beam, the
contour intervals and the vector scale are as in Fig.~15.}
\end{figure*}

\clearpage

\begin{figure*}
\hbox to \textwidth{
\includegraphics[bb = 39 163 567 667,width=8.8cm,clip]{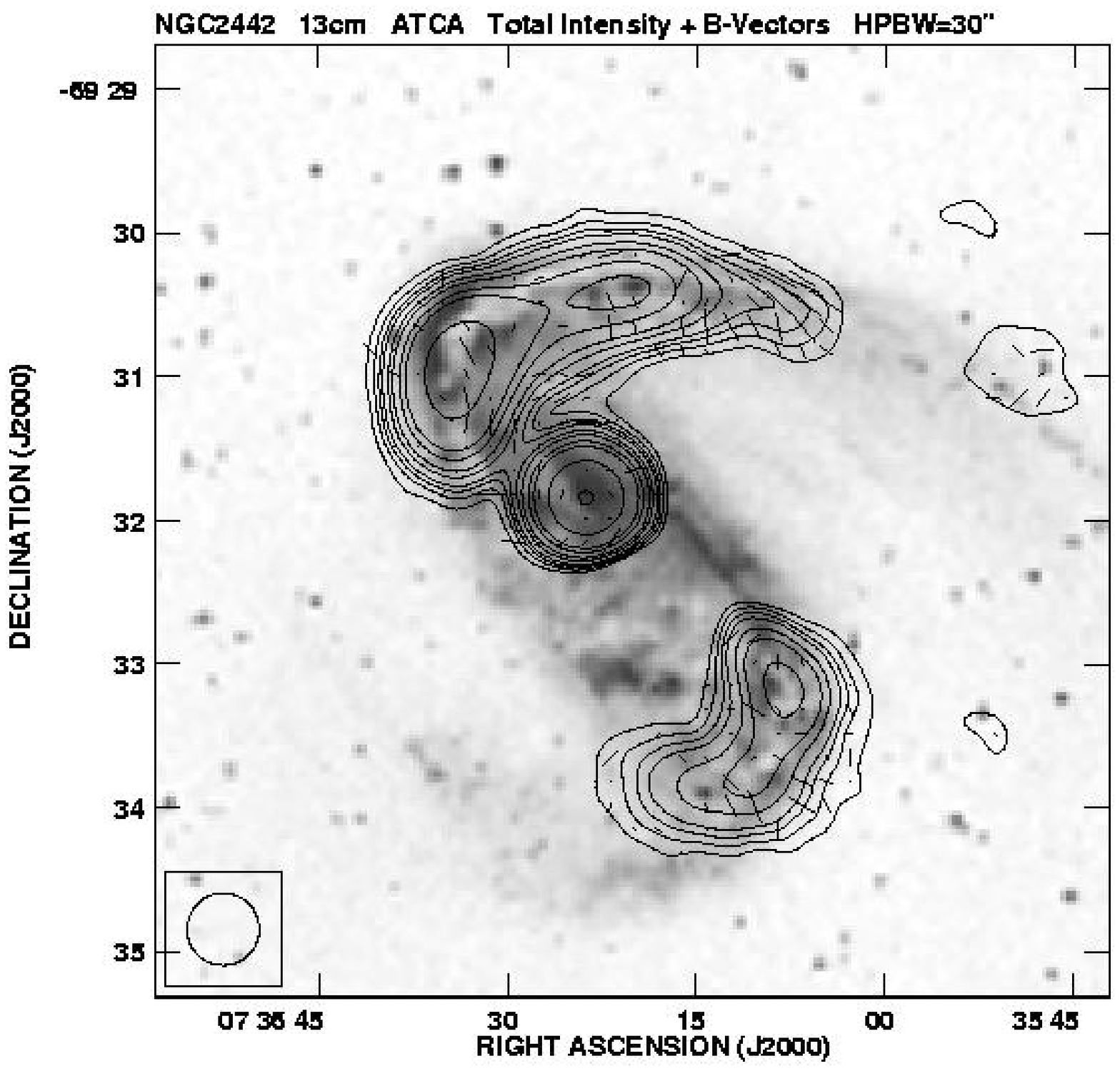}
\hfill
\includegraphics[bb = 39 163 567 667,width=8.8cm,clip]{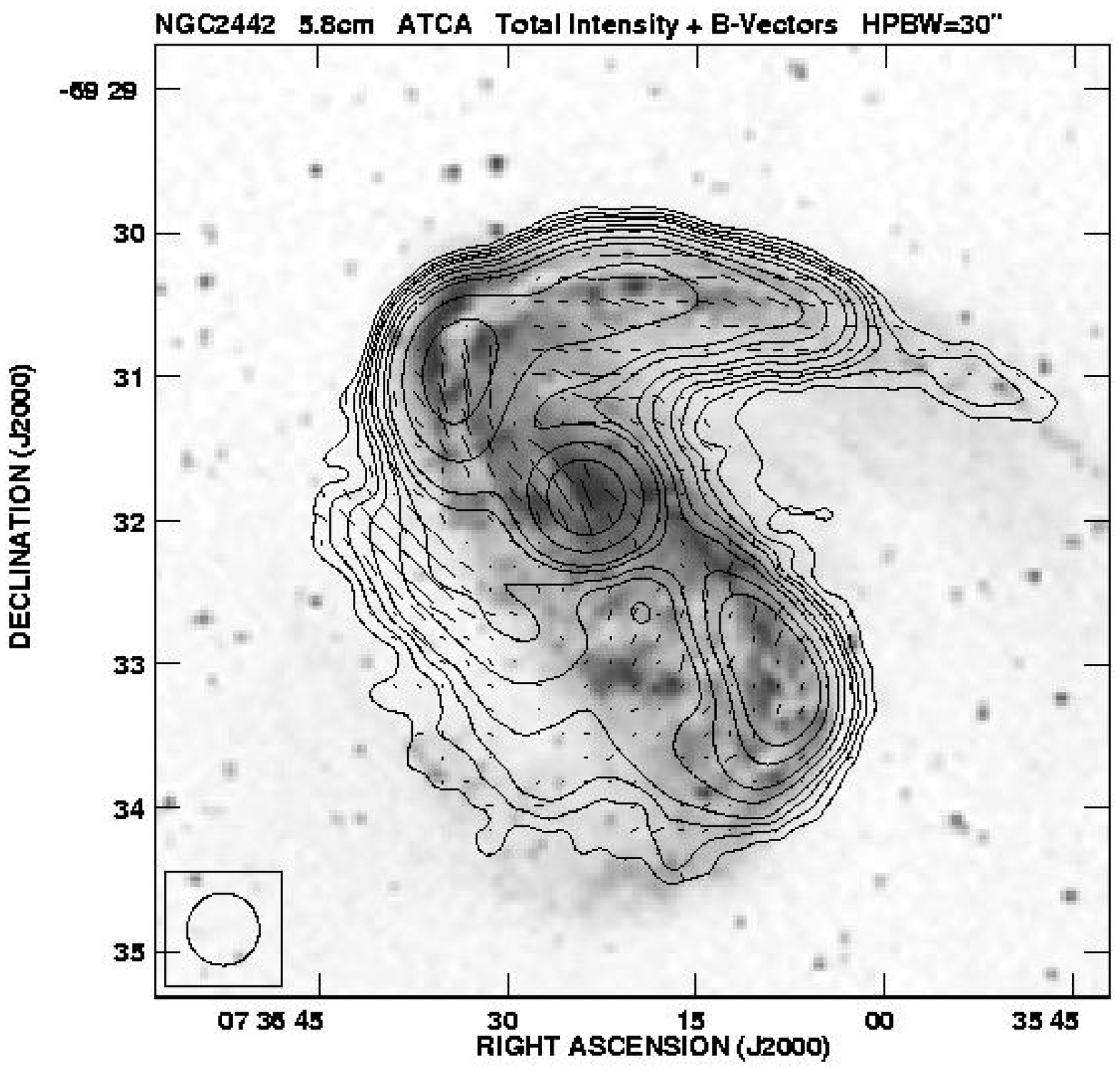}
}
\caption{Total intensity contours and the observed $B$-vectors of polarized emission
of NGC~2442, overlayed onto an optical image from the Digitized
Sky Surveys. The contours and the vector scale are as in Fig.~19.}
\end{figure*}


\begin{figure*}
\hbox to \textwidth{
\includegraphics[bb = 39 173 567 657,width=8.8cm,clip]{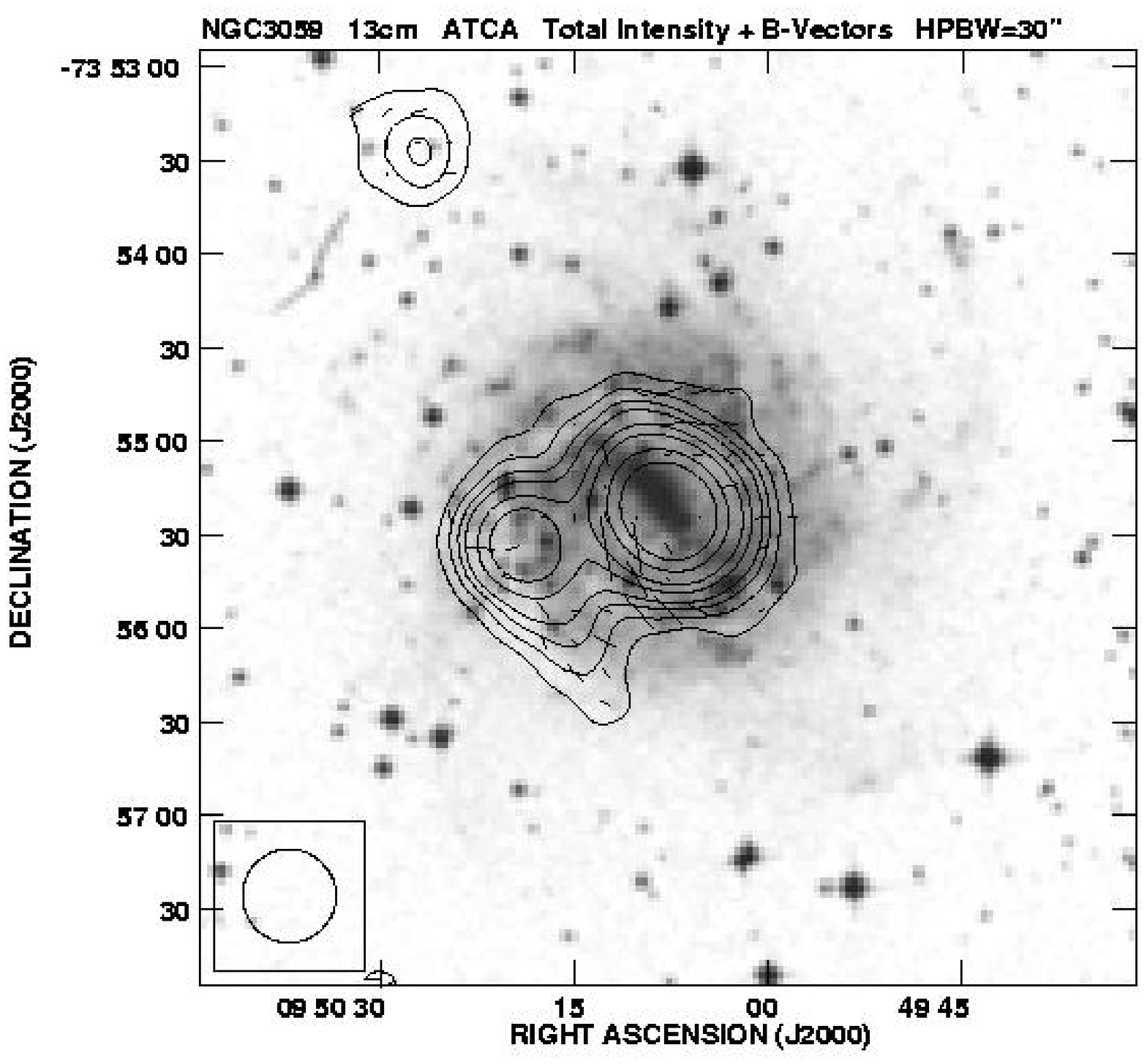}
\hfill
\includegraphics[bb = 39 173 567 657,width=8.8cm,clip]{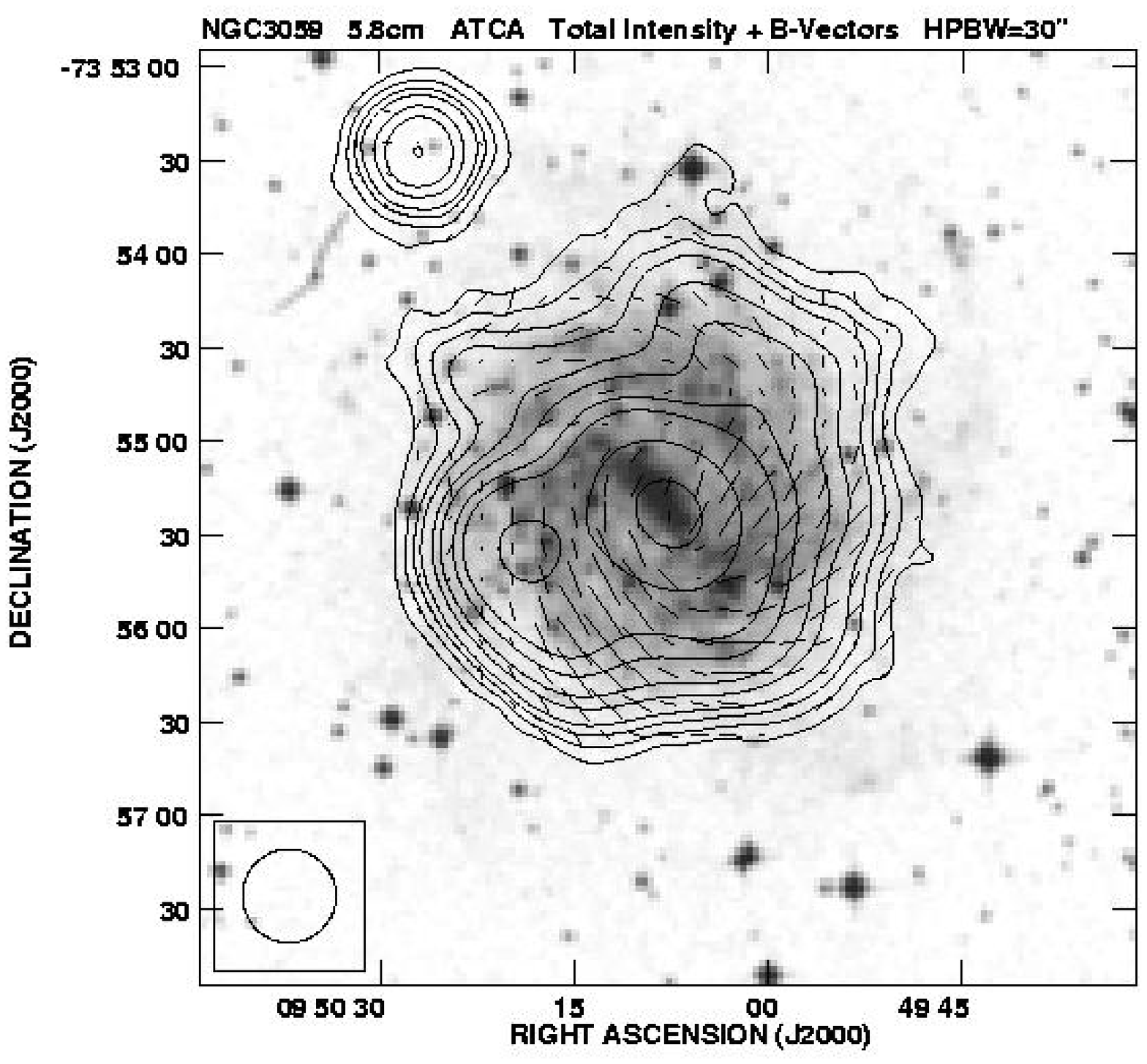}
}
\caption{Total intensity contours and the observed $B$-vectors of polarized emission
of NGC~3059, overlayed onto an optical image from the Digitized
Sky Surveys. The contours and the vector scale are as in Fig.~19.}
\end{figure*}


\begin{figure*}
\begin{minipage}[t]{8.8cm}
\includegraphics[bb = 39 173 567 657,width=8.8cm,clip]{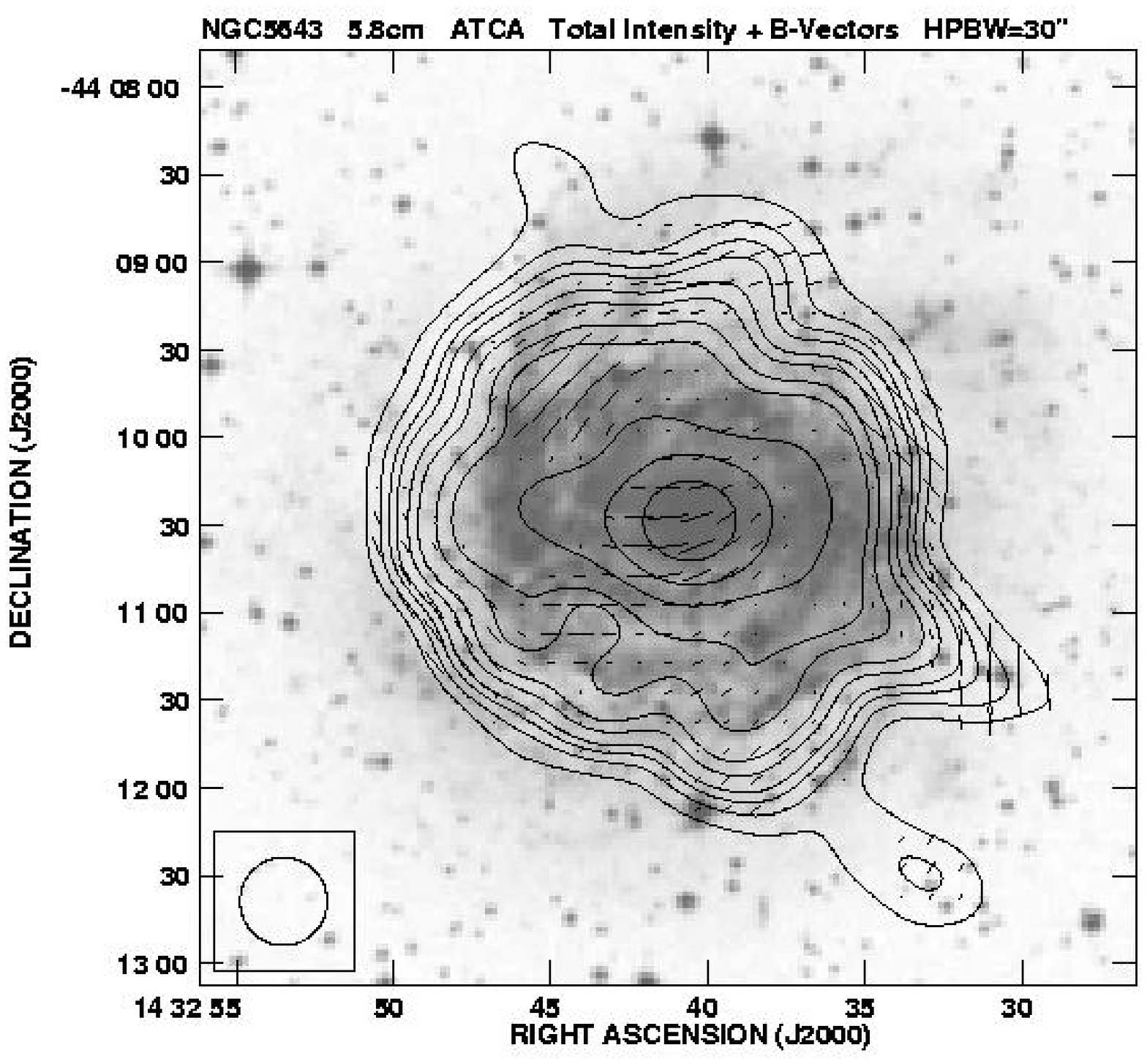}
\caption{Total intensity contours and the observed $B$-vectors of polarized emission
of NGC~5643, overlayed onto an optical image from the Digitized
Sky Surveys. The contours and the vector scale are as in Fig.~15.}
\end{minipage}\hfill
\begin{minipage}[t]{8.8cm}
\includegraphics[bb = 39 172 567 658,width=8.8cm,clip]{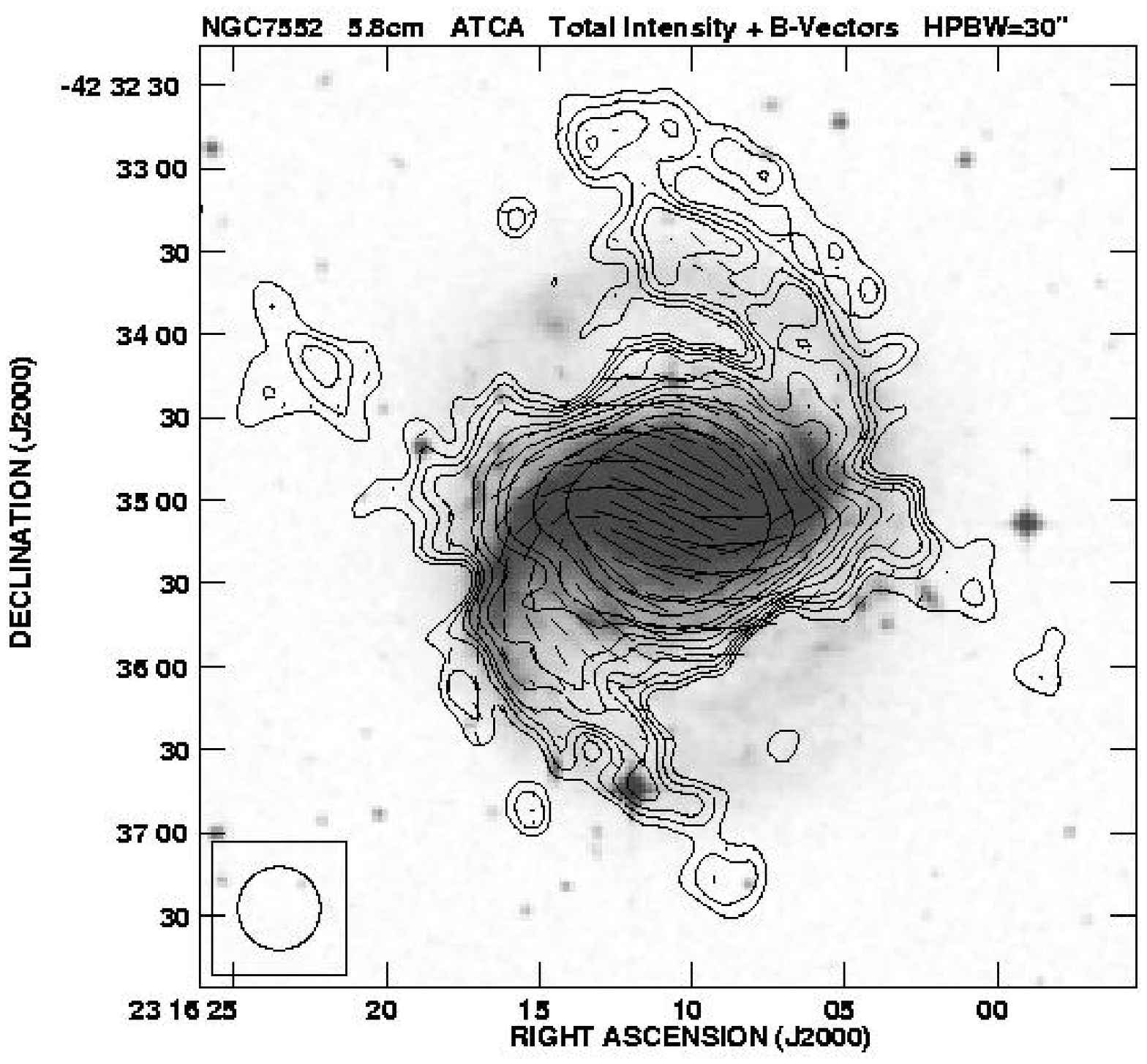}
\caption{Total intensity contours and the observed $B$-vectors of polarized emission
of NGC~7552, overlayed onto an optical image from the Digitized
Sky Surveys. The contours and the vector scale are as in Fig.~15.}
\end{minipage}
\end{figure*}

\clearpage

\begin{figure*}
\hbox to \textwidth{
\includegraphics[bb = 40 173 567 656,width=8.8cm,clip]{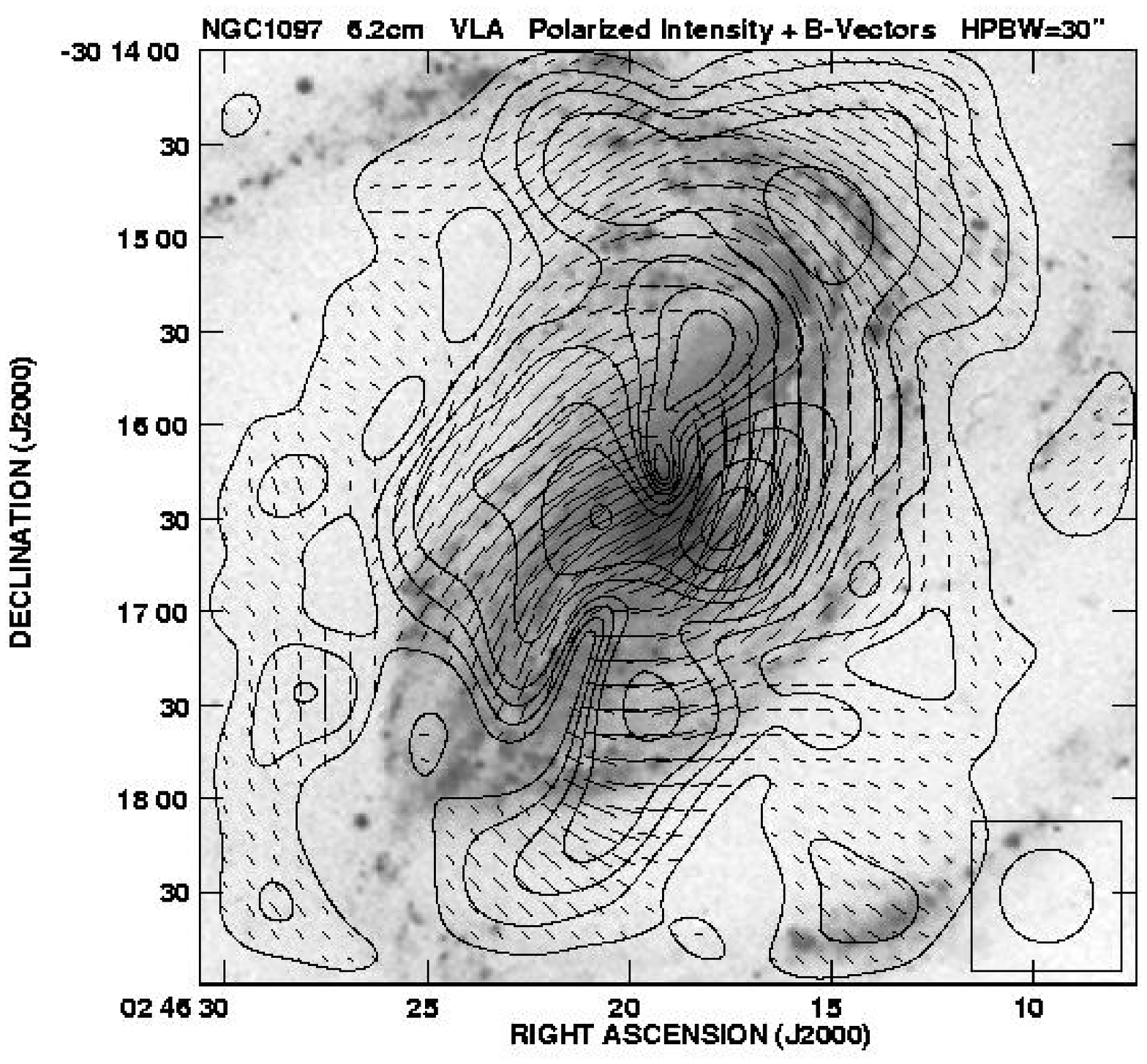}
\hfill
\includegraphics[bb = 40 173 567 656,width=8.8cm,clip]{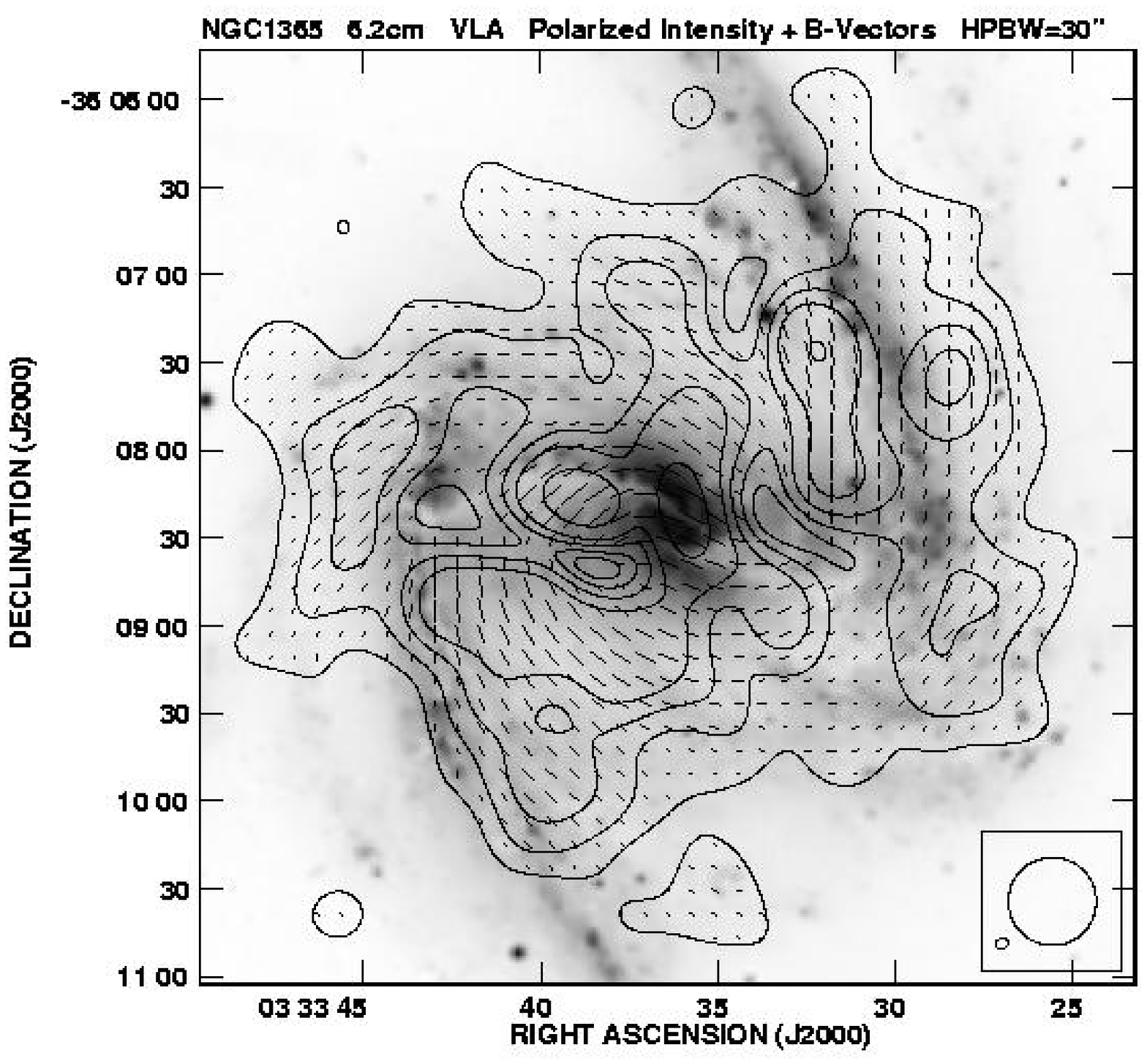}
}
\vspace{1.5cm}
\hbox to\textwidth{
\includegraphics[bb = 40 170 567 661,width=8.8cm,clip]{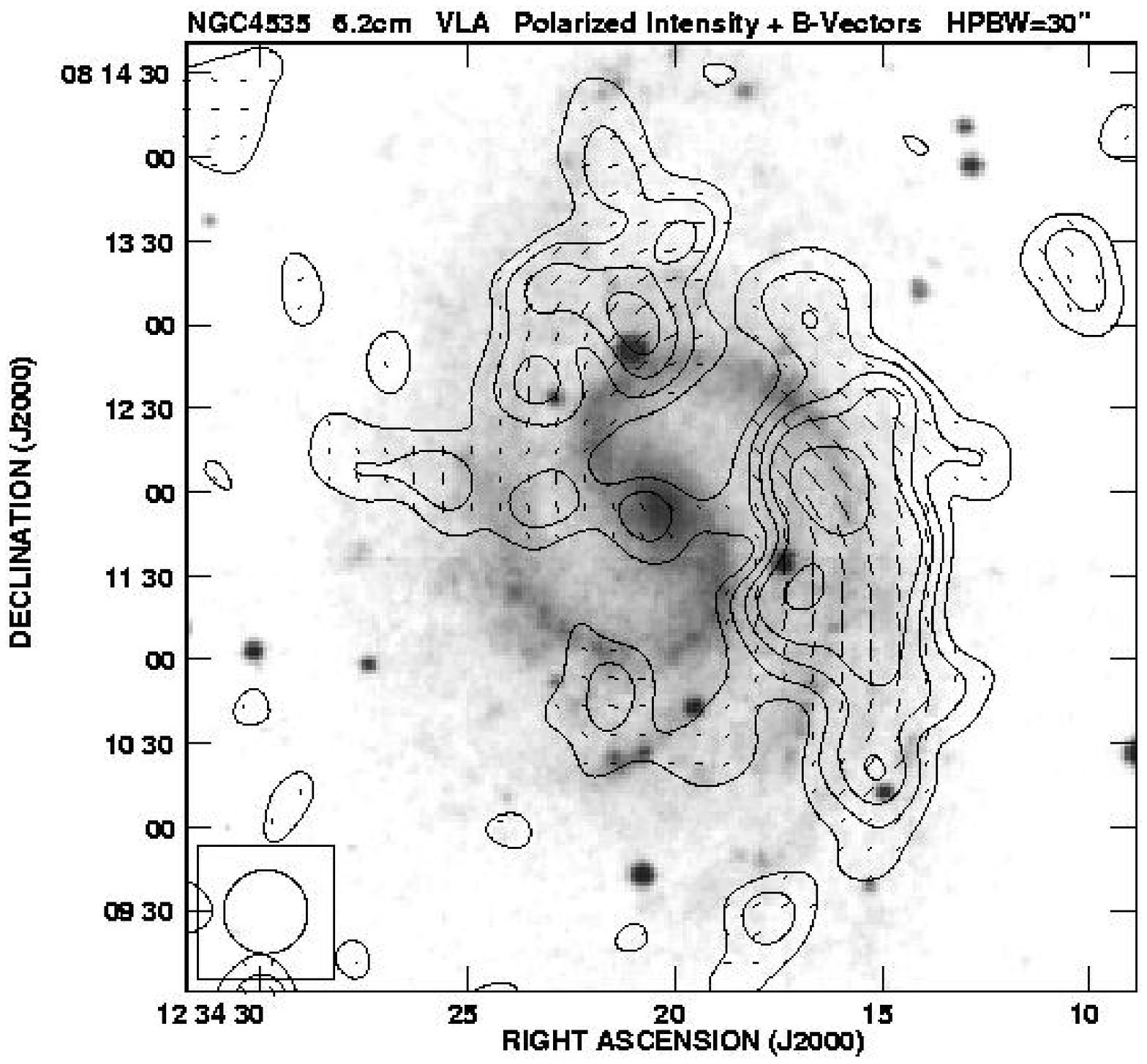}
\hfill
\includegraphics[bb = 40 139 569 692,width=8.8cm,clip]{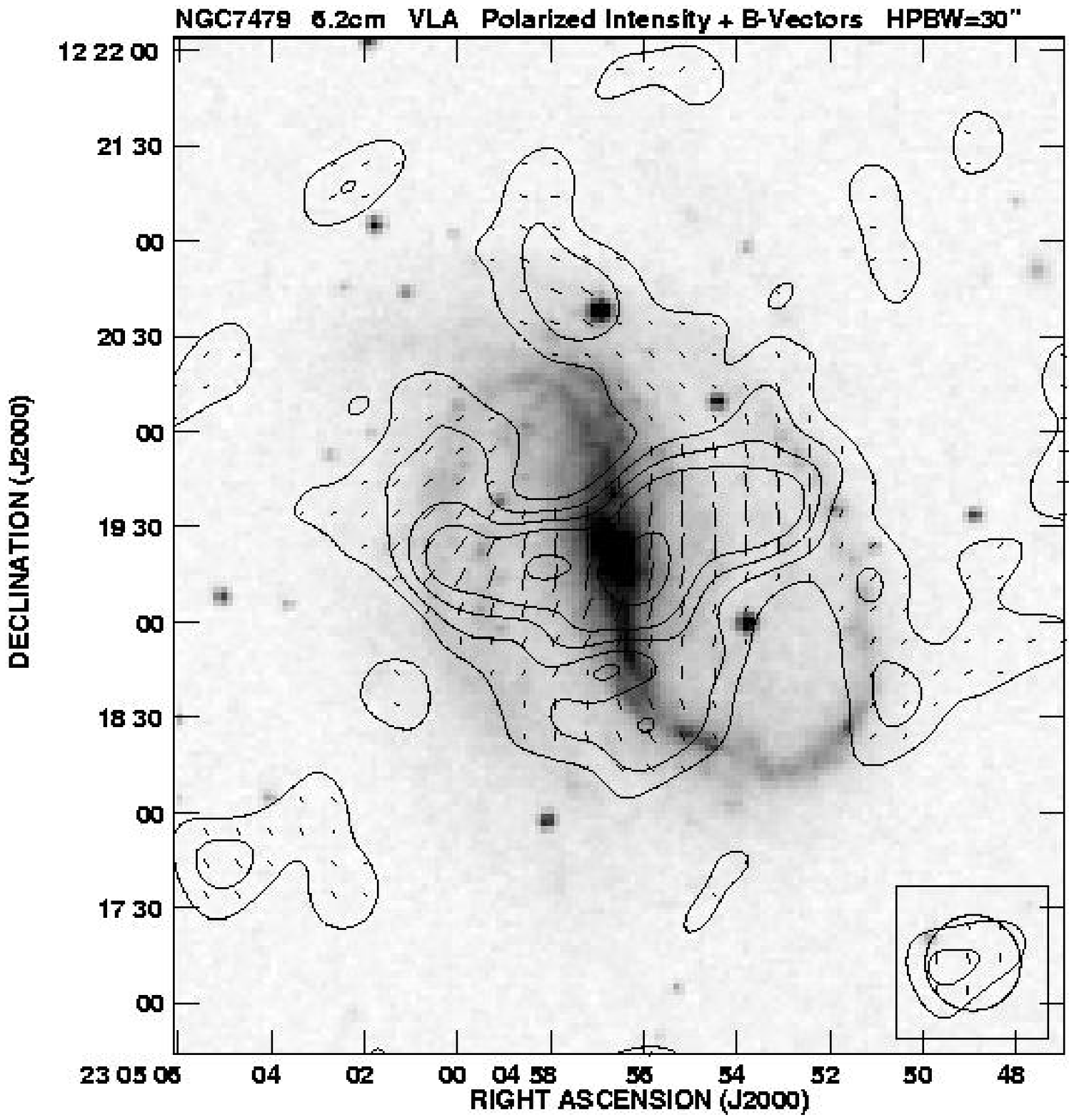}
}
\caption{Polarized intensity contours and the observed
$B$-vectors of NGC~1097, NGC~1365,
NGC~4535 and NGC~7479. The contour intervals are 1, 2, 3, 4, 6, 8, 12, 16
$\times$ the basic contour level, which is 50, 30, 30
and 30~$\mu$Jy/beam in the order of increasing NGC number.
A vector of 1\arcsec\ length corresponds to a
polarized intensity of 20~$\mu$Jy/beam area.}
\end{figure*}


\begin{figure*}
\hbox to \textwidth{
\includegraphics[bb = 40 173 567 656,width=8.8cm,clip]{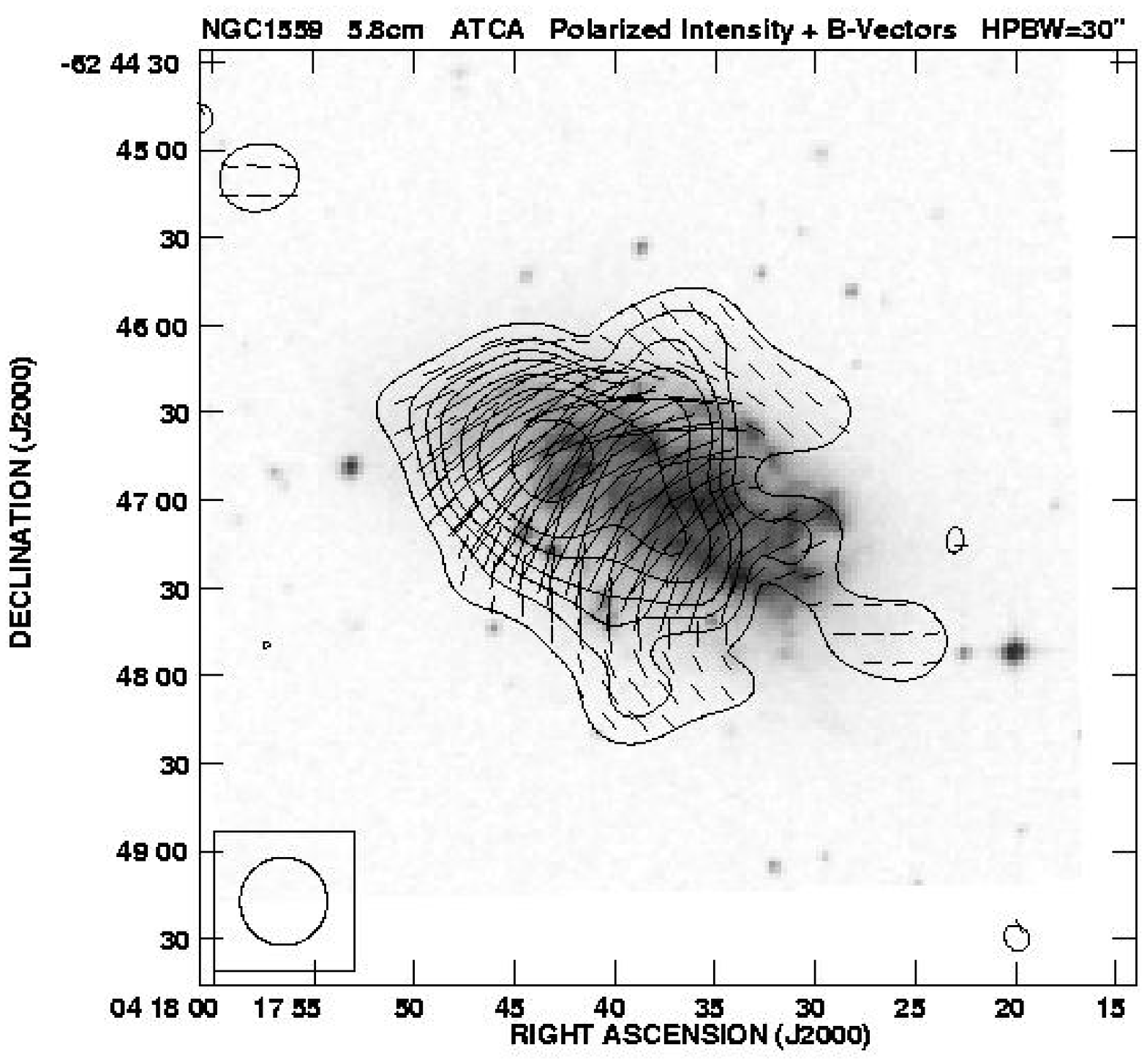}
\hfill
\includegraphics[bb = 40 173 571 656,width=8.8cm,clip]{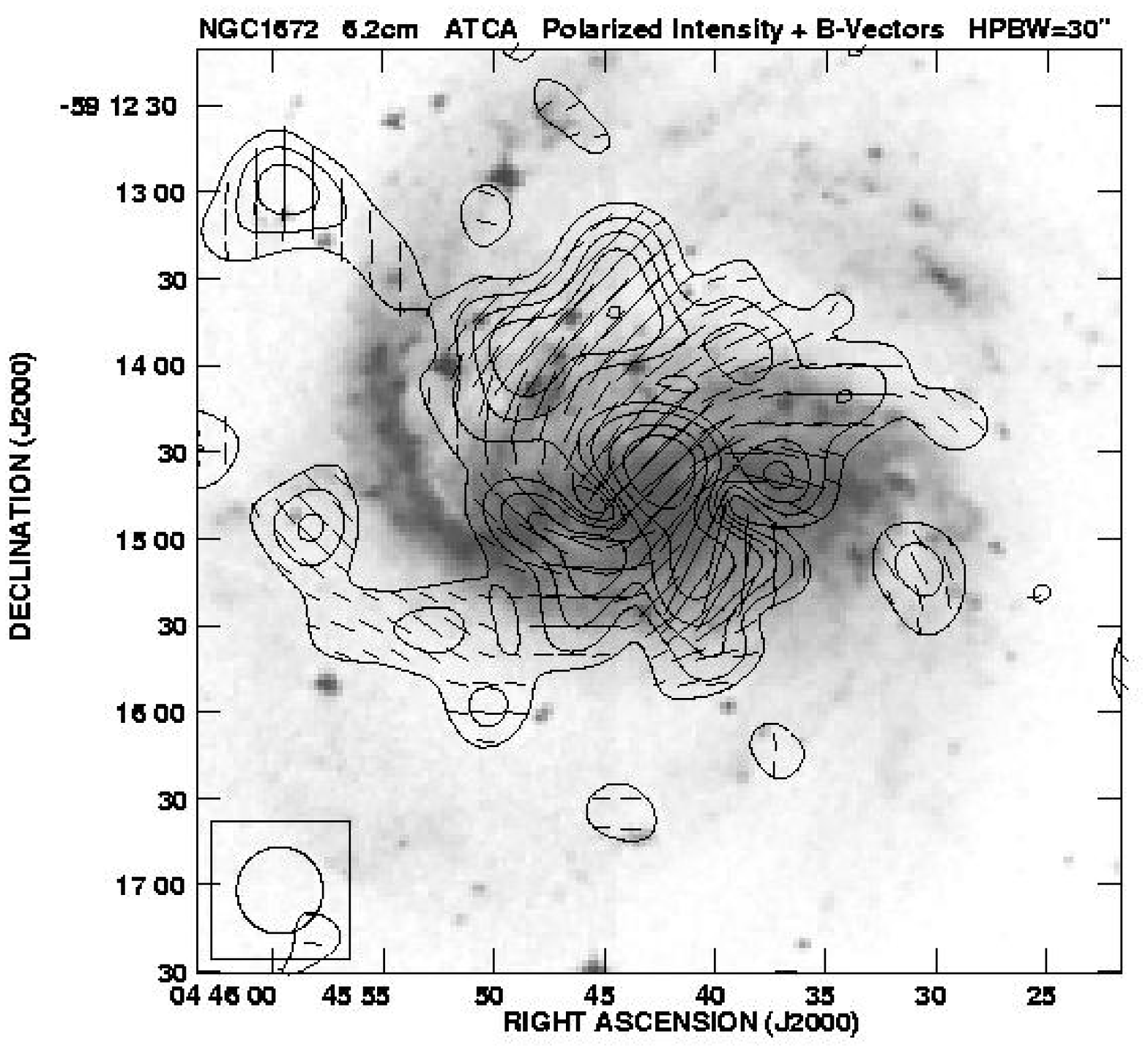}
}
\vspace{1.5cm}
\hbox to\textwidth{
\includegraphics[bb = 40 164 567 667,width=8.8cm,clip]{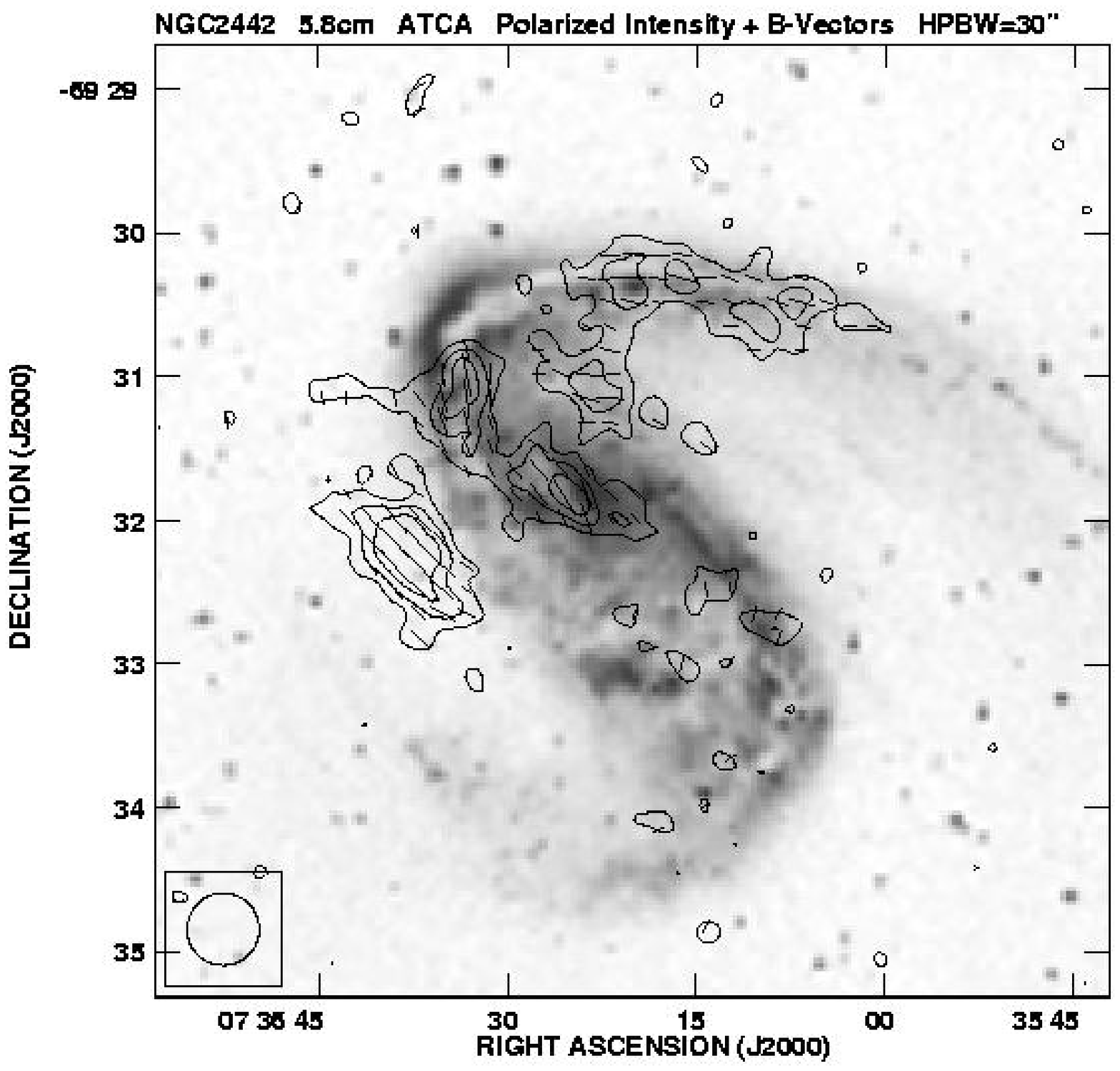}
\hfill
\includegraphics[bb = 40 172 567 657,width=8.8cm,clip]{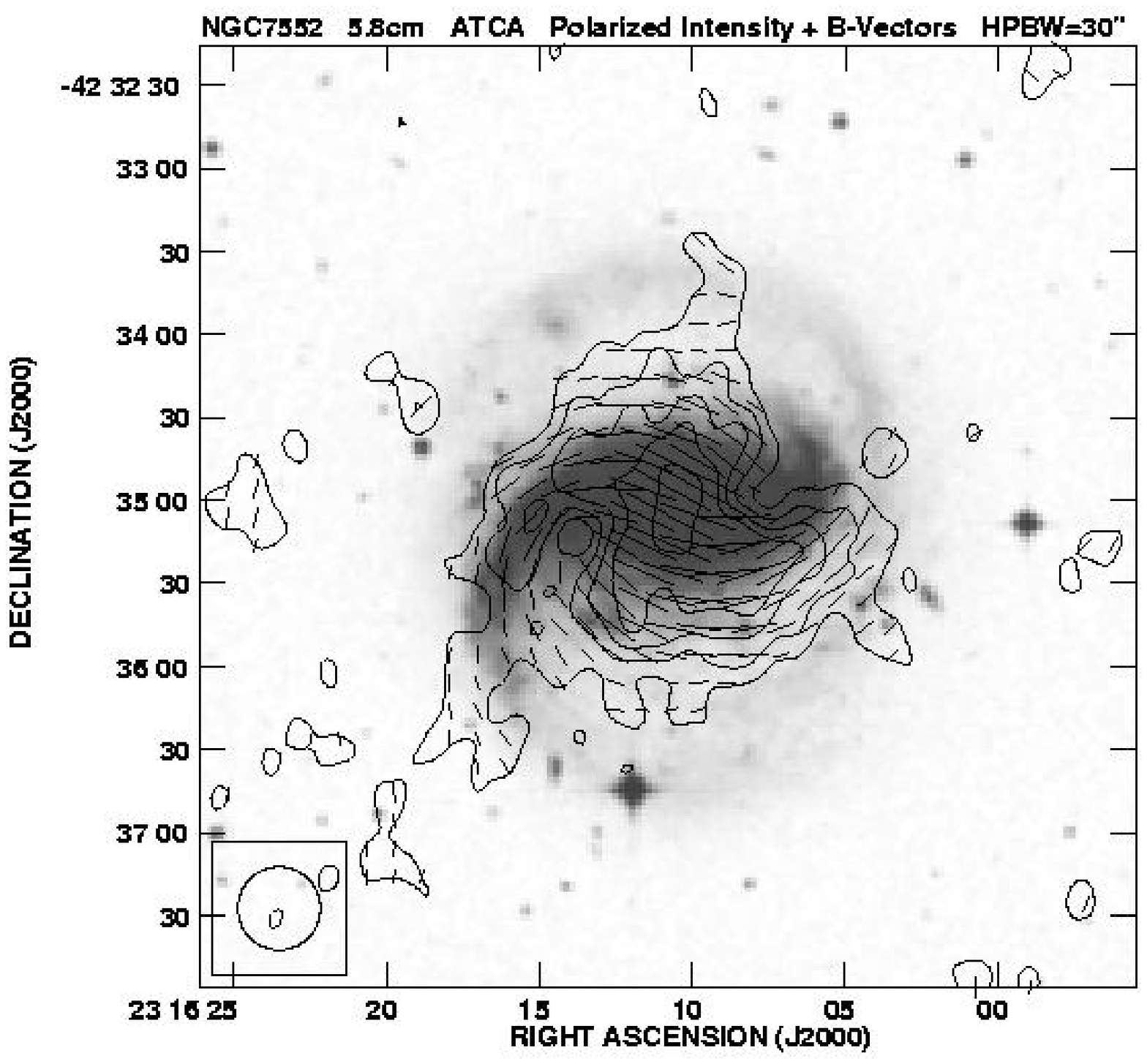}
}
\caption{Polarized intensity contours and the observed
$B$-vectors of NGC~1559, NGC~1672,
NGC~2442 and NGC~7552. The contour intervals are 1, 2, 3, 4, 6, 8, 12
$\times$ the basic contour level, which is 50~$\mu$Jy/beam.
A vector of 1\arcsec\ length corresponds to a polarized intensity
of 10~$\mu$Jy/beam area.}
\end{figure*}



\begin{thebibliography}{}
\bigskip

\bibitem[2000]{abraham+merrifield00}
Abraham, R.~G., \& Merrifield, M.~R. 2000, AJ, 120, 2835

\bibitem[1999]{aguerri99}
Aguerri, J.~A.~L. 1999, A\&A, 351, 43

\bibitem[1992]{atha92}
Athanassoula E. 1992, MNRAS 259, 345

\bibitem[2000]{beck00}
Beck, R. 2000, Phil.\ Trans.\ R.\ Soc.\ Lond.\ A, 358, 777

\bibitem[2001]{beck01}
Beck, R. 2001, in The Astrophysics of Galactic Cosmic Rays, eds. R.
Diehl, R. Kallenbach, E. Parizot, \& R. von Steiger, Space Sci. Rev.,
99, 243 (Dordrecht: Kluwer)

\bibitem[1996]{beck+hoernes96}
Beck, R., \& Hoernes, P. 1996, Nature, 379, 47

\bibitem[1996]{beck+96}
Beck, R., Brandenburg, A., Moss, D., Shukurov, A., \& Sokoloff, D. 1996,
ARA\&A, 34, 155

\bibitem[1999]{beck+99}
Beck, R., Ehle, M., Shoutenkov, V., Shukurov, A., \& Sokoloff, D. 1999,
Nature, 397, 324

\bibitem[1990]{bicay+helou90}
Bicay, M.~D., \& Helou, G. 1990, ApJ, 362, 59

\bibitem[2001]{block+01}
Block, D.~L., Puerari, I., Knapen, J.~H., et al. 2001, A\&A, 375, 761

\bibitem[2001]{buta+block01}
Buta, R., \& Block, D.~L. 2001, ApJ, 550, 243

\bibitem[1999]{chapelon+99}
Chapelon, S., Contini, T., \& Davoust, E. 1999, A\&A, 345, 81

\bibitem[1994]{chiba+lesch94}
Chiba, M., \& Lesch, H. 1994, A\&A, 284, 731

\bibitem[1981]{combes+sanders81}
Combes, F., \& Sanders, R.~H. 1981, A\&A, 96, 164

\bibitem[1987]{condon87}
Condon, J.~J. 1987, ApJS, 65, 485

\bibitem[1992]{condon92}
Condon, J.~J. 1992, ARA\&A, 30, 575

\bibitem[1996]{crocker+96}
Crocker, D.~A., Baugus, P.~D., Buta, R. 1996, ApJS, 105, 353

\bibitem[1985]{jong+85}
de Jong, T., Klein, U., Wielebinski, R., \& Wunderlich, E. 1985, A\&A, 147, L6

\bibitem[1991]{vaucouleurs+91}
de Vaucouleurs, G., de Vaucouleurs, A., Corwin, H.~G., et al.\ 1991,
Third Reference Catalogue of Bright Galaxies (New York: Springer)

\bibitem[1985]{elmegreen+elmegreen85}
Elmegreen, B.~G., \& Elmegreen, D.~M. 1985, ApJ, 288, 438

\bibitem[1996]{evans+96}
Evans, I.~N., Koratkar, A.~P., Storchi-Bergmann, T., et al.\ 1996, ApJS, 105, 93

\bibitem[1993]{fitt+alexander93}
Fitt, A.~J., \& Alexander, P. 1993, MNRAS, 261, 445

\bibitem[1994a]{forbes+94a}
Forbes, D.~A., Norris, R.~P., Williger, G.~M., \& Smith, R.~C.
1994a, AJ, 107, 984

\bibitem[1994b]{forbes+94b}
Forbes, D.~A., Kotilainen, J.~K., \& Moorwood, A.~F.~M. 1994b, ApJ, 433, L13

\bibitem[2001]{frick+01}
Frick, P., Beck, R., Berkhuijsen, E.~M., \& Patrickeyev, I. 2001, MNRAS, 327, 1145

\bibitem[1989]{fullmer+lonsdale89}
Fullmer, L., \& Lonsdale, C. 1989, Cataloged Galaxies and Quasars
Observed in the IRAS Survey, Version 2, JPL D--1932 (Pasadena: JPL)

\bibitem[1993]{garcia+93}
Garc\'{\i}a-Barreto, J.~A., Carrillo, R., Klein, U., \& Dahlem, M.
1993, Rev. Mex. Astron. Astrofis., 25, 31

\bibitem[1988]{gerin+88}
Gerin, M., Nakai, N., \& Combes, F. 1988, A\&A, 203, 44

\bibitem[2002]{harnett+02}
Harnett, J.~I., Ehle, M., Beck, R., Thierbach, M., Haynes, R.F. 2002,
in prep. ({\bf Paper III})

\bibitem[1998]{hoernes+98}
Hoernes, P., Berkhuijsen, E.~M., \& Xu, C. 1998, A\&A, 334, 57

\bibitem[1981]{hummel81}
Hummel, E. 1981, A\&A, 93, 93

\bibitem[1995]{hummel+beck95}
Hummel, E., \& Beck, R. 1995, A\&A, 303, 691

\bibitem[1987]{hummel+87}
Hummel, E., van der Hulst, J.~M., \& Keel, W.~C. 1987, A\&A, 172, 32

\bibitem[1988]{hummel+88}
Hummel, E., Davies, R.~D., Wolstencroft, R.~D., van der Hulst, J.~M., \&
Pedlar, A. 1988, A\&A, 199, 91

\bibitem[1995]{joersaeter+moorsel95}
J\"ors\"ater, S., \& van Moorsel, G.~A. 1995, AJ, 110, 2037

\bibitem[1997]{kristen+97}
Kristen, H., J\"ors\"ater, S., Lindblad, P.~O., \& Boksenberg, A. 1997, A\&A, 328, 483

\bibitem[1998]{laine+gottesman98}
Laine, S., \& Gottesman, S.~T. 1998, MNRAS, 297, 1041

\bibitem[1981]{laing81}
Laing, R. 1981, ApJ, 248, 87

\bibitem[2002]{laing02}
Laing, R. 2002, MNRAS, 329, 417

\bibitem[1976]{landau76}
Landau, L.~D., \& Lifshitz, E.~M. 1976, The Classical Theory
of Fields, \S41 (Oxford: Pergamon)

\bibitem[1999]{lindblad99}
Lindblad, P.~O. 1999, A\&A Rev, 9, 221

\bibitem[1997]{ma+97}
Ma, J., Peng, Q.-H., Chen, R., Ji, Z.-H., \& Tu, C.-P. 1997, A\&AS, 126,
503

\bibitem[1998]{ma+98}
Ma, J., Peng, Q.-H., \& Gu, Q.-S. 1998, A\&AS, 130, 449

\bibitem[1999]{macri+99}
Macri, L.~M., Huchra, J.~P., Stetson, P.~B., et al.\ 1999, ApJ, 521, 155

\bibitem[1998]{madore+98}
Madore, B.~F., Freedman, W.~L., Silbermann, N., et al.\ 1998, Nature, 395, 47

\bibitem[1996]{maoz+96}
Maoz, D., Barth, A.~J., Sternberg, A., et al.\ 1996, AJ, 111, 2248

\bibitem[1995]{martin95}
Martin, P. 1995, AJ, 109, 2428

\bibitem[1997]{martin+friedli97}
Martin, P., \& Friedli, D. 1997, A\&A, 326, 449

\bibitem[1997]{martinet+friedli97}
Martinet, L., \& Friedli, D. 1997, A\&A, 323, 363

\bibitem[1991]{mestel+91}
Mestel, L., \& Subramanian, K. 1991, MNRAS, 248, 677

\bibitem[1997]{mihos+bothun97}
Mihos, J.~C., \& Bothun, G.~D. 1997, ApJ, 481, 741

\bibitem[2001]{moellenhoff+01}
M\"ollenhoff, C., \& Heidt, J. 2001, A\&A, 368, 16

\bibitem[1998]{moss+98}
Moss, D., Korpi, M., Rautiainen, P., \& Salo, H. 1998, A\&A, 329, 895

\bibitem[2001]{moss+01}
Moss, D., Shukurov, A., Sokoloff, D., Beck, R., \& Fletcher, A. 2001,
A\&A, 380, 55 ({\bf Paper II})

\bibitem[1991]{neininger+91}
Neininger, N., Klein, U., Beck, R., \& Wielebinski, R. 1991, Nature,
352, 781

\bibitem[1995]{niklas95}
Niklas, S. 1995, PhD Thesis, University of Bonn

\bibitem[1997]{niklas97}
Niklas, S. 1997, A\&A, 322, 29

\bibitem[1997]{niklas+beck97}
Niklas, S., \& Beck, R. 1997, A\&A, 320, 54

\bibitem[1997]{niklas+97}
Niklas, S., Klein, U., \& Wielebinski, R. 1997, A\&A, 322, 19

\bibitem[1985]{ondrechen85}
Ondrechen, M.~P. 1985, AJ, 90, 1474

\bibitem[1983]{ondrechen+83}
Ondrechen, M.~P., \& van der Hulst, J.~M. 1983, ApJ, 269, L47

\bibitem[1989]{ondrechen+89}
Ondrechen, M.~P., van der Hulst, J.~M., \& Hummel, E. 1989, ApJ, 342, 39

\bibitem[1994]{quillen+94}
Quillen, A.~C., Frogel, J.~A., \& Gonz\'alez, R.~A. 1994, ApJ, 437, 162

\bibitem[1998]{reynaud+downes98}
Reynaud, D., \& Downes, D. 1998, A\&A, 337, 671

\bibitem[1999]{rohde+99}
Rohde, R., Beck, R., \& Elstner, D. 1999, A\&A, 350, 423

\bibitem[2001a]{roussel+01a}
Roussel, H., Vigroux, L., Bosma, A., et al.\ 2001a, A\&A, 369, 473

\bibitem[2001b]{roussel+01b}
Roussel, H., Sauvage, M., Vigroux, L., et al.\ 2001b, A\&A, 372, 406

\bibitem[2001]{ryder+01}
Ryder, S.~D., Koribalski, B., Staveley-Smith, L., et al.\ 2001, ApJ, 555, 232

\bibitem[1981]{sandage+tammann81}
Sandage, A., \& Tammann, G.~A. 1981, A Revised Shapley-Ames Catalog of
Bright Galaxies (Washington: Carnegie Inst.)

\bibitem[1995]{sandqvist+95}
Sandqvist, Aa., J\"ors\"ater, S., \& Lindblad, P.~O. 1995, A\&A, 295, 585

\bibitem[2001]{soida+01}
Soida, M., Urbanik, M., Beck, R., Wielebinski, R., \& Balkowski, C. 2001,
A\&A, 378,  40

\bibitem[1998]{sokoloff+98}
Sokoloff, D.D., Bykov, A.A., Shukurov, A., Berkhuijsen, E.M., Beck, R., \&
Poezd, A.D. 1998, MNRAS, 299, 189. Erratum: 1999, MNRAS, 303, 207

\bibitem[1997]{storch+97}
Storchi-Bergmann, T., Eracleous, M., Ruiz, M.~T., Livio, M., Wilson, A.~S.,
\& Filippenko, A.~V. 1997, ApJ, 489, 87

\bibitem[2000]{strong+00}
Strong, A.~W., Moskalenko, I.~V., \& Reimer, O. 2000, ApJ, 537, 763

\bibitem[1982]{tubbs82}
Tubbs, A.~D. 1982, ApJ, 255, 458

\bibitem[1989]{unger+89}
Unger, S.~W., Wolstencroft, R.~D., Pedlar, A., et al.\ 1989, MNRAS, 236, 425

\bibitem[1970]{whiteoak70}
Whiteoak, J.~B. 1970, Astrophys. Lett., 5, 29

\bibitem[2000]{wilke+00}
Wilke, K., M\"ollenhoff, C., \& Matthias, M. 2000, A\&A, 361, 507

\bibitem[1993]{young93}
Young, J.~S. 1993, in Star Formation, Galaxies and the Interstellar Medium,
eds. J.~Franco, F.~Ferrini \& G.~Tenorio-Tagle (Cambridge: Univ.~Press), 318

\bibitem[1989]{young+89}
Young, J.~S., Xie, S., Kenney, J.~D.~P., \& Rice, W.~L. 1989, ApJS, 70,
699

\bibitem[1997]{zaritsky+97}
Zaritsky, D., Smith, R., Frenk, C., \& White, S.~D.~M. 1997, ApJ, 478, 39


\end{thebibliography}
\end{document}